\documentclass[twocolumn]{aastex62}
\usepackage{float,graphicx,amsmath,multirow,listings}
\usepackage{color}
\usepackage[version=4]{mhchem}
\usepackage[T1]{fontenc}
\hypersetup{
    colorlinks = True,
    linkcolor=black
}
\usepackage{tabularx}
\usepackage{ragged2e}

\begin{document}

\title{2018 Census of Interstellar, Circumstellar, Extragalactic, Protoplanetary Disk, and Exoplanetary Molecules}
\author{Brett A. McGuire}
\altaffiliation{B.A.M. is a Hubble Fellow of the National Radio Astronomy Observatory.}
\affiliation{National Radio Astronomy Observatory, Charlottesville, VA 22903, USA}
\affiliation{Harvard-Smithsonian Center for Astrophysics, Cambridge, MA 02138, USA}

\begin{abstract}

To date, 204 individual molecular species, comprised of 16 different elements, have been detected in the interstellar and circumstellar medium by astronomical observations.  These molecules range in size from two atoms to seventy, and have been detected across the electromagnetic spectrum from cm-wavelengths to the ultraviolet.  This census presents a summary of the first detection of each molecular species, including the observational facility, wavelength range, transitions, and enabling laboratory spectroscopic work, as well as listing tentative and disputed detections. Tables of molecules detected in interstellar ices, external galaxies, protoplanetary disks, and exoplanetary atmospheres are provided.  A number of visual representations of this aggregate data are presented and briefly discussed in context.

\end{abstract}

\section{Introduction}
\label{intro}

Since the detection of the methylidyne (CH), the first molecule identified in the interstellar medium (ISM) \citep{Swings:1937dl,Dunham:1937nj,McKellar:1940io}, observations of molecules have played crucial roles in a wide range of applications, from broadening our understanding of interstellar chemical evolution \citep{Herbst:2009go} and the formation of planets \citep{Oberg:2011je} to providing exceptional astrophysical probes of physical conditions and processes \citep{Friesen:2013ii}.  Beginning in the early 1960s, the advent of radio astronomy enabled a boom in the detection of new molecules, a trend which has continued at a nearly linearly rate ever since.  

Yet, despite the remarkably steady (and perhaps even accelerating) pace of new molecular detections, the number of as yet unidentified spectral features attributable to molecules is staggering.  Indeed, these mysteries extend across the electromagnetic spectrum.  In the radio, high-sensitivity, broad-band spectral line surveys continue to reveal hundreds of features not assignable to transitions of molecules in spectroscopic databases \citep{Cernicharo:2013cc}, although there is evidence that a non-trivial number of these features may be due to transitions of vibrationally-excited or rare isotopic species that have not been completely catalogued \citep{Fortman:2012is}.

At shorter wavelengths, the presence of the unidentified infrared emission bands (UIRs) -- sharp, distinct emission features ubiquitous in ultraviolet (UV)-irradiated regions in our galaxy as well as seen in dozens of external galaxies -- continues to elude conclusive molecular identification.  A substantial body of literature now seems to favor the assignment of these emission features to polyclic aromatic hydrocarbons (PAHs) \citep{Tielens:2008fx}, or at the very least $sp^2$-hybridized aromatic carbon structures.  The possibility that these features arise from mixed aromatic/aliphatic organic nanoparticles has also been raised \citep{Kwok:2013gq}.  Regardless, it remains that no individual molecule has been definitely identified from UIR features.

A yet older mystery are the diffuse interstellar bands (DIBs), first discovered by Mary Lea Heger in 1922 \citep{McCall:2012db}.  Despite nearly a century of observation and laboratory work, these sharp features seen from the IR to the UV remain nearly completely unassigned.  \citet{Campbell:2015hp} recently reported the first attribution of a molecular carrier to a DIB that has not been disputed in the literature, that of \ce{C60+}.  Yet, hundreds of DIBs remain to be identified.

In the following sections, known, tentatively detected, and disputed interstellar, circumstellar, extragalactic, and exoplanetary molecules are catalogued and presented.  The major lists in this paper are as follows (names are in-document hyperlinks in most PDF viewers):

\begin{enumerate}
    \item \hyperref[known]{List of known interstellar and circumstellar molecules}
    \begin{enumerate}
        \item \hyperref[tentative]{List of tentative detections}
        \item \hyperref[disputed]{List of disputed detections}
    \end{enumerate}
    \item \hyperref[exgals]{List of molecules detected in external galaxies}
    \item \hyperref[ices]{List of molecules detected in interstellar ices}
    \item \hyperref[ppds]{List of molecules detected in protoplanetary disks}
\end{enumerate}

The primary list is that of the known interstellar and circumstellar molecules, and for each species in this list, a best-effort attempt was made to locate the first reported detection or detections of the species in the literature.  The detection source or sources and instrument are given, and for some species a short description of any particularly noteworthy attributes is given as well. When available, a reference to the enabling laboratory spectroscopy work cited in the detection paper is provided.  In the cases where the detection paper uses frequencies without apparent attribution to the laboratory spectroscopy, the most recent effort in the literature that would reasonably have been available at the time of publication of the detection is provided on a best-effort basis.

The classification of a molecule as detected, tentatively detected, or disputed was made as agnostically as possible.  Molecules are listed as detected if a literature source has claimed that detection, did not self-identify that detection as tentative, and no subsequent literature could be found that has disputes that claim.  Tentative detections are listed only if the source has claimed the detection as such.  Detections are considered disputed until a literature source has claimed that dispute has been resolved.

Subsets of the lists presented in this work are available both online and in numerous publications.  Three exceptional web-based resources are of particular note with respect to the list of interstellar and circumstellar detections: the Cologne Database for Molecular Spectroscopy's \emph{List of Molecules in Space} (H.S.P. M\"{u}ller)\footnote{\url{https://www.astro.uni-koeln.de/cdms/molecules}}, The Astrochymist's \emph{A Bibliography of Astromolecules} (D. Woon)\footnote{\url{http://www.astrochymist.org/astrochymist_ism.html}}, and the list of M. Araki\footnote{\url{http://www.rs.kagu.tus.ac.jp/tsukilab/research_seikanlist.html}}.  As of the time of publication, these resources are providing the most frequently updated lists of known interstellar and circumstellar species.  This publication is intended as a complement, rather than a supplement, to these resources.

This document is intended to be updated on an annual or semi-annual basis.\footnote{Suggested updates, corrections, and comments can be directed to bmcguire@nrao.edu.} A Python 3 script containing the full bibliographical and meta data information for all figures, tables, and molecules discussed here is available as Supplementary Information.  The latest version of the script, updated more frequently than this document, can always be found at \href{https://github.com/bmcguir2/astromolecule_census}{\url{https://github.com/bmcguir2/astromolecule_census}}.   A brief description of the script and some of its function is provided in Appendix~\ref{app:script}.

Tables \ref{two_seven} and \ref{eight_more} summarize the current list of known interstellar and circumstellar molecules.  The column headers and individual molecule entries are active in-document hyperlinks in most PDF viewers.  A list of commonly-used abbreviations for facilities is given in Table~\ref{abbrevs}.  In \S\ref{statistics}, some aggregate analysis of the data is presented.  These figures are available as high-quality supplemental data files in PDF and PNG formats. 

\begin{table}[h!]
    \centering
    \scriptsize
    \caption{Commonly-used facility abbreviations.}
    \begin{tabular*}{\columnwidth}{l @{\extracolsep{\fill}} l}
    \hline\hline
    Abbreviation    &   Description \\
    \hline
    ALMA            &   Atacama Large Millimeter/submillimeter Array \\
    APEX            &   Atacama Pathfinder Experiment           \\
    ARO             &   Arizona Radio Observatory               \\
    ATCA            &   Australian Telescope Compact Array      \\
    BIMA            &   Berkeley-Illinois-Maryland Array        \\
    CSO             &   Caltech Submillimeter Observatory       \\
    FCRAO           &   Five College Radio Astronomy Observatory           \\
    FUSE            &   \emph{Far Ultraviolet Spectroscopic Explorer}   \\
    GBT             &   Green Bank Telescope                    \\
    IRAM            &   Institut de Radioastronomie Millim\'{e}trique \\
    IRTF            &   Infrared Radio Telescope Facility       \\
    ISO             &   \emph{Infrared Space Observatory}       \\
    KPNO            &   Kitt Peak National Observatory          \\
    MWO             &   Millimeter-wave Observatory             \\   
    NRAO            &   National Radio Astronomy Observatory    \\   
    OVRO            &   Owens Valley Radio Observatory          \\
    PdBI            &   Plateau de Bure Interferometer          \\
    SEST            &   Swedish-ESO Submillimeter Telescope     \\
    SMA             &   Sub-millimeter Array                    \\
    SMT             &   Sub-millimeter Telescope                \\
    SOFIA           &   \emph{Stratospheric Observatory for Infrared Astronomy} \\
    UKIRT           &   United Kingdom Infrared Telescope       \\
    \hline
    \end{tabular*}

    \label{abbrevs}
\end{table}

\begin{table*}
\centering
\caption{List of detected interstellar molecules with two to seven atoms, categorized by number of atoms, and vertically ordered by detection year.  Column headers and molecule formulas are in-document hyperlinks in most PDF viewers.}
\begin{tabular*}{\textwidth}{l l @{\extracolsep{\fill}}  l @{\extracolsep{\fill}}  l  l   l @{\extracolsep{\fill}}  l @{\extracolsep{\fill}} l}
\hline\hline
\multicolumn{2}{l}{\hyperref[2atoms]{2 Atoms}} &\multicolumn{2}{l}{\hyperref[3atoms]{3 Atoms}}& \hyperref[4atoms]{4 Atoms} & \hyperref[5atoms]{5 Atoms} & \hyperref[6atoms]{6 Atoms} & \hyperref[7atoms]{7 Atoms} \\
\hline
\hyperref[CH]{\ce{CH}}	&	\hyperref[CP]{\ce{CP}}	&	\hyperref[H2O]{\ce{H2O}}	&	\hyperref[N2O]{\ce{N2O}}	&	\hyperref[NH3]{\ce{NH3}}	&	\hyperref[HC3N]{\ce{HC3N}}	&	\hyperref[CH3OH]{\ce{CH3OH}}	&	\hyperref[CH3CHO]{\ce{CH3CHO}}	\\
\hyperref[CN]{\ce{CN}}	&	\hyperref[NH]{\ce{NH}}	&	\hyperref[HCO+]{\ce{HCO+}}	&	\hyperref[MgCN]{\ce{MgCN}}	&	\hyperref[H2CO]{\ce{H2CO}}	&	\hyperref[HCOOH]{\ce{HCOOH}}	&	\hyperref[CH3CN]{\ce{CH3CN}}	&	\hyperref[CH3CCH]{\ce{CH3CCH}}	\\
\hyperref[CH+]{\ce{CH+}}	&	\hyperref[SiN]{\ce{SiN}}	&	\hyperref[HCN]{\ce{HCN}}	&	\hyperref[H3+]{\ce{H3+}}	&	\hyperref[HNCO]{\ce{HNCO}}	&	\hyperref[CH2NH]{\ce{CH2NH}}	&	\hyperref[NH2CHO]{\ce{NH2CHO}}	&	\hyperref[CH3NH2]{\ce{CH3NH2}}	\\
\hyperref[OH]{\ce{OH}}	&	\hyperref[SO+]{\ce{SO+}}	&	\hyperref[OCS]{\ce{OCS}}	&	\hyperref[SiCN]{\ce{SiCN}}	&	\hyperref[H2CS]{\ce{H2CS}}	&	\hyperref[NH2CN]{\ce{NH2CN}}	&	\hyperref[CH3SH]{\ce{CH3SH}}	&	\hyperref[CH2CHCN]{\ce{CH2CHCN}}	\\
\hyperref[CO]{\ce{CO}}	&	\hyperref[CO+]{\ce{CO+}}	&	\hyperref[HNC]{\ce{HNC}}	&	\hyperref[AlNC]{\ce{AlNC}}	&	\hyperref[C2H2]{\ce{C2H2}}	&	\hyperref[H2CCO]{\ce{H2CCO}}	&	\hyperref[C2H4]{\ce{C2H4}}	&	\hyperref[HC5N]{\ce{HC5N}}	\\
\hyperref[H2]{\ce{H2}}	&	\hyperref[HF]{\ce{HF}}	&	\hyperref[H2S]{\ce{H2S}}	&	\hyperref[SiNC]{\ce{SiNC}}	&	\hyperref[C3N]{\ce{C3N}}	&	\hyperref[C4H]{\ce{C4H}}	&	\hyperref[C5H]{\ce{C5H}}	&	\hyperref[C6H]{\ce{C6H}}	\\
\hyperref[SiO]{\ce{SiO}}	&	\hyperref[N2]{\ce{N2}}	&	\hyperref[N2H+]{\ce{N2H+}}	&	\hyperref[HCP]{\ce{HCP}}	&	\hyperref[HNCS]{\ce{HNCS}}	&	\hyperref[SiH4]{\ce{SiH4}}	&	\hyperref[CH3NC]{\ce{CH3NC}}	&	\hyperref[c-C2H4O]{\ce{c-C2H4O}}	\\
\hyperref[CS]{\ce{CS}}	&	\hyperref[CF+]{\ce{CF+}}	&	\hyperref[C2H]{\ce{C2H}}	&	\hyperref[CCP]{\ce{CCP}}	&	\hyperref[HOCO+]{\ce{HOCO+}}	&	\hyperref[c-C3H2]{\ce{c-C3H2}}	&	\hyperref[HC2CHO]{\ce{HC2CHO}}	&	\hyperref[CH2CHOH]{\ce{CH2CHOH}}	\\
\hyperref[SO]{\ce{SO}}	&	\hyperref[PO]{\ce{PO}}	&	\hyperref[SO2]{\ce{SO2}}	&	\hyperref[AlOH]{\ce{AlOH}}	&	\hyperref[C3O]{\ce{C3O}}	&	\hyperref[CH2CN]{\ce{CH2CN}}	&	\hyperref[H2C4]{\ce{H2C4}}	&	\hyperref[C6H-]{\ce{C6H-}}	\\
\hyperref[SiS]{\ce{SiS}}	&	\hyperref[O2]{\ce{O2}}	&	\hyperref[HCO]{\ce{HCO}}	&	\hyperref[H2O+]{\ce{H2O+}}	&	\hyperref[l-C3H]{\ce{l-C3H}}	&	\hyperref[C5]{\ce{C5}}	&	\hyperref[C5S]{\ce{C5S}}	&	\hyperref[CH3NCO]{\ce{CH3NCO}}	\\
\hyperref[NS]{\ce{NS}}	&	\hyperref[AlO]{\ce{AlO}}	&	\hyperref[HNO]{\ce{HNO}}	&	\hyperref[H2Cl+]{\ce{H2Cl+}}	&	\hyperref[HCNH+]{\ce{HCNH+}}	&	\hyperref[SiC4]{\ce{SiC4}}	&	\hyperref[HC3NH+]{\ce{HC3NH+}}	&	\hyperref[HC5O]{\ce{HC5O}}	\\
\hyperref[C2]{\ce{C2}}	&	\hyperref[CN-]{\ce{CN-}}	&	\hyperref[HCS+]{\ce{HCS+}}	&	\hyperref[KCN]{\ce{KCN}}	&	\hyperref[H3O+]{\ce{H3O+}}	&	\hyperref[H2CCC]{\ce{H2CCC}}	&	\hyperref[C5N]{\ce{C5N}}	&		\\
\hyperref[NO]{\ce{NO}}	&	\hyperref[OH+]{\ce{OH+}}	&	\hyperref[HOC+]{\ce{HOC+}}	&	\hyperref[FeCN]{\ce{FeCN}}	&	\hyperref[C3S]{\ce{C3S}}	&	\hyperref[CH4]{\ce{CH4}}	&	\hyperref[HC4H]{\ce{HC4H}}	&		\\
\hyperref[HCl]{\ce{HCl}}	&	\hyperref[SH+]{\ce{SH+}}	&	\hyperref[SiC2]{\ce{SiC2}}	&	\hyperref[HO2]{\ce{HO2}}	&	\hyperref[c-C3H]{\ce{c-C3H}}	&	\hyperref[HCCNC]{\ce{HCCNC}}	&	\hyperref[HC4N]{\ce{HC4N}}	&		\\
\hyperref[NaCl]{\ce{NaCl}}	&	\hyperref[HCl+]{\ce{HCl+}}	&	\hyperref[C2S]{\ce{C2S}}	&	\hyperref[TiO2]{\ce{TiO2}}	&	\hyperref[HC2N]{\ce{HC2N}}	&	\hyperref[HNCCC]{\ce{HNCCC}}	&	\hyperref[c-H2C3O]{\ce{c-H2C3O}}	&		\\
\hyperref[AlCl]{\ce{AlCl}}	&	\hyperref[SH]{\ce{SH}}	&	\hyperref[C3]{\ce{C3}}	&	\hyperref[CCN]{\ce{CCN}}	&	\hyperref[H2CN]{\ce{H2CN}}	&	\hyperref[H2COH+]{\ce{H2COH+}}	&	\hyperref[CH2CNH]{\ce{CH2CNH}}	&		\\
\hyperref[KCl]{\ce{KCl}}	&	\hyperref[TiO]{\ce{TiO}}	&	\hyperref[CO2]{\ce{CO2}}	&	\hyperref[SiCSi]{\ce{SiCSi}}	&	\hyperref[SiC3]{\ce{SiC3}}	&	\hyperref[C4H-]{\ce{C4H-}}	&	\hyperref[C5N-]{\ce{C5N-}}	&		\\
\hyperref[AlF]{\ce{AlF}}	&	\hyperref[ArH+]{\ce{ArH+}}	&	\hyperref[CH2]{\ce{CH2}}	&	\hyperref[S2H]{\ce{S2H}}	&	\hyperref[CH3]{\ce{CH3}}	&	\hyperref[CNCHO]{\ce{CNCHO}}	&	\hyperref[HNCHCN]{\ce{HNCHCN}}	&		\\
\hyperref[PN]{\ce{PN}}	&	\hyperref[NS+]{\ce{NS+}}	&	\hyperref[C2O]{\ce{C2O}}	&	\hyperref[HCS]{\ce{HCS}}	&	\hyperref[C3N-]{\ce{C3N-}}	&	\hyperref[HNCNH]{\ce{HNCNH}}	&	\hyperref[SiH3CN]{\ce{SiH3CN}}	&		\\
\hyperref[SiC]{\ce{SiC}}	&		&	\hyperref[MgNC]{\ce{MgNC}}	&	\hyperref[HSC]{\ce{HSC}}	&	\hyperref[PH3]{\ce{PH3}}	&	\hyperref[CH3O]{\ce{CH3O}}	&		&		\\
	&		&	\hyperref[NH2]{\ce{NH2}}	&	\hyperref[NCO]{\ce{NCO}}	&	\hyperref[HCNO]{\ce{HCNO}}	&	\hyperref[NH3D+]{\ce{NH3D+}}	&		&		\\
	&		&	\hyperref[NaCN]{\ce{NaCN}}	&		&	\hyperref[HOCN]{\ce{HOCN}}	&	\hyperref[H2NCO+]{\ce{H2NCO+}}	&		&		\\
	&		&		&		&	\hyperref[HSCN]{\ce{HSCN}}	&	\hyperref[NCCNH+]{\ce{NCCNH+}}	&		&		\\
	&		&		&		&	\hyperref[HOOH]{\ce{HOOH}}	&	\hyperref[CH3Cl]{\ce{CH3Cl}}	&		&		\\
	&		&		&		&	\hyperref[l-C3H+]{\ce{l-C3H+}}	&		&		&		\\
	&		&		&		&	\hyperref[HMgNC]{\ce{HMgNC}}	&		&		&		\\
	&		&		&		&	\hyperref[HCCO]{\ce{HCCO}}	&		&		&		\\
	&		&		&		&	\hyperref[CNCN]{\ce{CNCN}}	&		&		&		\\
\hline
\end{tabular*}
\label{two_seven}
\end{table*}

\begin{table*}
\centering
\caption{List of detected interstellar molecules with eight or more atoms, categorized by number of atoms, and vertically ordered by detection year.  Column headers and molecule formulas are in-document hyperlinks in most PDF viewers.}
\begin{tabular*}{\textwidth}{l @{\extracolsep{\fill}}  l @{\extracolsep{\fill}}  l  @{\extracolsep{\fill}} l   @{\extracolsep{\fill}} l @{\extracolsep{\fill}}  l @{\extracolsep{\fill}} l}
\hline\hline
\hyperref[8atoms]{8 Atoms} & \hyperref[9atoms]{9 Atoms} & \hyperref[10atoms]{10 Atoms} & \hyperref[11atoms]{11 Atoms} & \hyperref[12atoms]{12 Atoms} & \hyperref[13atoms]{13 Atoms} & \hyperref[fullerenes]{Fullerenes}\\
\hline
\hyperref[HCOOCH3]{\ce{HCOOCH3}}	&	\hyperref[CH3OCH3]{\ce{CH3OCH3}}	&	\hyperref[acetone]{\ce{(CH3)2CO}}	&	\hyperref[HC9N]{\ce{HC9N}}	&	\hyperref[C6H6]{\ce{C6H6}}	&	\hyperref[C6H5CN]{\ce{c-C6H5CN}}	&	\hyperref[C60]{\ce{C60}}	\\
\hyperref[CH3C3N]{\ce{CH3C3N}}	&	\hyperref[CH3CH2OH]{\ce{CH3CH2OH}}	&	\hyperref[HOCH2CH2OH]{\ce{HO(CH2)2OH}}	&	\hyperref[CH3C6H]{\ce{CH3C6H}}	&	\hyperref[n-C3H7CN]{\ce{n-C3H7CN}}	&		&	\hyperref[C60+]{\ce{C60+}}	\\
\hyperref[C7H]{\ce{C7H}}	&	\hyperref[CH3CH2CN]{\ce{CH3CH2CN}}	&	\hyperref[CH2CH2CHO]{\ce{CH2CH2CHO}}	&	\hyperref[CH3CH2OCHO]{\ce{CH3CH2OCHO}}	&	\hyperref[i-C3H7CN]{\ce{i-C3H7CN}}	&		&	\hyperref[C70]{\ce{C70}}	\\
\hyperref[CH3COOH]{\ce{CH3COOH}}	&	\hyperref[HC7N]{\ce{HC7N}}	&	\hyperref[CH3C5N]{\ce{CH3C5N}}	&	\hyperref[CH3COOCH3]{\ce{CH3COOCH3}}	&		&		&		\\
\hyperref[H2C6]{\ce{H2C6}}	&	\hyperref[CH3C4H]{\ce{CH3C4H}}	&	\hyperref[CH3CHCH2O]{\ce{CH3CHCH2O}}	&		&		&		&		\\
\hyperref[glycolaldehyde]{\ce{CH2OHCHO}}	&	\hyperref[C8H]{\ce{C8H}}	&	\hyperref[CH3OCH2OH]{\ce{CH3OCH2OH}}	&		&		&		&		\\
\hyperref[HC6H]{\ce{HC6H}}	&	\hyperref[acetamide]{\ce{CH3CONH2}}	&		&		&		&		&		\\
\hyperref[CH2CHCHO]{\ce{CH2CHCHO}}	&	\hyperref[C8H-]{\ce{C8H-}}	&		&		&		&		&		\\
\hyperref[CH2CCHCN]{\ce{CH2CCHCN}}	&	\hyperref[CH2CHCH3]{\ce{CH2CHCH3}}	&		&		&		&		&		\\
\hyperref[NH2CH2CN]{\ce{NH2CH2CN}}	&	\hyperref[CH3CH2SH]{\ce{CH3CH2SH}}	&		&		&		&		&		\\
\hyperref[CH3CHNH]{\ce{CH3CHNH}}	&	\hyperref[HC7O]{\ce{HC7O}}	&		&		&		&		&		\\
\hyperref[CH3SiH3]{\ce{CH3SiH3}}	&		&		&		&		&		&		\\
\hline
\end{tabular*}
\label{eight_more}
\end{table*}

\section{Detection Techniques}
\label{detecting}

As will be seen in the following pages, the vast majority of new molecule detections ($\sim$80\%) have been made using radio astronomy techniques in the cm, mm, and sub-mm wavelength ranges.  The manuscript, and indeed this section, is therefore inherently biased toward radio astronomy.  To understand this bias, it is helpful to review the techniques, and their limitations, used to detect molecules across various wavelength ranges.  In the end, it comes down the energy required, and that which is available, to populate molecular energy levels for undergoing transitions at various wavelengths.  This in turn dictates the environments and sightlines in which detections can be accomplished.

\subsection{Radio Astronomy}
\label{detecting:radio}

Here, radio astronomical observations are roughly defined as those stretching from centimeter to far-infrared (100's of $\mu$m) wavelengths, or below $\sim$2~THz (in frequency).  The primary molecular signal arising in these regions is that of the rotational motion of molecules.  For a linear molecule (symmetric and asymmetric tops follow similar, if more complex patterns), the rotational energy levels $J$ are given, to first order, by Equation~\ref{linear_energy}:
\begin{equation}
E_J = BJ(J+1),
\label{linear_energy}
\end{equation}
where $B$ is the rotational constant of the molecule.  $B$ is inversely proportional to the moment of inertia of the molecule ($I$):
\begin{equation}
B = \frac{h^2}{8\pi^2I},
\label{linear_B}
\end{equation}
where $I$ is related to the reduced mass of the molecule ($\mu$) and the radial extent of the mass ($r$):
\begin{equation}
I = \mu r^2.
\end{equation}
Thus, lighter, smaller molecules will have large values of $B$ (e.g. $B$[NH] = 490~GHz; \citealt{Klaus:1997ku}) and heavier, larger molecules will have very small values of $B$ (e.g. $B$[\ce{HC9N}] =  0.29~GHz; \citealt{McCarthy:2000eo}).  The frequencies of rotational transitions are given by the difference in energy levels (Equation~\ref{linear_energy}). Small molecules with large values of $B$, and correspondingly widely spaced energy levels, will have transitions at higher frequencies.  Larger molecules, with smaller values of $B$, have closely-spaced energy levels with transitions at lower frequencies.

Assuming \ce{NH} and \ce{HC9N} are reasonable examples of the range of sizes of typical interstellar molecules, the ground-state rotational transitions ($J = 1\rightarrow0$) therefore fall between $\sim$1~THz (NH) and $\sim$500~MHz (\ce{HC9N}).   As a result, observations of the pure-rotational transitions of molecules are most often (nearly exclusively) conducted with radio astronomy facilities.

The absolute energies of these rotational levels also dictate the manner and environments in which they may be observed.  The equivalent $kT$ energy of most rotational levels of interstellar species is typically $<1000$~K.  In sufficiently dense environments, the distribution of energy between these levels is governed by collisions with other gas particles, a condition commonly referred to as local thermodynamic equilibrium (LTE), and is thus proportional (or equal to) to kinetic temperature of the gas.  For warm regions such as hot cores, sufficient energy is available to produce a population distribution (excitation temperature, $T_{ex}$) well above the background radiation field, permitting the observation of emission from molecules undergoing rotational transitions (see, e.g. \citealt{Belloche:2013eba}).  Even in very cold regions, like TMC-1 ($T_{ex}$~=~5--10~K), sufficient energy is still present to populate the lowest few rotational energy levels, and emission can still be seen (see, e.g., \citealt{Kaifu:2004tk}).  Conversely, when $T_{ex}$ is below the background radiation field temperature (typically black/greybody dust emission), molecules can be observed in absorption against this continuum (see, e.g. SH; \citealt{Neufeld:2012gz}).    Thus, a major advantage of radio observations is the ability to observe molecules in virtually any source, either in emission (due to the low energy needed to populate rotational levels) or in absorption against background continuum.

The major weakness of this technique is that it is blind to completely symmetric molecules (and largely blind to those that are highly symmetric).  The strength of a rotational transition is proportional to the square of the permanent electric dipole moment of the molecule, and thus highly-symmetric molecules often possess either very weak or no allowed pure rotational transitions (e.g. \ce{CH4}).  Further, as the molecules increase in size and complexity, the number of accessible energy levels increases substantially, and thus the partitioning of population among these levels results in the number of molecules emitting/absorbing at a given frequency being decreased dramatically.  Finally, molecules trapped in the condensed phase cannot undergo free rotation, making their detection by rotational spectroscopy in the radio impossible.

\subsection{Infrared Astronomy}
\label{detecting:infrared}

Pure rotational transitions of molecules in the radio occur within a single vibrational state of a molecule.  This is often, but not exclusively, the ground vibrational state (see, e.g., \citealt{Goldsmith:1983bc}).  Transitions can also occur between vibrational states, the energies of which normally fall between 100s and many 1000s of cm$^{-1}$ (see, e.g., \ce{CO2} \citealt{Stull:1962cl}), and the resulting wavelengths of light between a few and a few hundred $\mu$m (i.e. the near to far infrared).  Because these energy levels are substantially higher than those of pure rotational transitions, observations of emission from vibrational transitions usually requires exceptionally warm regions or a radiative pumping mechanism to populate the levels.  A common example of this effect is the aforementioned UIR features, which are often attributed to the infrared vibrational transitions of PAHs that have been pumped by an external, enhanced UV radiation field \citep{Tielens:2008fx}.  

Absent these extraordinary excitation conditions, however, observation of vibrational transitions of molecules in the IR requires a background radiation source for the molecules to be seen against in absorption.  This places a number of limitations on the breadth of environments that can be probed with IR astronomy, as these sightlines must not only fortuitously contain a background source, but also be optically thin enough to allow for the transmission of that light through the source. Even given these limitations, IR astronomy has been an extraordinary useful tool for molecular astrophysics, particularly in the identification of molecules which are highly-symmetric and lack strong or non-zero permanent dipole moments.

Beyond the identification of new species, however, a substantial benefit of IR astronomy is the ability to observe molecules condensed into interstellar ices.  While condensed-phase molecules cannot undergo free rotation, their vibrational transitions remain accessible.  These are largely seen in absorption, as the energy required to excite a transition into emission is likely to simply cause the molecule to desorb into the gas phase.  As well, the vibrational motions are not unaffected by the condensed-phase environment.  Both the physical structure of the solid and the molecular content of the surrounding material can alter the frequencies and lineshapes of vibrational modes in small and large ways (see, e.g., \citealt{Cooke:2016hw}).  That said, these changes are often readily observable and quantifiable in the laboratory, and as a result, IR astronomy has provided a unique window into the molecular content of condensed-phase materials in the ISM.

\subsection{Visible and Ultraviolet Astronomy}
\label{detecting:uvvis}

The least well-represented wavelength ranges are the visible and ultraviolet.  Transitions arising in these regions typically involve population transfer between electronic levels of molecules and atoms (see, e.g. \ce{N2}; \citealt{Lofthus:1977jw}).  As with vibrational transitions in the infrared, no permanent dipole moment is required, enabling the detection of species otherwise blind to radio astronomy.  The energy requirements for driving electronic transitions into emission are great enough that they are typically seen only in stellar atmospheres and other extremely energetic environments (see, e.g., emission from atomic Fe in Eta Carinae; \citealt{Aller:1966fs}).  In these conditions, many molecules will simply dissociate into their constituent atoms.

Thus, like the infrared, the most likely avenue of molecular detection in the visible and UV is through absorption spectroscopy against a background source.  This imposes similar restrictions to the infrared, but with the added challenge of the comparatively higher opacity of interstellar clouds at these wavelengths.  As a result, the few detections in these ranges have all been in diffuse, line of sight clouds to bright background continuum sources.

\subsection{Summary}
\label{detecting:summary}

In summary, the detections presented below will be heavily biased toward radio astronomy.  The primary molecular signal arising in the radio is that of transitions between rotational energy levels.  The energetics of the environments in which most molecules are found is very well matched to that required to populate these rotational energy levels, as opposed to the vibrational and especially electronic levels dominating the IR and UV/Vis regimes.  The resulting ability to observe in both emission and absorption, over a wide range of environments, substantially increases the opportunities for detection.

\section{Known Interstellar Molecules}
\label{known}

This section covers those species with published detections that have not been disputed.  This list includes tentative detections and disputed detections that were later confirmed.  Molecules are ordered first by number of atoms, then by year detected.  A common name is provided after the molecular formula for most species.  Note that for simplicity, due to differences in beam sizes, pointing centers, and nomenclature through time, detections toward sub-regions within the Sgr B2 and Orion sources have not been differentiated here. The exception to this is the Orion Bar photon-dominated/photo-dissociation region (PDR).

    
    \begin{center}
        \large
        \textsc{\textbf{Two-Atom Molecules}}
        \label{2atoms}
        \normalsize
    \end{center}        

    \subsection{\ce{CH} (methylidyne)}
    \label{CH}
    
    \citet{Swings:1937dl} suggested that an observed line at $\lambda = 4300$~\AA~by \citet{Dunham:1937nj} using the Mount Wilson Observatory in diffuse gas toward a number of supergiant B stars might have been due to the $^2\Delta \leftarrow ^2\Pi$ transition of CH, reported in the laboratory by \citet{Jevons:1932id}.  \citet{McKellar:1940io} later identified several additional transitions in observational data.  The first radio identification was reported by \citet{Rydbeck:1973dp} at 3335~MHz with the Onsala telescope toward more than a dozen sources using estimated fundamental rotational transition frequencies from \citet{Shklovskii:1953me}, \citet{Goss:1966mn}, and \citet{Baird:1971yb}.  The first direct measurement of the CH rotational spectrum was reported by \citet{Brazier:1983jh}. 

    \subsection{\ce{CN} (cyano radical)}
    \label{CN}
    
    Lines of CN were first reported using the Mount Wilson Observatory around $\lambda = 3875$~\AA~by \citet{McKellar:1940io} and \citet{Adams:1941tj} based on the laboratory work of \citet{Jenkins:1938go}. Later, \citet{Jefferts:1970ld} reported the first detection of rotational emission from CN by observation of the $J = 1-0$ transition at 113.5~GHz in Orion and W51 using the NRAO 36-ft telescope.  Identification was made prior to laboratory observation of the rotational transitions of CN, and was based on rotational constants derived from electronic transitions measured by \citet{Poletto:1965uf} and dipole moments measured by \citet{Thomson:1968jl}.  The laboratory rotational spectra were first measured seven years later by \citet{Dixon:1977yc}. 
    
    \subsection{\ce{CH^+} (methylidyne cation)}
    \label{CH+}
    
    \citet{Swings:1937dl} suggested that a set of three lines at $\lambda =$~4233, 3958, and 3745~\AA~seen with the Mount Wilson Observatory in diffuse gas belonged to a diatomic cation.  \citet{Douglas:1941uc} confirmed the assignment to \ce{CH^+} by laboratory observation of these lines in the $^1\Pi \leftarrow ^1\Sigma$ transition.  The first observation of rotational transitions of \ce{CH^+} was reported by \citet{Cernicharo:1997il} toward NGC 7027 using ISO, and based on rotational constants derived from rovibronic transitions measured by \citet{Carrington:1982ep}.  
    
    \subsection{\ce{OH} (hydroxyl radical)}
    \label{OH}

    \citet{Weinreb:1963lf} reported the first detection of OH through observation of its ground state, $^2\Pi _{3/2}$, $J = 3/2$, $\Lambda$-doubled, $F = 2\rightarrow2$ and $F = 1\rightarrow1$ hyperfine transitions at 1666~MHz toward Cas A using the Millstone Hill Observatory.  The frequencies were measured in the laboratory by \citet{Ehrenstein:1959jj}.
    
    \subsection{\ce{CO} (carbon monoxide)}
    \label{CO}
    
    \citet{Wilson:1970jm} reported the observation of the $J = 1\rightarrow0$ transition of CO toward Orion using the NRAO 36-ft telescope at a (velocity-shifted) frequency of 115267.2~MHz, based on a rest frequency of 115271.2~MHz from \citet{Cord:1968kw}.
    
    \subsection{\ce{H2} (molecular hydrogen)}
    \label{H2}
    
    Molecular hydrogen was reported by \citet{Carruthers:1970my} toward $\xi$~Per using a spectrograph on an Aerobee-150 rocket launched from White Sands Missile Range.  The $B^1\Sigma _u \leftarrow X^1\Sigma _g$ Lyman series in the range of 1000--1400~\AA~was observed and identified by direct comparison to a absorption cell observed with the same instrument used for the rocket observations.
    
    \subsection{\ce{SiO} (silicon monoxide)}
    \label{SiO}
    
    The $J=3\rightarrow2$ transition of SiO at 130246~MHz was observed by \citet{Wilson:1971yn} toward Sgr~B2 using the NRAO 36-ft telescope.  The transition frequency was calculated based on the laboratory work of \citet{Raymonda:1970dt} and \citet{Torring:1968pd}.  SiO was the first confirmed silicon-containing species in the ISM.
    
    \subsection{\ce{CS} (carbon monosulfide)}
    \label{CS}
    
    \citet{Penzias:1971kw} reported the detection of the $J = 3\rightarrow2$ transition of CS at 146969~MHz toward Orion, W51, IRC+10216, and DR21 using the NRAO 36-ft telescope.  The laboratory microwave spectrum was first reported by \citet{Mockler:1955wl}. CS was the first confirmed sulfur-containing species in the ISM.  
    
    \subsection{\ce{SO} (sulfur monoxide)}
    \label{SO}
    
    An unidentified line in NRAO 36-ft observations of Orion was assigned to the $J_K = 3_2\rightarrow2_1$ transition of SO by \citet{Gottlieb:1973md} based on laboratory data from \citet{Winnewisser:1964kk}.  The authors subsequently observed SO emission toward numerous other sources, as well as identifying the $4_3\rightarrow3_2$ transition using the 16-ft telescope at McDonald Observatory.  SO was the first molecule detected in space in a $^3\Sigma$ ground electronic state via radio astronomy.
    
    \subsection{\ce{SiS} (silicon monosulfide)}
    \label{SiS}
    
    \citet{Morris:1975im} reported the detection of the $J = 6\rightarrow5$ and $5\rightarrow4$ transitions of SiS at 108924.6~MHz and 90771.85~MHz, respectively, based on the laboratory work of \citet{Hoeft:1965fh}.  The observations were conducted toward IRC+10216 using the NRAO 36-ft telescope.
    
    \subsection{\ce{NS} (nitrogen sulfide)}
    \label{NS}
    
    \citet{Gottlieb:1975ke} and \citet{Kuiper:1975er} simultaneously and independently reported the detection of NS.  \citet{Gottlieb:1975ke} observed the $J=5/2\rightarrow3/2$ transition in the $^2\Pi_{1/2}$ electronic state at 115154~MHz \citep{Amano:1969yc} toward Sgr B2 using the 16-ft antenna at the University of Texas Millimeter Wave Observatory over three periods in 1973--1974, and confirmed the detection using the NRAO 36-ft telescope in May 1975.  \citet{Kuiper:1975er} observed the same transitions toward Sgr B2 using the NRAO 36-ft telescope in February 1975.
    
    \subsection{\ce{C2} (dicarbon)}
    \label{C2}
    
    The detection of \ce{C2} was reported by \citet{Souza:1977mk} toward Cygnus OB2 No. 12 using the Smithsonian Institution's Mount Hopkins Observatory.  The detected lines at 10140~\AA~were measured in the laboratory by \citet{Phillips:1948jp}.
    
    \subsection{\ce{NO} (nitric oxide)}
    \label{NO}
    
    \citet{Liszt:1978ep} reported the identification of NO in the ISM toward Sgr B2 using the NRAO 36-ft telescope.  The detection was based on the laboratory analysis of the hyperfine-splitting in the $^2\Pi_{1/2}$ $J = 3/2\rightarrow1/2$ transition near 150.1~GHz as reported in \citet{Gallagher:1956om}.
    
    \subsection{\ce{HCl} (hydrogen chloride)}
    \label{HCl}
    
    Interstellar H$^{35}$Cl was reported by \citet{Blake:1985pf} who used the Kuiper Airborne Observatory to observe the $J = 1\rightarrow0$ transition at 625918.8~MHz toward Orion.  The laboratory transition frequencies were reported in \citet{deLucia:1971lq}.
    
    \subsection{\ce{NaCl} (sodium chloride)}
    \label{NaCl}
    
    The detection of NaCl was reported by \citet{Cernicharo:1987wp} toward IRC+10216 using IRAM 30-m observations of six transitions between 91--169~GHz, as well as the $J=8\rightarrow7$ transition of \ce{Na^{37}Cl} at 101961.9~MHz.  The laboratory frequencies were obtained from \citet{Lovas:1974co}. 
    
    \subsection{\ce{AlCl} (aluminum chloride)}
    \label{AlCl}
    
    The detection of AlCl was reported by \citet{Cernicharo:1987wp} toward IRC+10216 using IRAM 30-m observations of four transitions between 87--160~GHz, as well as the $J=10\rightarrow9$ and $11\rightarrow10$ transitions of \ce{Al^{37}Cl} at 142322.5~MHz and 156546.8~MHz, respectively.  The laboratory frequencies were obtained from \citet{Lovas:1974co}.     
    
    \subsection{\ce{KCl} (potassium chloride)}
    \label{KCl}
    
    The detection of KCl was reported by \citet{Cernicharo:1987wp} toward IRC+10216 using IRAM 30-m observations of six transitions between 100--161~GHz.  The laboratory frequencies were obtained from \citet{Lovas:1974co}.      
    
    \subsection{\ce{AlF} (aluminum fluoride)}
    \label{AlF}
    
    \citet{Cernicharo:1987wp} claimed a tentative detection of AlF toward IRC+10216 using IRAM 30-m observations of three transitions between 99--165~GHz.  The laboratory frequencies were obtained from \citet{Lovas:1974co}. The detection was confirmed by \citet{Ziurys:1994kw} using CSO observations of IRC+10216.  Three additional lines, the $J=7\rightarrow6$, $8\rightarrow7$, and $10\rightarrow9$ were observed at 230793, 263749, and 329642~MHz, respectively, based on laboratory data from \citet{Wyse:1970ex}.
    
    \subsection{\ce{PN} (phosphorous mononitride)}
    \label{PN}
    
    \citet{Sutton:1985vs} first suggested that a feature observed in their line survey data using the OVRO 10.4-m telescope toward Orion at 234936~MHz could be attributed to the $J = 5\rightarrow4$ transition of PN, using the laboratory data reported in \citet{Wyse:1972go}.  \citet{Turner:1987fg} and \citet{Ziurys:1987vx} later simultaneously confirmed the detection of PN.  \citet{Turner:1987fg} observed the $2\rightarrow1$, $3\rightarrow2$, and $5\rightarrow4$ transitions at 93980, 140968, and 234936~MHz, respectively, toward Orion, W51, and Sgr B2 using the NRAO 12-m telescope. \citet{Ziurys:1987vx} observed the  $2\rightarrow1$, $3\rightarrow2$, and $5\rightarrow4$ transitions toward Orion using the FCRAO 14-m telescope, and the $6\rightarrow5$ using the NRAO 12-m telescope.  The $2\rightarrow1$ transition was also observed toward Sgr B2 and W51.  PN was the first phosphorous-containing molecule detected in the ISM.
    
    \subsection{\ce{SiC} (silicon carbide)}
    \label{SiC}
    
    \citet{Cernicharo:1989mw} reported the detection of the SiC radical in IRC+10216 using the IRAM 30-m telescope.  Fine structure and $\Lambda$-doubled lines were detected in the $J = 2\rightarrow1$, $4\rightarrow3$, and $6\rightarrow5$ transitions in the $^3\Pi$ electronic ground state near 81, 162, and 236~GHz, respectively, based on laboratory data presented in the same manuscript.  The authors make note of unidentified lines suggestive of the SiC radical in earlier data toward IRC+10216 beginning in 1976, both in their own observations and in those of I. Dubois (unpublished) using the NRAO 36-ft telescope.
    
    \subsection{\ce{CP} (carbon monophosphide)}
    \label{CP}
    
    \citet{Saito:1989wp} measured the rotational spectrum of CP in the laboratory and conducted an astronomical search for the $N = 1\rightarrow0$ and $2\rightarrow1$ transitions at 48 and 96~GHz, respectively, using the Nobeyama 45-m telescope across several sources, resulting in non-detections.  \citet{Guelin:1990iw} subsequently reported the successful detection of the $2\rightarrow1$ and $5\rightarrow4$ (239~GHz) transitions toward IRC+10216 using the IRAM 30-m telescope.
    
    \subsection{\ce{NH} (imidogen radical)}
    \label{NH}
    
    \citet{Meyer:1991ft} reported the detection of the $A^3\Pi-X^3\Sigma$ (0,0) $R_1$(0) line of NH in absorption toward $\xi$ Per and HD 27778 at 3358~\AA~using the KPNO 4-m telescope.  The laboratory rest frequencies were obtained from \citet{Dixon:1959dp}.  The earliest reported detection of rotational transitions of NH appears to be that of \citet{Cernicharo:2000rt}, who observed the $N_J = 2_3\rightarrow1_2$ transition at 974~GHz toward Sgr B2 with ISO.  Although no laboratory reference is given, the frequencies used were presumably those of \citet{Klaus:1997ku}.
    
    \subsection{\ce{SiN} (silicon nitride)}
    \label{SiN}
    
    \citet{Turner:1992kw} reported the detection of the $N=2\rightarrow1$ and $6\rightarrow5$ transitions of SiN at 87 and 262~GHz, respectively, toward IRC+10216 using the NRAO 12-m telescope.  The $2\rightarrow1$ transition was measured by \citet{Saito:1983hc}, and the frequency for the $6\rightarrow5$ was calculated from the constants given therein.  
    
    \subsection{\ce{SO^+} (sulfur monoxide cation)}
    \label{SO+}
    
    The detection of \ce{SO^+} was reported by \citet{Turner:1992cj} toward IC 443G using the NRAO 12-m telescope.  The $\Lambda$-doubled $^2\Pi_{1/2}$, $J = 5/2\rightarrow3/2$ and $9/2\rightarrow7/2$ transitions at 116 and 209~GHz, respectively, were observed.  The frequencies of the $5/2\rightarrow3/2$ transitions were measured by \citet{Amano:1991hq}, while those of the $9/2\rightarrow7/2$ transitions were calculated from the constants given therein.
    
    \subsection{\ce{CO^+} (carbon monoxide cation)}
    \label{CO+}
    
    \ce{CO^+} was first observed toward M17SW and NGC 7027 using the NRAO 12-m telescope by \citet{Latter:1993nb}. The $N=2\rightarrow1$, $J = 3/2\rightarrow1/2$ and $5/2\rightarrow3/2$ near 236~GHz and the $3\rightarrow2$, $5/2\rightarrow3/2$ were observed based on the laboratory work of \citet{Sastry:1981lk}. An earlier reported detection of the $2\rightarrow1$, $5/2\rightarrow3/2$ transition in Orion by \citet{Erickson:1981bn}, based on the laboratory data of \citet{Dixon:1975iw}, was later definitely assigned to transitions of \ce{^{13}CH3OH} by \citet{Blake:1984mk}.  
    
    \subsection{\ce{HF} (hydrogen fluoride)}
    \label{HF}
    
    \citet{Neufeld:1997td} reported the detection of the $J = 2\rightarrow1$ transition of HF at 2.5~THz toward Sgr B2 using ISO.  The rest frequencies were measured in the laboratory by \citet{Nolt:1987gd}.
    
    \subsection{\ce{N2} (nitrogen)}
    \label{N2}
    
    \citet{Knauth:2004vp} observed the c$^{\prime}_4$$^1$$\Sigma_u^+$--$X^1$$\Sigma_g^+$ and c$_3$$^1$$\Sigma_u^+$--$X^1$$\Sigma_g^+$ transitions of \ce{N2} at 958.6 and 960.3~\AA~, respectively, using FUSE observations toward HD 124314.  The oscillator strengths of the c$^{\prime}_4$$^1 $$\Sigma_u^+$--$X^1\Sigma_g^+$ transition were obtained from \citet{Stark:2000fe}, and the c$_3$~$^1$$ \Sigma_u^+$--$X^1$$\Sigma_g^+$ transition from a private communication with G. Stark.
    
    \subsection{\ce{CF^+} (fluoromethylidynium cation)}
    \label{CF+}
    
    The detection of \ce{CF^+} was reported by \citet{Neufeld:2006ej} using IRAM 30-m and APEX 12-m observations of the $J = 1\rightarrow0$, $2\rightarrow1$, and $3\rightarrow2$ transitions at 102.6, 205.2, and 307.7~GHz, respectively, toward the Orion Bar.  The laboratory frequencies were reported by \citet{Plummer:1986jn}.
    
    \subsection{\ce{PO} (phosphorous monoxide)}
    \label{PO}
    
    \citet{Tenenbaum:2007po} reported the detection of the $\Lambda$-doubled $J = 11/2\rightarrow9/2$ and $13/2\rightarrow11/2$ transitions of PO at 240 and 284~GHz, respectively, using SMT 10-m observations of VY Canis Majoris.  Although not cited, the frequencies were presumably obtained from the laboratory work of \citet{Kawaguchi:1983co} and \citet{Bailleux:2002hl}.
    
    \subsection{\ce{O2} (oxygen)}
    \label{O2}
    
    A tentative detection of molecular oxygen was reported by \citet{Goldsmith:2002vy} in SWAS observations of $\rho$ Oph, with further evidence provided by \citet{Larsson:2007dx} using observations of the $N_J = 1_1\rightarrow1_0$ transition at 118750 MHz with the Odin satellite toward $\rho$ Oph A.  Although no citation is provided, the frequencies were presumably obtained from the laboratory work of \citet{Endo:1982hj}.  Later, \citet{Goldsmith:2011dj} reported the observation of the $N_J = 3_3\rightarrow1_2$, $5_4\rightarrow3_4$, and $7_6\rightarrow5_6$ transitions at 487, 774, and 1121~GHz, respectively, using \emph{Herschel}/HIFI observations of Orion.  Although there appears to be no citation to the laboratory data, these transition frequencies were likely taken from \citet{Drouin:2010fz}.    
    
    \subsection{\ce{AlO} (aluminum monoxide)}
    \label{AlO}
    
    \citet{Tenenbaum:2009ba} reported the detection of the $N = 7\rightarrow6$, $6\rightarrow5$, and $4\rightarrow3$ transitions of AlO at 268, 230, and 153~GHz, respectively, toward VY Canis Majoris using the SMT 10-m telescope.  The laboratory frequencies were reported in \citet{Yamada:1990gz}.
    
    \subsection{\ce{CN^-} (cyanide anion)}
    \label{CN-}
    
    The detection of \ce{CN^-} was reported by \citet{Agundez:2010ks} in observations of IRC+10216 using the IRAM 30-m telescope.  The $J = 1\rightarrow0$, $2\rightarrow1$, and $3\rightarrow2$ lines at 112.3, 224.5, and 336.8~GHz were observed, based on the laboratory work of \citet{Amano:2008ij}.
    
    \subsection{\ce{OH^+} (hydroxyl cation)}
    \label{OH+}
    
    \ce{OH^+} was detected by \citet{Wyrowski:2010eh} using APEX 12-m observations of Sgr B2.  The $N = 1\rightarrow0$, $J = 0\rightarrow1$ transitions were detected in absorption at 909~GHz based on the laboratory work of \citet{Bekooy:1985ku}.  Nearly simultaneously, \citet{Benz:2010ei} reported the detection of the $1\rightarrow0$, $3/2\rightarrow1/2$ transitions at 1033~GHz toward W3 IRS5 using \emph{Herschel}/HIFI, and \citet{Gerin:2010bg} reported a detection along the line of sight to W31C, also with \emph{Herschel}/HIFI.
    
    \subsection{\ce{SH^+} (sulfanylium cation)}
    \label{SH+}
    
    \citet{Benz:2010ei} reported the detection of the $N_J = 1_2\rightarrow0_1$ transition of \ce{SH^+} at 526~GHz using \emph{Herschel}/HIFI observations of W3. Although only a reference to the CDMS database is given, the frequency was almost certainly based on the laboratory work of \citet{Hovde:1987jn} and \citet{Brown:2009ee}. Although published after \citet{Benz:2010ei}, the first \emph{submitted} detection appears to be that of \citet{Menten:2011hk} who reported the detection of \ce{SH^+} in absorption toward Sgr B2, using the APEX 12-m telescope.  The $N_J = 1_1\rightarrow0_1$ transition at 683~GHz was observed based on the laboratory work of \citet{Hovde:1987jn} and \citet{Brown:2009ee}. \citet{Benz:2010ei} acknowledge the earlier submission of \citet{Menten:2011hk} in their work.
    
    \subsection{\ce{HCl^+} (hydrogen chloride cation)}
    \label{HCl+}
    
    The detection of \ce{HCl^+} was reported by \citet{DeLuca:2012cv} in \emph{Herschel}/HIFI observations of W31C and W49N.  Hyperfine and $\Lambda$-doubling structure was observed in the $^2\Pi_{3/2}$ $J = 5/2\rightarrow3/2$ transition of \ce{H^{35}Cl^+} at 1.444 THz toward both sources, and the same transition of \ce{H^{37}Cl^+} was also observed toward W31C.  The laboratory work was described in \citet{Gupta:2012ga}.
    
    \subsection{\ce{SH} (mercapto radical)}
    \label{SH}
    
    \citet{Neufeld:2012gz} reported the detection of the $^2\Pi_{3/2}$ $J = 5/2\rightarrow3/2$ $\Lambda$-doubled transition of SH at 1383~GHz along the line of sight to W49A using the GREAT instrument on SOFIA.  The transition frequencies relied upon the work of \citet{Morino:1995oi} and \citet{Klisch:1996mn}.
    
    \subsection{\ce{TiO} (titanium monoxide)}
    \label{TiO}
    
    TiO was detected in VY Canis Majoris by \citet{Kaminski:2013gk} using a combination of SMA and PdBI observations.  Several fine structure components in the $J = 11\rightarrow10$, $J = 10\rightarrow9$, $J = 9\rightarrow8$, and $J = 7\rightarrow6$ transitions were observed between 221 and 352~GHz based on the laboratory work of \citet{Namiki:1998id}.
    
    \subsection{\ce{ArH^+} (argonium)}
    \label{ArH+}
    
    Although a feature was initially observed in several \emph{Herschel} datasets near 618~GHz \citep{Neill:2014cb}, the attribution to \ce{^{36}ArH^+} was not readily apparent.  On Earth, \ce{^{40}Ar} is 300 times more abundant than  \ce{^{36}Ar} \citep{Lee:2006bc}, and as no signal from \ce{^{40}ArH^+} was visible, \ce{^{36}ArH^+} seemed an unlikely carrier.  \citet{Barlow:2013dr}, however, recognized that \ce{^{36}Ar} is the dominant isotope in the ISM \citep{Cameron:1970dw}, and identified the $J = 1\rightarrow0$ and $2\rightarrow1$ rotational lines of \ce{^{36}ArH^+} at 617.5 and 1234.6~GHz, respectively, in \emph{Herschel}/SPIRE spectra of the Crab Nebula.  A citation is only given to the CDMS database, as no laboratory data appears to exist for \ce{^{36}ArH^+}.  Instead, it is derived in the database to observational accuracy using isotopic scaling factors from other isotopologues.  
    
    \subsection{\ce{NS^+} (nitrogen sulfide cation)}
    \label{NS+}
    
    \citet{Cernicharo:2018bv} reported both the laboratory spectroscopy and astronomical identification of \ce{NS^+}.  The $J = 2\rightarrow1$ line has been observed with the IRAM 30-m telescope toward numerous sightlines, with confirming observations of the $3\rightarrow2$ and $5\rightarrow4$ in several.  
    
    
    \begin{center}
        \large
        \textsc{\textbf{Three-Atom Molecules}}
        \label{3atoms}
        \normalsize
    \end{center}        
    
    \subsection{\ce{H2O} (water)}
    \label{H2O}
    
    \citet{Cheung:1969mh} reported the detection of the $J_{K_a,K_c} = 6_{1,6}\rightarrow5_{2,3}$ transition of \ce{H2O} at 22.2~GHz using the Hat Creek Observatory toward Sgr B2, Orion, and W49.  No citation to the laboratory data appears to be given, but presumably was obtained from the work of \citet{Golden:1948jh}.
    
    \subsection{\ce{HCO^+} (formylium cation)}
    \label{HCO+}
    
    \citet{Buhl:1970rd} first reported the discovery of a bright emission feature at 89190~MHz toward Orion, W51, W3(OH), L134, and Sgr A in observations with the NRAO 36-ft telescope, and named the carrier `X-ogen.'  Shortly thereafter, \citet{Klemperer:1970uh} suggested the attribution of this line to \ce{HCO^+}.  The detection and attribution was confirmed by the laboratory observation of \ce{HCO^+} by \citet{Woods:1975vi}.
    
    \subsection{\ce{HCN} (hydrogen cyanide)}
    \label{HCN}
    
    The first reported detection of HCN was that of \citet{Snyder:1971ps} who observed the ground state $J = 1\rightarrow0$ transition at 88.6~GHz toward W3(OH), Orion, Sgr A, W49, W51, and DR 21 using the NRAO 36-ft telescope.  The enabling laboratory spectroscopy was reported by \citet{deLucia:1969hg}.
    
    \subsection{\ce{OCS} (carbonyl sulfide)}
    \label{OCS}
    
    Emission from the $J = 9\rightarrow8$ transition of OCS at 109463~MHz toward Sgr B2 was reported by \citet{Jefferts:1971hy} using the NRAO 36-ft telescope.  The laboratory spectroscopy was reported in \citet{King:1954bn}.
    
    \subsection{\ce{HNC} (hydrogen isocyanide)}
    \label{HNC}
    
    Both \citet{Zuckerman:1972cm} and \citet{Snyder:1972jh} observed unidentified emission signal at 90.7~GHz using the NRAO 36-ft telescope toward W51 and NGC 2264, respectively.  \citet{Snyder:1972jh} suggested this line to be due to HNC, which was confirmed four years later with the first laboratory measurement of the $J = 1\rightarrow0$ transition in the laboratory by \citet{Blackman:1976kf}.
    
    \subsection{\ce{H2S} (hydrogen sulfide)}
    \label{H2S}
    
    \citet{Thaddeus:1972oh} reported the detection of the $J_{K_a,K_c} = 1_{1,0}\rightarrow1_{0,1}$ transition of \emph{ortho}-\ce{H2S} at 168.7~GHz toward a number of sources with the NRAO 36-ft telescope.  The laboratory transition frequencies were measured by \citet{Cupp:1968yg}.
    
    \subsection{\ce{N2H^+} (protonated nitrogen)}
    \label{N2H+}
    
    \citet{Turner:1974jk} first reported an unidentified emission line at 93.174~GHz in NRAO 36-ft telescope toward a number of sources, including Sgr B2, DR 21(OH), NGC 2264, and NGC 6334N.  In a companion letter, \citet{Green:1974gc} suggested the carrier was \ce{N2H^+}, based on theoretical calculations.  \citet{Thaddeus:1975uh} claim the following year to have confirmed the detection by observing an exceptional match to the predicted \ce{^{14}N} hyperfine splitting.  Laboratory work by \citet{Saykally:1976ke} solidified the detection.
    
    \subsection{\ce{C2H} (ethynyl radical)}
    \label{C2H}
    
    The detection of \ce{C2H} was reported by \citet{Tucker:1974fs} through observation of four $\lambda$-doubled, hyperfine components of the $N = 1\rightarrow0$ transition near 87.3~GHz in NRAO 36-ft telescope observations of Orion and numerous other sources.  The assignment was made based on a calculated rotational constant under the assumption that the \ce{C\bond{1}H} and \ce{C\bond{3}C} bonds had the same length as those already known in acetylene (\ce{C2H2}).  A linear structure was assumed based on the laboratory work of \citet{Cochran:1964dy} and \citet{Graham:1974kx}.  The confirming laboratory microwave spectroscopy was later reported by \citet{Sastry:1981kj}.
    
    \subsection{\ce{SO2} (sulfur dioxide)}
    \label{SO2}
    
    \ce{SO2} was reported in NRAO 36-ft telescope observations toward Orion and Sgr B2 by \citet{Snyder:1975cr}.  The $J_{K_a,K_c} = 8_{1,7}\rightarrow8_{0,8}$, $8_{3,5}\rightarrow9_{2,8}$, and $7_{3,5}\rightarrow8_{2,6}$ transitions at 83.7, 86.6, and 97.7~GHz, respectively, were observed in emission.  The enabling laboratory microwave spectroscopy work was performed by \citet{Steenbeckeliers:1968kj} with additional computation analysis performed by \citet{Kirchhoff:1972er}.
    
    \subsection{\ce{HCO} (formyl radical)}
	\label{HCO}
	
	\citet{Snyder:1976hg} reported the detection of the $N_{K_-,K_+} = 1_{0,1} - 0_{0,0}$, $J = 3/2\rightarrow1/2$, $F = 2\rightarrow1$ transition of HCO at 86671~MHz toward W3, NGC 2024, W51, and K3-50 using the NRAO 36-ft telescope.  The laboratory microwave spectroscopy was reported by \citet{Saito:1972oi}.
	
	\subsection{\ce{HNO} (nitroxyl radical)}
	\label{HNO}
	
	The detection of HNO was first reported by \citet{Ulich:1977by} who observed the $J_{K_a,K_c} = 1_{0,1}\rightarrow0_{0,0}$ transition at 81477~MHz toward Sgr B2 and NGC 2024 using the NRAO 36-ft telescope.  The laboratory frequency was measured by \citet{Saito:1973nb}.  The initial detection was subject to significant controversy in the literature (see \citealt{Snyder:1993po}).  Subsequent observation of additional lines by \citet{Hollis:1991ua} and \citet{Ziurys:1994dc} confirmed the initial detection, aided by the laboratory work of \citet{Sastry:1984jk}.
	
	\subsection{\ce{HCS^+} (protonated carbon monosulfide)}
	\label{HCS+}
	
	\citet{Thaddeus:1981uy} reported the detection of four harmonically spaced, unidentified lines in NRAO 36-ft and Bell 7-m telescope observations of Orion and Sgr B2.  Based on the harmonic pattern, the lines were assigned to the $J = 2\rightarrow1$, $3\rightarrow2$, $5\rightarrow4$, and $6\rightarrow5$ transitions of \ce{HCS^+} at 85, 128, 213, and 256~GHz, respectively. The confirming laboratory measurements were presented in a companion letter by \citet{Gudeman:1981fy}. 
	
	\subsection{\ce{HOC^+} (hydroxymethyliumylidene) }
	\label{HOC+}
	
	\ce{HOC^+} was reported by \citet{Woods:1983va} in FCRAO and Onsala observations toward Sgr B2.  The $J = 1\rightarrow0$ transition at 89478~MHz was identified based on the laboratory work of \citet{Gudeman:1982yt}.  The detection was confirmed by \citet{Ziurys:1995mh} who observed the $J = 2\rightarrow1$, and $3\rightarrow2$ transitions at 179 and 268~GHz, respectively.
	
	\subsection{\ce{c-SiC2} (silacyclopropynylidene)}
	\label{SiC2}
	
	\citet{Thaddeus:1984fn} reported the detection of \ce{c-SiC2} in NRAO 36-ft and Bell 7-m telescope observations of IRC+10216.  Nine transitions between 93--171~GHz were observed and assigned based on the laboratory work of \citet{Michalopoulos:1984iy} who derived rotational constants from rotationally resolved optical transitions.  \citet{Thaddeus:1984fn} note that a number of these transitions had previously been observed in IRC+10216 and classified as unidentified by various other researchers as early as 1976. \citet{Suenram:1989co} and \citet{Gottlieb:1989cr} subsequently reported the laboratory observation of the pure rotational spectrum. \ce{c-SiC2} was the first molecular ring molecule identified in the ISM.
	
	\subsection{\ce{C2S} (dicarbon sulfide)}
	\label{C2S}
	
	The detection of \ce{C2S} ($^3\Sigma ^-$) was reported toward TMC-1 and Sgr B2 by \citet{Saito:1987fa} using the Nobeyama 45-m telescope.  The laboratory measurements were also performed by \citet{Saito:1987fa}. Although published a month earlier than \citet{Saito:1987fa}, \citet{Cernicharo:1987jh} reported the detection of \ce{C2S} in IRC+10216 using the laboratory results of \citet{Saito:1987fa} from a preprint article.
	
	\subsection{\ce{C3} (tricarbon)}
	\label{C3}
	
	\citet{Hinkle:1988yc} reported the detection of \ce{C3} in KPNO 4-m telescope observations of IRC+10216.  The $\nu_3$ band of \ce{C3} was observed and assigned based on combination differences from the work of \citet{Gausset:1965ea} near 2030~cm$^{-1}$.  
	
	\subsection{\ce{CO2} (carbon dioxide)}
	\label{CO2}
	
	\citet{deHendecourt:1989iu} reported the observation of the $\nu_2$ bending mode of \ce{CO2} ice in absorption at 15.2~$\mu$m using archival spectra from the IRAS database toward AFGL 961, AFGL 989, and AFGL 890.  The assignment was based on laboratory spectroscopy of mixed \ce{CO2} ices \citep{dHendecourt:1986wx}. Gas-phase \ce{CO2} was later observed toward numerous sightlines with ISO as a sharp absorption feature near 15~$\mu$m superimposed on a broad solid-phase absorption feature near the same frequency \citep{vanDischoeck:1996nd}.  The frequencies were obtained from \citet{Paso:1980ec}, while the band strengths used were taken from \citet{ReichleJr:1972hg}.
	
	\subsection{\ce{CH2} (methylene)}
	\label{CH2}
	
	\ce{CH2} was first detected by \citet{Hollis:1989nh} in NRAO 12-m telescope observations of Orion.  Several hyperfine components of $N_{K_a,K_c} = 4_{0,4}\rightarrow3_{1,3}$ transition were observed at 68 and 71~GHz and were assigned based on the laboratory work of \citet{Lovas:1983cp}.  The detection was later confirmed by \citet{Hollis:1995ng}.
	
	\subsection{\ce{C2O} (dicarbon monoxide)}
	\label{C2O}
	
	The detection of \ce{C2O} was reported by \citet{Ohishi:1991yt} toward TMC-1 using the Nobeyama 45-m telescope.  The $N_J = 1_2\rightarrow0_1$ and $2_3\rightarrow1_2$ transitions at 22 and 46~GHz, respectively, were assigned based on the laboratory data of \citet{Yamada:1985re}.
	
	\subsection{\ce{MgNC} (magnesium isocyanide)}
	\label{MgNC}
	
	\citet{Kawaguchi:1993ft} successfully assigned a series of three, unidentified harmonically spaced doublet emission lines detected by \citet{Guelin:1986mq} in IRAM 30-m observations of IRC+10216 to the $N = 7\rightarrow6$, $8\rightarrow7$, and $9\rightarrow8$ transitions of MgNC based on their own laboratory spectroscopy of the species.  MgNC was the first magnesium-containing molecule to be detected in the ISM.
	
	\subsection{\ce{NH2} (amidogen radical)}
	\label{NH2}
	
	The detection of \ce{NH2} was reported by \citet{vanDishoeck:1993ds} toward Sgr B2 using CSO observations of five components of the $J = 3/2\rightarrow3/2$, $1/2\rightarrow1/2$, and $3/2\rightarrow1/2$ transitions at 426, 469, and 461~GHz, respectively, based on the laboratory work of \citet{Burkholder:1988gm} and \citet{Charo:1981wh}.  Some hyperfine splitting was resolved.
	
	\subsection{\ce{NaCN} (sodium cyanide)}
	\label{NaCN}
	
    The detection of NaCN was reported by \citet{Turner:1994es} using NRAO 12-m telescope observations of IRC+10216.  The $J_{K_a,K_c} = 5_{0,5}\rightarrow4_{0,4}$, $6_{0,6}\rightarrow5_{0,5}$, $7_{0,7}\rightarrow6_{0,6}$, and $9_{0,9}\rightarrow8_{0,8}$ transitions at 78, 93, 108, and 139~GHz, respectively, were observed.  The assignments were based on predictions from the rotational constants derived in \citet{vanVaals:1984gt}.
    
    \subsection{\ce{N2O} (nitrous oxide)}
	\label{N2O}
	
	\citet{Ziurys:1994hd} reported the detection of the $J = 3\rightarrow2$, $4\rightarrow3$, $5\rightarrow4$, and $6\rightarrow5$ transitions of \ce{N2O} at 75, 100, 125, and 150~GHz, respectively, in NRAO 12-m observations toward Sgr B2.  The laboratory frequencies were obtained from \citet{Lovas:1978hr}.
	
	\subsection{\ce{MgCN} (magnesium cyanide radical)}
	\label{MgCN}
	
	The detection of the MgCN radical was reported by \citet{Ziurys:1995pf} in NRAO 12-m and IRAM 30-m observations of IRC+10216.  The $N = 11\rightarrow10$, $10\rightarrow9$, and $9\rightarrow8$ transitions were observed, some with resolved spin-rotation splitting, based on the accompanying laboratory work of \citet{Anderson:1994if}.
	
	\subsection{\ce{H3^+}}
	\label{H3+}
	
	First suggested as a possible interstellar molecule in \citet{Martin:1961wr}, \ce{H3+} was detected 35 years later in absorption toward GL 2136 and W33A by \citet{Geballe:1996iw} using UKIRT to observe three transitions of the $\nu_2$ fundamental band near 3.7~$\mu$m.  The laboratory work was performed by \citet{Oka:1980tr}.
	
	\subsection{\ce{SiCN} (silicon monocyanide radical)}
	\label{SiCN}
	
	\citet{Guelin:2000kl} reported the detection of the SiCN radical in IRAM 30-m observations of IRC+10216.  Three $\Lambda$-doubled transitions at 83, 94, and 105~GHz were detected based on the laboratory work by \citet{Apponi:2000re}.
	
	\subsection{\ce{AlNC} (aluminum isocyanide)}
	\label{AlNC}
	
	The detection of AlNC toward IRC+10216 with the IRAM 30-m telescope was reported by \citet{Ziurys:2002ds}.  Five transitions between 132--251~GHz were assigned based on the laboratory work of \citet{Robinson:1997bh}.
	
	\subsection{\ce{SiNC} (silicon monoisocyanide radical)}
	\label{SiNC}
	
	\citet{Guelin:2004cg} reported the detection of the SiNC radical in IRAM 30-m telescope observations of IRC+10216.  Four rotational transitions between 83--134~GHz were observed and assigned based on the laboratory work of \citet{Apponi:2000re}.
	
	\subsection{\ce{HCP} (phosphaethyne)}
	\label{HCP}
	
	\citet{Agundez:2007ns} reported the detection of four transitions ($J = 4\rightarrow3$ through $7\rightarrow6$) of HCP toward IRC+10216 using the IRAM 30-m telescope.  The enabling laboratory work was performed by \citet{Bizzocchi:2001ij} and references therein.
	
	\subsection{\ce{CCP} (dicarbon phosphide radical)}
	\label{CCP}
	
	The detection of the CCP radical in ARO 12-m observations of IRC+10216 was reported by \citet{Halfen:2008dw}.  Five transitions with  partially resolved hyperfine splitting were identified and assigned based on laboratory work described in the same manuscript.
	
	\subsection{\ce{AlOH} (aluminum hydroxide)}
	\label{AlOH}
	
	AlOH was detected in ARO 12-m and SMT observations of VY Canis Majoris by \citet{Tenenbaum:2010hp}. The $J = 9\rightarrow8$, $7\rightarrow6$, and $5\rightarrow4$ transitions at 283, 220, and 157~GHz, respectively, were identified based on the enabling laboratory work of \citet{Apponi:1993gf}.
	
	\subsection{\ce{H2O^+} (oxidaniumyl cation)}
	\label{H2O+}
	
	\ce{H2O^+} was identified in \emph{Herschel}/HIFI spectra toward DR21, Sgr B2, and NGC 6334 by \citet{Ossenkopf:2010es}.  Six hyperfine components of the $N_{K_a,K_c} = 1_{1,1}\rightarrow0_{0,0}$, $J = 3/2\rightarrow1/2$ transition in the $^2B_1$ electronic ground state at 1.115~THz were observed.  The assignments were based on a synthesis, re-analysis, and extrapolation of the laboratory work by \citet{Strahan:1986du} and \citet{Murtz:1998br}  Nearly simultaneously, \citet{Gerin:2010bg} reported the observation of \ce{H2O+} along the line of sight to W31C, also using \emph{Herschel}/HIFI.
	
	\subsection{\ce{H2Cl^+} (chloronium cation)}
	\label{H2Cl+}
	
	\citet{Lis:2010ff} reported the observation of the $J_{K_a,K_c} = 2_{1,2}\rightarrow1_{0,1}$ transitions of \emph{ortho}-\ce{H2^{35}Cl^+} and \emph{ortho}-\ce{H2^{37}Cl^+} near 781~GHz, and the $J_{K_a,K_c} = 1_{1,1}\rightarrow0_{0,0}$ transition of \emph{para}-\ce{H2^{35}Cl^+} near 485~GHz in absorption toward NGC 6334I and Sgr B2 using \emph{Herschel}/HIFI.  The frequencies were derived from the rotational constants determined in the laboratory work of \citet{Araki:2001dh}.
	
	\subsection{\ce{KCN} (potassium cyanide)}
	\label{KCN}
	
	The detection of KCN was reported by \citet{Pulliam:2010he} using ARO 12-m, IRAM 30-m, and SMT observations of IRC+10216.  More than a dozen transitions between 85--250~GHz were assigned based on the laboratory work of \citet{Torring:1980ef}.
	
	\subsection{\ce{FeCN} (iron cyanide)}
	\label{FeCN}
	
	\citet{Zack:2011jx} reported the detection of six transitions of FeCN between 84--132~GHz in ARO 12-m observations of IRC+10216.  The enabling laboratory spectroscopy was reported in \citet{Flory:2011ew}, which was not yet published at the time of the publication of \citet{Zack:2011jx}.
	
	\subsection{\ce{HO2} (hydroperoxyl radical)}
	\label{HO2}
	
	The detection of \ce{HO2} was reported in IRAM 30-m and APEX observations toward $\rho$ Oph A by \citet{Parise:2012hh}.  Several hyperfine components were observed in the $N_{K_a,K_c} = 2_{0,2}\rightarrow1_{0,1}$ and $4_{0,4}\rightarrow3_{0,3}$ transitions at 130 and 261~GHz, respectively.  The laboratory work was performed by \citet{Beers:1975ei}, \citet{Saito:1977cg}, and \citet{Charo:1982he}.
	
	\subsection{\ce{TiO2} (titanium dioxide)}
	\label{TiO2}
	
	\ce{TiO2} was detected by \citet{Kaminski:2013gk} in SMA and PdBI observations of VY Canis Majoris.  More than two dozen transitions of \ce{TiO2} were observed between 221--351~GHz based on the laboratory work of \citet{Brunken:2008rt} and \citet{Kania:2011gp}.
	
	\subsection{\ce{CCN} (cyanomethylidyne)}
	\label{CCN}
	
	\citet{Anderson:2014kt} reported the detection of CCN in ARO 12-m and SMT observations of IRC+10216.  The $J = 9/2\rightarrow7/2$, $13/2\rightarrow11/2$, and $19/2\rightarrow17/2$ $\Lambda$-doubled transitions near 106, 154, and 225~GHz, respectively, were identified and assigned based on the laboratory work of \citet{Anderson:2015op} which was published several months later.
	
	\subsection{\ce{SiCSi} (disilicon carbide)}
	\label{SiCSi}
	
	112 transitions of SiCSi were observed in IRAM 30-m telescope observations of IRC+10216 between 80--350~GHz as reported by \citet{Cernicharo:2015cw}.  The transition frequencies were derived based on the contemporaneous laboratory work of \citet{McCarthy:2015do}.
	
	\subsection{\ce{S2H} (hydrogen disulfide)}
	\label{S2H}
	
	\citet{Fuente:2017gd} reported the detection of \ce{S2H} in IRAM 30-m observations of the Horsehead PDR region.  Although \citet{Fuente:2017gd} do not not provide quantum numbers, they detect four lines corresponding to the unresolved hyperfine doublets of the $J_{K_a,K_c} = 6_{0,6}\rightarrow5_{0,5}$ and $7_{0,7}\rightarrow6_{0,6}$ transitions near 94 and 110~GHz, respectively.  The laboratory analysis was described in \citet{Tanimoto:2000im}.
	
	\subsection{\ce{HCS} (thioformyl radical)}
	\label{HCS}
	
	The detection of HCS was reported by \citet{Agundez:2018kh} in IRAM 30-m observations of L483.  Five resolved hyperfine components of the $N_{K_a,K_c} = 2_{0,2}\rightarrow1_{0,1}$ transition of HCS near 81~GHz were identified based on the laboratory work of \citet{Habara:2002gd}.
	
	\subsection{\ce{HSC} (sulfhydryl carbide radical)}
	\label{HSC}
	
	The detection of HSC was reported by \citet{Agundez:2018kh} in IRAM 30-m observations of L483.  Two resolved hyperfine components of the $N_{K_a,K_c} = 2_{0,2}\rightarrow1_{0,1}$ transition of HSC near 81~GHz were identified based on the laboratory work of \citet{Habara:2000dy}.
	
	\subsection{\ce{NCO} (isocyanate radical)}
	\label{NCO}
	
	\citet{Marcelino:2018cp} reported the detection of the NCO radical in IRAM 30-m observations of L483.  Several hyperfine components of the $J = 7/2\rightarrow5/2$ and $9/2\rightarrow7/2$ transitions near 81.4 and 104.6~GHz were identified and assigned based on the laboratory work of \citet{Saito:1970if} and \citet{Kawaguchi:1985hf}.
    
    
    \begin{center}
        \large
        \textsc{\textbf{Four-Atom Molecules}}
        \label{4atoms}
        \normalsize
    \end{center}        
    
    \subsection{\ce{NH3} (ammonia)}
	\label{NH3}
	
	The first detection of \ce{NH3} was reported by \citet{Cheung:1968yr} by observation of its ($J,K$) = (1,1) and (2,2) inversion transitions in the Galactic Center using a 20-m radio telescope located at Hat Creek Observatory.  No references to the laboratory data are given, but the frequencies were likely derived from the accumulated works of \citet{Cleeton:1934gy}, \citet{GuntherMohr:1954cw}, \citet{Kukolich:1965ey}, and \citet{Kukolich:1967nd}. 
	
	\subsection{\ce{H2CO} (formaldehyde)}
	\label{H2CO}
	
	\citet{Snyder:1969km} reported the detection of \ce{H2CO} in NRAO 140-ft telescope observations of more than a dozen interstellar sources.  The $J_{K_a,K_c} = 1_{1,1}\rightarrow1_{1,0}$ transition at 4830~MHz was observed without hyperfine splitting, based on the laboratory data of \citet{Shinegari:1967bl}.
	
	\subsection{\ce{HNCO} (isocyanic acid)}
	\label{HNCO}

    HNCO was observed in NRAO 36-ft observations of Sgr B2 by \citet{Snyder:1972tq}.  The $J_{K_a,K_c} = 4_{0,4}\rightarrow3_{0,3}$ transition at 87925~MHz was detected and assigned based on the laboratory work of \citet{Kewley:1963cn}.  The detection was confirmed in a companion article with the observation of the $1_{0,1}\rightarrow0_{0,0}$ transition by \citet{Buhl:1972bj}.
    
    \subsection{\ce{H2CS} (thioformaldehyde)}
	\label{H2CS}
	
	\citet{Sinclair:1973ft} reported the detection of \ce{H2CS} through Parkes 64-m observations of the Sgr B2.  The $J_{K_a,K_c} = 2_{1,1}\rightarrow2_{1,2}$ transition at 3139~MHz was identified and assigned based on the laboratory work of \citet{Johnson:1970yd}.
	
	\subsection{\ce{C2H2} (acetylene)}
	\label{C2H2}
	
	The detection of \ce{C2H2} was reported by \citet{Ridgway:1976il} in Mayall 4-m telescope observations of IRC+10216.  The $\nu_1 + \nu_5$ combination band of acetylene at 4091~cm$^{-1}$ was identified based on the laboratory work of \citet{Baldacci:1973ju}.
	
	\subsection{\ce{C3N} (cyanoethynyl radical)}
	\label{C3N}
	
	\citet{Guelin:1977kd} reported the tentative detection of \ce{C3N} in NRAO 36-ft observations of IRC+10216.  They observed two sets of doublet transitions, with the center frequencies at 89 and 99~GHz.  Based on calculated rotational constants, they tentatively assigned the emission to \ce{C3N}. \citet{Friberg:1980lc} later confirmed the detection by observing the predicted $N  = 3\rightarrow2$ transitions near 30~GHz.  The laboratory spectra were subsequently measured by \citet{Gottlieb:1983yd}.
	
	\subsection{\ce{HNCS} (isothiocyanic acid)}
	\label{HNCS}
	
	HNCS was detected by observation of the $J = 11\rightarrow10$, $9\rightarrow8$, and $8\rightarrow7$ transitions, among others, at 192, 106, and 94~GHz, respectively, in Bell 7-m and NRAO 36-ft telescope observations of Sgr B2 by \citet{Frerking:1979yd}.  The laboratory frequencies were measured by \citet{Kewley:1963cn}. 
	
	\subsection{\ce{HOCO^+} (protonated carbon dioxide)}
	\label{HOCO+}
	
	\citet{Thaddeus:1981uy} reported the observation of a series of three lines in Bell 7-m observations of Sgr B2 they assigned to the $J = 6\rightarrow5$, $5\rightarrow4$, and $4\rightarrow3$ transitions of \ce{HOCO+} at 128, 107, and 86~GHz, respectively, based on \emph{ab initio} calculations by \citet{Green:1976cq}.  The detection was later confirmed by laboratory observation of the rotational spectrum by \citet{Bogey:1984vk}.
	
	\subsection{\ce{C3O} (tricarbon monoxide)}
	\label{C3O}
	
	The detection of \ce{C3O} was reported by \citet{Matthews:1984nb} through NRAO 140-ft telescope observations of TMC-1.  The $J = 2\rightarrow1$ transition at 19244~MHz was observed and assigned based on laboratory work by \citet{Brown:1983kl}.  Three further transitions up to $J = 9\rightarrow8$ were observed in a follow-up study in the same source by \citet{Brown:1985nw}. The detection was confirmed by \citet{Kaifu:2004tk} with the observation of numerous additional lines in Nobeyama 45-m observations of TMC-1.  
	
	\subsection{\ce{l-C3H} (propynylidyne radical)}
	\label{l-C3H}
	
	The $l$-\ce{C3H} radical was detected toward TMC-1 and IRC+10216 through a combination of observations with the Bell 7-m, University of Massachusetts 14-m, NRAO 36-ft, and Onsala 20-m telescopes as reported in \citet{Thaddeus:1985vb}.  A number of transitions between 33--103~GHz were identified based on the accompanying laboratory work of \citet{Gottlieb:1985gh}.
	
	\subsection{\ce{HCNH^+} (protonated hydrogen cyanide)}
	\label{HCNH+}
	
	\ce{HCNH^+} was reported in NRAO 12-m and MWO 4.9-m telescope observations of Sgr B2 by \citet{Ziurys:1986cw}.  The $J = 1\rightarrow0$, $2\rightarrow1$, and $3\rightarrow2$ transitions were targeted at 74, 148, and 222~GHz, respectively, based on frequencies estimated from the laboratory work of \citet{Altman:1984gk} and confirmed during the course of publication by the laboratory work of \citet{Bogey:1985fb}.
	
	\subsection{\ce{H3O^+} (hydronium cation)}
	\label{H3O+}
	
	\citet{Wootten:1986tg} and \citet{Hollis:1986fs} nearly simultaneously reported the detection of \ce{H3O^+}.  \citet{Wootten:1986tg} used the NRAO 12-m telescope to search for the $J_K = 1_1\rightarrow2_1$ transition at 307.2~GHz in Orion and Sgr B2, reporting a detection in each.  \citet{Hollis:1986fs} also observed the same transition in both Orion and Sgr B2 with the NRAO 12-m telescope.  Later, \citet{Wootten:1991js} reported the detection of a confirming line in Orion and Sgr B2 with the identification of the $3_2\rightarrow2_2$ transition of \emph{ortho}-\ce{H3O^+} using the CSO.   These detections were based on the laboratory work of \citet{Plummer:1985im}, \citet{Bogey:1985hy}, and \citet{Liu:1985gf}. 
	
	\subsection{\ce{C3S} (tricarbon monosulfide radical)}
	\label{C3S}
	
	The detection of \ce{C3S} was reported by \citet{Yamamoto:1987jd}.  The authors measured the laboratory rotational spectrum of the molecule, and assigned the $^1\Sigma$ $J = 4\rightarrow3$, $7\rightarrow6$, and $8\rightarrow7$ lines at 23, 40, and 46~GHz, respectively to previously unidentified lines in the observations of \citet{Kaifu:1987cf} toward TMC-1 using the Nobeyama 45-m telescope.
	
	\subsection{\ce{c-C3H} (cyclopropenylidene radical)}
	\label{c-C3H}
	
	\citet{Yamamoto:1987we} reported the detection of the hyperfine components of the $J = 5/2\rightarrow3/2$ and $3/2\rightarrow1/2$ transitions at 91.5 and 91.7~GHz, respectively, in Nobeyama 45-m observations of TMC-1.  The enabling laboratory spectroscopy was presented in the same manuscript.
	
	\subsection{\ce{HC2N} (cyanocarbene radical)}
	\label{HC2N}
	
	\citet{Guelin:1991uy} reported the detection of nine lines of \ce{HC2N} in IRAM 30-m observations of IRC+10216 between 72--239~GHz.  The enabling laboratory work was presented in \citet{Saito:1984ju} and \citet{Brown:1990gh}.
	
	\subsection{\ce{H2CN} (methylene amidogen radical)}
	\label{H2CN}
	
	The detection of \ce{H2CN} was reported in NRAO 12-m observations of TMC-1 by \citet{Ohishi:1994hk} who observed two hyperfine components of the $N_{K_a,K_c} = 1_{0,1}\rightarrow0_{0,0}$ transition at 73.3~GHz.  The laboratory spectroscopy was conducted by \citet{Yamamoto:1992fx}.
	
	\subsection{\ce{SiC3} (silicon tricarbide)}
	\label{SiC3}
	
	\citet{Apponi:1999jd} reported the detection of \ce{SiC3} in NRAO 12-m observations of IRC+10216.  Nine transitions of \ce{SiC3} between 81--103~GHz were observed and assigned based on laboratory work by the same group, published shortly thereafter in \citet{McCarthy:1999fs} and \citet{Apponi:1999iz}.
	
	\subsection{\ce{CH3} (methyl radical)}
	\label{CH3}
	
	The detection of \ce{CH3} was reported by \citet{Feuchtgruber:2000hh} using ISO observations of the $\nu_2$ $Q$-branch line at 16.5~$\mu$m and the $R(0)$ line at 16.0~$\mu$m toward Sgr~A*.  The enabling laboratory work was performed by \citet{Yamada:1981dk}.  The first ground-based detection of \ce{CH3} was described later by \citet{Knez:2009gp}.
	
	\subsection{\ce{C3N^-} (cyanoethynyl anion)}
	\label{C3N-}
	
	Both the laboratory spectroscopy and astronomical detection of \ce{C3N^-} were presented in \citet{Thaddeus:2008ux}.  The anion was detected in IRAM 30-m observations of IRC+10216 in the $J = 10\rightarrow9$, $11\rightarrow10$, $14\rightarrow13$, and $15\rightarrow14$ transitions at 97, 107, 136, and 146~GHz, respectively.
	
	\subsection{\ce{PH3} (phosphine)}
	\label{PH3}
	
	\citet{Agundez:2008kh} and \citet{Tenenbaum:2008ke} simultaneously and independently reported the tentative detection of \ce{PH3}.  \citet{Tenenbaum:2008ke} observed emission from the $J_K = 1_0\rightarrow0_0$ transition at 267~GHz toward IRC+10216 and CRL~2688 using the SMT telescope.  \citet{Agundez:2008kh} observed the same transition toward IRC+10216 using the IRAM 30-m telescope, and searched for the $J = 3\rightarrow2$ transition at 801~GHz using the CSO, unsuccessfully.  These detections were confirmed by \citet{Agundez:2014lw} through successful observation of the 534~GHz $J = 2\rightarrow1$ line toward IRC+10216 with \emph{Herschel}/HIFI.  The laboratory spectroscopy was presented in \citet{Cazzoli:2006dr}, \citet{SousaSilva:2013kr}, and \citet{Muller:2013fc}.
	
	\subsection{\ce{HCNO} (fulminic acid)}
	\label{HCNO}
	
	The discovery of HCNO was reported by \citet{Marcelino:2009jj} through observation of the $J = 5\rightarrow4$ and $4\rightarrow3$ transitions at 92 and 115~GHz, respectively, in IRAM 30-m observations of TMC-1, L1527, B1-b, L1544, and L183.  The laboratory rotational spectra were reported in \citet{Winnewisser:1971fh}.
	
	\subsection{\ce{HOCN} (cyanic acid)}
	\label{HOCN}
	
    The laboratory rotational spectrum and tentative astronomical identification of HOCN was reported by \citet{Brunken:2009ew} toward Sgr B2 using archival survey data.  The $J_{K_a,K_c} = 5_{0,5}\rightarrow4_{0,4}$ and $6_{0,6}\rightarrow5_{0,5}$ transitions at 104.9 and 125.8~GHz, respectively, were observed in Bell 7-m observations by \citet{Cummins:1986yt}. The $4_{0,4}\rightarrow3_{0,3}$ transition at 83.9~GHz was observed in the survey of \citet{Turner:1989rt} using the NRAO 36-ft telescope, but mis-assigned in that work to a transition of \ce{CH3OD}. These detections were confirmed shortly thereafter by \citet{Brunken:2010hp} using IRAM 30-m observations of Sgr B2 to self-consistently observe the $4_{0,4}\rightarrow3_{0,3}$ through $8_{0,8}\rightarrow7_{0,7}$ transitions.
    
    \subsection{\ce{HSCN} (thiocyanic acid)}
	\label{HSCN}
	
	\citet{Halfen:2009it} reported the detection of HSCN in ARO 12-m observations of Sgr B2.  Six lines between 69--138~GHz were observed and assigned to the $J_{K_a,K_c} = 6_{0,6}\rightarrow5_{0,5}$ through $12_{0,12}\rightarrow11_{0,11}$ transitions based on the laboratory work of \citet{Brunken:2009fh}.
	
	\subsection{\ce{HOOH} (hydrogen peroxide)}
	\label{HOOH}
	
	The detection of HOOH was reported by \citet{Bergman:2011dy} in APEX 12-m observations toward $\rho$ Oph A.  Three lines corresponding to the $J_{K_a,K_c} = 3_{0,3}\rightarrow2_{1,1}$ $\tau = 4\rightarrow2$, $6_{1,5}\rightarrow5_{0,5}$ $2\rightarrow4$, and $5_{0,5}\rightarrow4_{1,3}$ $4\rightarrow2$ torsion-rotation transitions at 219, 252, and 318~GHz, respectively, were assigned based on the laboratory work of \citet{Helminger:1981nd} and \citet{Petkie:1995dc}.
	
	\subsection{\ce{l-C3H^+} (cyclopropynylidynium cation)}
	\label{l-C3H+}
	
	\citet{Pety:2012cp} first reported the detection of a series of eight harmonically spaced unidentified lines in IRAM 30-m observations of the Horsehead PDR region which seemed to correspond to the $J = 4\rightarrow3$ through $12\rightarrow11$ transitions of a linear or quasi-linear molecule.  Based on the comparison of their derived rotational constants to \emph{ab initio} values calculated by \citet{Ikuta:1997it}, they tentatively assigned the carrier of these lines to the $l$-\ce{C3H^+} cation.  Using the derived spectroscopic parameters, \citet{McGuire:2013hc} subsequently detected the $1\rightarrow0$ and $2\rightarrow1$ transitions at 22.5 and 45~GHz, respectively, in absorption toward Sgr B2 using the GBT 100-m telescope.  Contemporaneously, \citet{Huang:2013fo} and \citet{Fortenberry:2013ew} challenged the attribution to $l$-\ce{C3H^+}, suggesting that the \ce{C3H^-} anion was a more likely candidate, based on better agreement between the distortion constants from their high-level quantum chemical calculations to those derived by \citet{Pety:2012cp}.  This assertion was challenged by \citet{McGuire:2014gt}, who argued that the formation and destruction chemistry in the high-UV environments the molecule was found strongly favored $l$-\ce{C3H+} over \ce{C3H^-}.  The assignment to $l$-\ce{C3H+} was shortly thereafter confirmed by the laboratory observation and characterization of the rotational spectrum of the cation by \citet{Brunken:2014cf}.
	
	\subsection{\ce{HMgNC} (hydromagnesium isocyanide)}
	\label{HMgNC}
	
	\citet{Cabezas:2013bo} reported the laboratory and astronomical identification of HMgNC in IRAM 30-m observations of IRC+10216. The $J = 8\rightarrow7$ through $13\rightarrow12$ transitions between 88 and 142~GHz were observed and assigned based on predicted values from the laboratory observations of the hyperfine-split $1\rightarrow0$ and $2\rightarrow1$ transitions at 11 and 22~GHz, respectively.
	
	\subsection{\ce{HCCO} (ketenyl radical)}
	\label{HCCO}
	
	The detection of HCCO was reported by \citet{Agundez:2015di} using IRAM 30-m observations of Lupus-1A and L483.  Four fine and hyperfine components of the $N = 4\rightarrow3$ transition of HCCO were resolved near 86.7~GHz and assigned based on the laboratory work of \citet{Endo:1987bt} and \citet{Ohshima:1993kp}.

	\subsection{\ce{CNCN} (isocyanogen)}
	\label{CNCN}
	
	\citet{Agundez:2018tm} reported the detection of \ce{CNCN} in L483 using IRAM 30-m observations.  The $J = 8\rightarrow7$, $9\rightarrow8$, $10\rightarrow9$ transitions at 82.8, 93.1, and 103.5~GHz were observed and assigned based on the laboratory work of \citet{Winnewisser:1992ed} and \citet{Gerry:1990kj}.  A tentative detection in TMC-1 was also reported in the same work.

    
    \begin{center}
        \large
        \textsc{\textbf{Five-Atom Molecules}}
        \label{5atoms}
        \normalsize
    \end{center}        

	\subsection{\ce{HC3N} (cyanoacetylene)}
	\label{HC3N}
	
	\citet{Turner:1971kj} reported the detection of the $F = 2\rightarrow1$ and $1\rightarrow1$ hyperfine components of the $J = 1\rightarrow0$ transition of \ce{HC3N} at 9098 and 9097~MHz, respectively, in NRAO 140-ft telescope observations of Sgr B2.  The assignment was based on the laboratory work of \citet{Tyler:1963hu}.  The detection was confirmed the following year by \citet{Dickinson:1972rt} through observation of two hyperfine components of the $J = 2\rightarrow1$ transition at 18~GHz in Haystack Observatory 120-ft telescope observations of Sgr B2.
	
	\subsection{\ce{HCOOH} (formic acid)}
	\label{HCOOH}
	
	The detection of HCOOH was reported by \citet{Zuckerman:1971de} in Sgr B2 using the NRAO 140-ft telescope.  The $J_{K_a,K_c} = 1_{1,1}\rightarrow1_{1,0}$ transition at 1639~MHz was identified and assigned based on laboratory work presented in the same manuscript.  Later, \citet{Winnewisser:1975we} confirmed the detection by observation of the $2_{1,1}\rightarrow2_{1,2}$ transition at 4.9~GHz in Sgr B2 using the MPIfR 100-m telescope.  This identification was based on the laboratory work of \citet{Bellet:1971fp} and \citet{Bellet:1971eo} (in French).
	
	\subsection{\ce{CH2NH} (methanimine)}
	\label{CH2NH}
	
	\citet{Godfrey:1973ui} reported the detection of \ce{CH2NH} in Sgr B2 using the Parkes 64-m telescope.  The $J = 1_{1,0}\rightarrow1_{1,1}$ transition at 5.2~GHz was observed and assigned based on laboratory work presented in the same manuscript, and building on the earlier laboratory efforts of \citet{Johnson:1972ij}.  \ce{CH2NH} is also referred to as methylenimine.
	
	\subsection{\ce{NH2CN} (cyanamide)}
	\label{NH2CN}
	
	\citet{Turner:1975pf} described the detection of \ce{NH2CN} in Sgr B2 by observation of its $J_{K_a,K_c} = 5_{1,4}\rightarrow4_{1,3}$ and $4_{1,3}\rightarrow3_{1,2}$ transitions at 100.6 and 80.5~GHz, respectively, using the NRAO 36-foot telescope.  The assignment was based on the work of \citet{Lide:1962hd}, \citet{Millen:1962ba}, and \citet{Tyler:1972bd}, and subsequently confirmed in the laboratory by \citet{Johnson:1976ew}.
	
	\subsection{\ce{H2CCO} (ketene)}
	\label{H2CCO}
	
	The detection of \ce{H2CCO} was reported by \citet{Turner:1977ig} in NRAO 36-foot telescope observations of Sgr B2.  The $J_{K_a,K_c} = 5_{1,4}\rightarrow4_{1,3}$, $5_{1,5}\rightarrow4_{1,4}$, and $4_{1,3}\rightarrow3_{1,2}$ transitions at 102, 100, and 82~GHz, respectively, were observed and assigned based on the laboratory work of \citet{Johnson:1952dp} and \citet{Johns:1972hd}.
	
	\subsection{\ce{C4H} (butadiynyl radical)}
	\label{C4H}
	
	\citet{Guelin:1978sd} reported the tentative detection of \ce{C4H} using NRAO 36-ft observations of IRC+10216 in combination with the previously published observations of \citet{Scoville:1978nq} and \citet{Liszt:1978jl} in complementary frequency ranges.  In total, four sets of spin doublet lines were observed between 86--114~GHz, and assigned to the $N = 9\rightarrow8$ through $N = 12\rightarrow11$ transitions of \ce{C4H} based on comparison to the theoretical rotational constants calculated by \citet{Wilson:1977rd}.  The detection was subsequently confirmed through the laboratory efforts of \citet{Gottlieb:1983yd}.
	
	\subsection{\ce{SiH4} (silane)}
	\label{SiH4}
	
	The detection of \ce{SiH4} was reported by \citet{Goldhaber:1984mh} in IRTF observations toward IRC+10216.  Thirteen absorption lines in the $\nu_4$ band near 917~cm$^{-1}$ were assigned based on laboratory work described in the same manuscript.
	
	\subsection{\ce{c-C3H2} (cyclopropenylidene)}
	\label{c-C3H2}
	
	The laboratory spectroscopy and interstellar detection of \ce{c-C3H2} was described in a pair of papers by \citet{Thaddeus:1985hw} and \citet{Vrtilek:1987iu}.  After the report of \citet{Thaddeus:1985hw}, the molecule began to be detected in numerous sources, with the best detections, as reported in \citet{Vrtilek:1987iu} being in Sgr B2, Orion, and TMC-1 with the Bell 7-m telescope.  A total of 11 transitions between 18--267~GHz were initially assigned.  
	
	\subsection{\ce{CH2CN} (cyanomethyl radical)}
	\label{CH2CN}
	
	\citet{Irvine:1988ej} reported the detection of the \ce{CH2CN} radical in TMC-1 and Sgr B2 using a combination of observations from the FCRAO 14-m, NRAO 140-ft, Onsala 20-m, and Nobeyama 45-m telescopes.  Numerous hyperfine-resolved components of the $N_{K_a,K_c} = 1_{0,1}\rightarrow0_{0,0}$ and $2_{0,2}\rightarrow1_{0,1}$ transitions at 20 and 40~GHz, respectively were observed in TMC-1, as well as an unresolved detection of the $4_{0,4}\rightarrow4_{0,3}$ transition at 101~GHz.  The assignments were made based on laboratory work presented in a companion paper from \citet{Saito:1988ba}.  Five partially-resolved components were also detected in Sgr B2 between 40--101~GHz.
	
	\subsection{\ce{C5} (pentacarbon)}
	\label{C5}
	
	The detection of \ce{C5} was reported by \citet{Bernath:1989md} in KPNO 4-m telescope observations of IRC+10216.  More than a dozen $P$- and $R$-branch transitions of the $\nu_3$ asymmetric stretching mode near 2164~cm$^{-1}$ were detected, guided by the laboratory work of \citet{Vala:1989ba}.
	
	\subsection{\ce{SiC4} (silicon tetracarbide)}
	\label{SiC4}
	
	\citet{Ohishi:1989ic} reported the detection of \ce{SiC4} in Nobeyama 45-m observations of IRC+10216.  Their search was initially guided by quantum chemical calculations they carried out, the results of which led to the assignment of five astronomically observed features between 37--89~GHz to the $J = 12\rightarrow11$, $13\rightarrow12$, $14\rightarrow13$, $16\rightarrow15$, and $28\rightarrow27$ transitions of \ce{SiC4}.  In the same work, they confirmed the detection by conducting laboratory microwave spectroscopy of the $J = 42\rightarrow41$ through $48\rightarrow47$ transitions.
	
	\subsection{\ce{H2CCC} (propadienylidene)}
	\label{H2CCC}
	
	The detection of \ce{H2CCC} was reported by \citet{Cernicharo:1991iv} in IRAM 30-m telescope observations and archival Effelsberg 100-m telescope observations \citep{Cernicharo:1987mj} of TMC-1.  The $J_{K_a,K_c} = 1_{0,1}\rightarrow0_{0,0}$ transition of \emph{para}-\ce{H2CCC} at 21~GHz, and three transitions of \emph{ortho}-\ce{H2CCC} between 103--147~GHz were identified based on the laboratory work of \citet{Vrtilek:1990gr}.
	
	\subsection{\ce{CH4} (methane)}
	\label{CH4}
	
	\ce{CH4} was detected in both the gas phase and solid phase in IRTF observations toward NGC 7538 IRS 9 by \citet{Lacy:1991fb}.  The gas-phase $R(0)$ line and two blended $R(2)$ lines at 1311.43, 1322.08, and 1322.16~cm$^{-1}$ were observed and assigned based on the laboratory work of \citet{Champion:1989cd}.  The solid-phase absorption of the $\nu_4$ mode  near 7.7~$\mu$m was observed and assigned based on the laboratory work of \citet{dHendecourt:1986wx}.
	
	\subsection{\ce{HCCNC} (isocyanoacetylene)}
	\label{HCCNC}
	
	\citet{Kawaguchi:1992bl} reported the detection of \ce{HCCNC}, also known as isocyanoethyne and  ethynyl isocyanide, in Nobeyama 45-m observations of TMC-1.  The $J = 4\rightarrow3$, $5\rightarrow4$, and $9\rightarrow8$ transitions at 40, 50, and 89~GHz, respectively, were identified and assigned based on the laboratory work of \citet{Krueger:2010gi}.
	
	\subsection{\ce{HNCCC}}
	\label{HNCCC}
	
	\citet{Kawaguchi:1992pw} reported both the laboratory spectroscopy and astronomical identification of HNCCC in Nobeyama 45-m telescope observations of TMC-1.  The $J = 3\rightarrow2$, $4\rightarrow3$, and $5\rightarrow4$ transitions at 28, 37, and 47~GHz, respectively, were assigned and identified.
	
	\subsection{\ce{H2COH^+} (protonated formaldehyde)}
	\label{H2COH+}
	
	The detection of \ce{H2COH^+} was reported by \citet{Ohishi:1996hc} in Nobeyama 45-m and NRAO 12-m telescope observations of Sgr B2, Orion, and W51.  Six transitions between 32--174~GHz were observed and assigned based on the laboratory work of \citet{Chomiak:1994bd}.
	
	\subsection{\ce{C4H^-} (butadiynyl anion)}
	\label{C4H-}
	
	\citet{Cernicharo:2007id} reported the detection of \ce{C4H^-} in IRAM 30-m observations of IRC+10216 through observation of five transitions between 84--140~GHz.  The detection was enabled by the laboratory work of \citet{Gupta:2007vc}.
	
	\subsection{\ce{CNCHO} (cyanoformaldehyde)}
	\label{CNCHO}
	
	\ce{CNCHO} was detected in GBT 100-m observations of Sgr B2(N) by \citet{Remijan:2008wn}.  A total of seven transitions between 2.1 and 41~GHz were identified and assigned based on the laboratory work of \citet{Bogey:1988bv} and \citet{Bogey:1995ci}.  \ce{CNCHO} is also referred to as formyl cyanide.
	
	\subsection{\ce{HNCNH} (carbodiimide)}
	\label{HNCNH}
	
	\citet{McGuire:2012jf} reported the detection of HNCNH in GBT 100-m observations of Sgr B2.  Four sets of transitions were identified to arise from maser action, while several other stronger transitions incapable of masing were not observed.  The assignments were made based on the laboratory work of \citet{Birk:1980yr}, \citet{Wagener:1995wu}, and \citet{Jabs:1997tr}.
	
	\subsection{\ce{CH3O} (methoxy radical)}
	\label{CH3O}
	
	The detection of \ce{CH3O} was reported in IRAM 30-m observations of B1-b by \citet{Cernicharo:2012eq}. Several hyperfine components of the $N = 1\rightarrow0$ transition near 82.4~GHz were identified and assigned based on the laboratory work of \citet{Endo:1984ej} and \citet{Momose:1988bu}.  Components of the $2\rightarrow1$ transition are shown, but no frequency information is provided other than that the transitions are at 2~mm.
	
	\subsection{\ce{NH3D^+} (deuterated ammonium cation)}
	\label{NH3D+}
	
	The detection of \ce{NH3D^+} was reported toward Orion and B1-b in IRAM 30-m observations by \citet{Cernicharo:2013uw}.  A single line with unresolved hyperfine structure was observed and assigned to the $J_K = 1_0\rightarrow0_0$ transition of \ce{NH3D^+} based on laboratory results presented in the same manuscript.  A companion article by \citet{Domenech:2013kl} detailed the laboratory results.
	
	\subsection{\ce{H2NCO^+} (protonated isocyanic acid)}
	\label{H2NCO+}
	
	\citet{Gupta:2013ky} reported a tentative detection of \ce{H2NCO^+} in GBT 100-m observations of Sgr B2.  Based on laboratory work presented in the same manuscript, they identified and assigned the $J_{K_a,K_c} = 0_{0,0}\rightarrow1_{0,1}$ and $1_{1,0}\rightarrow2_{1,1}$ transitions of \emph{para}- and \emph{ortho}-\ce{H2NCO^+} at 20.2 and 40.8~GHz, respectively.  The detection was confirmed by \citet{Marcelino:2018cp} in IRAM 30-m observations of L483.  They observed six additional transitions of \ce{H2NCO^+} between 80--116~GHz.
	
	\subsection{\ce{NCCNH^+} (protonated cyanogen)}
	\label{NCCNH+}
	
	\citet{Agundez:2015ko} reported the detection of \ce{NCCNH^+} in TMC-1 and L483 using IRAM 30-m and Yerbes 40-m observations.  The $J = 5\rightarrow4$ and $10\rightarrow9$ transitions at 44.3 and 88.8~GHz, respectively, were identified and assigned based on the laboratory work of \citet{Amano:1991jr} and \citet{Gottlieb:2000nc}.
	
	\subsection{\ce{CH3Cl} (chloromethane)}
	\label{CH3Cl}
	
	\ce{CH3Cl}, also called methyl chloride, was detected in ALMA observations of IRAS 16293-2422, and in ROSINA mass spectrometry measurements of comet 67P/Churyumov-Gerasimenko, by \citet{Fayolle:2017bg}.  Both the \ce{^{35}Cl} and \ce{^{37}Cl} isotopologues were detected through observation and assignment of the $J_K = 13_K\rightarrow12_K$ $K = 0$ to 4 transitions between 340--360~GHz.  The laboratory work was presented by \citet{Wlodarczak:1986bj}.
	
    
    \begin{center}
        \large
        \textsc{\textbf{Six-Atom Molecules}}
        \label{6atoms}
        \normalsize
    \end{center}        
    
    \subsection{\ce{CH3OH} (methanol)}
	\label{CH3OH}
	
	\citet{Ball:1970gi} reported the detection of \ce{CH3OH} in NRAO 140-ft observations of Sgr A and Sgr B2.  The $J_{K_a,K_c} = 1_{1,0}\rightarrow1_{1,1}$ transition at 834~MHz was observed and assigned based on laboratory work described in the same manuscript.  
	
	\subsection{\ce{CH3CN} (methyl cyanide)}
	\label{CH3CN}
	
	\ce{CH3CN} was detected by \citet{Solomon:1971by} in Sgr B2 and Sgr A using the NRAO 36-foot telescope.  Two fully resolved, and two blended $K$ components of the $J = 6\rightarrow5$ transition near 110.4~GHz were identified and assigned based on the laboratory data catalogued in \citet{Cord:1968kw}.  It is likely that the frequencies listed in \citet{Cord:1968kw} were obtained from, or relied heavily upon, the prior work by \citet{Kessler:1950jk}.  \ce{CH3CN} is also referred to as acetonitrile.
	
	\subsection{\ce{NH2CHO} (formamide)}
	\label{NH2CHO}
	
	The detection of \ce{NH2CHO} was reported by \citet{Rubin:1971hg} in NRAO 140-ft observations of Sgr B2.  The $J_{K_a,K_c} = 2_{1,1}\rightarrow2_{1,2}$, $F = 2\rightarrow2$ and $1\rightarrow1$ hyperfine components were resolved, while the $3\rightarrow3$ was detected as part of a blend with the H112$\alpha$ transition.  The enabling laboratory spectroscopy was reported in the same manuscript. \ce{NH2CHO} was the first interstellar molecule detected containing H, C, N, and O.
	
	\subsection{\ce{CH3SH} (methyl mercaptan)}
	\label{CH3SH}
	
	\citet{Linke:1979dc} reported the detection of \ce{CH3SH} in Bell 7-m telescope observations of Sgr B2.  Six transitions between 76--102~GHz were identified and assigned based on the laboratory work of \citet{Kilb:1955ge} and unpublished data of D.R. Johnson.  This latter dataset was analyzed and formed the foundation of the more extensive work of \citet{Lees:1980nv}.
	
	\subsection{\ce{C2H4} (ethylene)}
	\label{C2H4}
	
	The detection of \ce{C2H4} was reported by \citet{Betz:1981if} toward IRC+10216 using the 1.5-m McMath Solar Telescope.  Three rotationally-resolved ro-vibrational transitions were observed in the $\nu_7$ state: the $J_{K_a,K_c} = 5_{1,5}\rightarrow5_{0,5}$ at 28.5~THz was assigned based on the laboratory work of \citet{Lambeau:1980iw}, while the $1_{1,0}\rightarrow0_{0,0}$ and $8_{0,8}\rightarrow8_{1,8}$ were assigned based on laboratory work described in \citet{Betz:1981if}.
	
	\subsection{\ce{C5H} (pentynylidyne radical)}
	\label{C5H}
	
	A tentative detection of \ce{C5H} was reported by \citet{Cernicharo:1986gh} in IRAM 30-m observations of IRC+10216.  They identify a series of lines that could be assigned to a radical species in a $^2\Pi$ ground state with quantum numbers $J = 31/2\rightarrow29/2$ and $35/2\rightarrow33/2$ through $43/2\rightarrow41/2$ between 74--103~GHz.  Based on comparison of the derived rotational constant ($B_0 = 2387$~MHz) to a calculated rotational constant ($B_0 = 2375$~MHz) described in the same manuscript, they tentatively assigned the emission to the \ce{C5H} radical.  The subsequent laboratory work by \citet{Gottlieb:1986ew} confirmed the assignment, and the interstellar identifications were extended shortly thereafter by \citet{Cernicharo:1986ue} and \citet{Cernicharo:1987lw}.
	
	\subsection{\ce{CH3NC} (methyl isocyanide)}
	\label{CH3NC}
	
	A tentative detection of \ce{CH3NC} was reported by \citet{Cernicharo:1988kj} in IRAM 30-m observations of Sgr B2.  Somewhat blended signal was observed near the frequencies corresponding to the $J = 4\rightarrow3$, $5\rightarrow4$, and $7\rightarrow6$ transitions at 80.4, 100.5, and 140.7~GHz, respectively.  Although no reference is given for these frequencies, they were presumably enabled by the laboratory work of \citet{Kukolich:1972iq} and \citet{Ring:1947bn}.  The detection was confirmed nearly two decades later in GBT 100-m observations of Sgr B2 by \citet{Remijan:2005kd} who observed the $J_K = 1_0\rightarrow0_0$ transition in absorption at 20.1~GHz and by \citet{Gratier:2013gm} who observed three lines in the $J = 5\rightarrow4$ transition at 100.5~GHz with the IRAM 30-m toward the Horsehead PDR.
	
	\subsection{\ce{HC2CHO} (propynal)}
	\label{HC2CHO}
	
	\citet{Irvine:1988dw} reported the detection of \ce{HC2CHO} in NRAO 140-ft and Nobeyama 45-m observations of TMC-1.  The $J_{K_a,K_c} = 2_{0,2}\rightarrow1_{0,1}$ and $4_{0,4}\rightarrow3_{0,3}$ transitions at 18.7 and 37.3~GHz, respectively, were assigned based on the laboratory work of \citet{Winnewisser:1973gc}.
	
	\subsection{\ce{H2C4} (butatrienylidene)}
	\label{H2C4}
	
	\citet{Cernicharo:1991ck} reported the detection of the \ce{H2C4} carbene molecule in IRAM 30-m observations of IRC+10216.  Eight transitions with $J$ values between 8--15 where identified and assigned between 80--135~GHz based on the laboratory spectroscopy of \citet{Killian:1990di}.
	
	\subsection{\ce{C5S} (pentacarbon monosulfide radical)}
	\label{C5S}
	
	The \ce{C5S} radical was tentatively detected by \citet{Bell:1993ir} in NRAO 140-ft telescope observations of IRC+10216.  A weak line corresponding to the $J = 13\rightarrow12$ transition of \ce{C5S} was identified near 23960~MHz based on the laboratory work of \citet{Kasai:1993ds}.  The detection was confirmed in \citet{Agundez:2014gm} with IRAM 30-m observations of IRC+10216.  The $J = 44\rightarrow43$, $45\rightarrow44$, and $46\rightarrow45$ transitions at 81.2, 83.0, and 84.9~GHz, respectively, were identified and assigned based on the laboratory work of \citet{Gordon:2001pd}.
	
	\subsection{\ce{HC3NH+} (protonated cyanoacetylene)}
	\label{HC3NH+}
	
	\citet{Kawaguchi:1994ez} reported the detection of \ce{HC3NH^+} in Nobeyama 45-m observations of TMC-1.  The $J = 4\rightarrow3$ and $5\rightarrow4$ transitions at 34.6 and 43.3~GHz, respectively, were identified and assigned based on the frequencies reported by \citet{Lee:1987dk} from infrared difference frequency laser spectroscopy.  Using the astronomically detected pure rotational lines as a guide, the rotational spectrum was then measured in the laboratory by \citet{Gottlieb:2000nc}.
	
	\subsection{\ce{C5N} (cyanobutadiynyl radical)}
	\label{C5N}
	
	The \ce{C5N} radical was detected by \citet{Guelin:1998kf} in Effelsberg 100-m and IRAM 30-m observations of TMC-1.  Two sets of hyperfine components of the $J = 17/2\rightarrow15/2$ and $65/2\rightarrow63/2$ transitions of \ce{C5N} at 25.2 and 89.8~GHz, respectively, were identified and assigned based on the laboratory work of \citet{Kasai:1997gx}.
	
	\subsection{\ce{HC4H} (diacetylene)}
	\label{HC4H}
	
	\citet{Cernicharo:2001mw} reported the detection of \ce{HC4H} in ISO observations of CRL 618 near 15.9~$\mu$m.  The observed absorption signal was identified and assigned to the $\nu_8$ fundamental bending mode of \ce{HC4H} based on the laboratory work of \citet{Arie:1992gh}.
	
	\subsection{\ce{HC4N}}
	\label{HC4N}
	
	The \ce{HC4N} radical was detected by \citet{Cernicharo:2004eh} in IRAM 30-m observations of IRC+10216.  A dozen transitions beteween 83--97~GHz were identified and assigned based on the laboratory work of \citet{Tang:1999yt}.
	
	\subsection{\ce{c-H2C3O} (cyclopropenone)}
	\label{c-H2C3O}
	
	\citet{Hollis:2006ih} reported the detection of \ce{c-H2C3O} in GBT 100-m observations of Sgr B2.  Six transitions between 9.3--44.6~GHz were identified and assigned based on the laboratory work of \citet{Benson:1973bs} and \citet{Guillemin:1990hp}.
	
	\subsection{\ce{CH2CNH} (ketenimine)}
	\label{CH2CNH}
	
	The detection of \ce{CH2CNH} was reported by \citet{Lovas:2006pj} using GBT 100-m telescope observations of Sgr B2.  The $J_{K_a,K_c} = 9_{1,8}\rightarrow10_{0,10}$, $8_{1,7}\rightarrow9_{0,9}$, and $7_{1,6}\rightarrow8_{0,8}$ transitions at 4.9, 23.2, and 41.5~GHz, respectively, were identified and assigned based on the laboratory work of \citet{Rodler:1984jm} and \citet{Rodler:1986hb}.
	
	\subsection{\ce{C5N^-} (cyanobutadiynyl anion)}
	\label{C5N-}
	
    \citet{Cernicharo:2008wi} reported the detection of \ce{C5N^-} in IRAM 30-m observations of IRC+10216.  They identify and assign 11 lines to the $J = 29\rightarrow28$ through $40\rightarrow39$ transitions of \ce{C5N^-} by comparison to the calculated rotational constants presented in \citet{Botschwina:2008bw}.  To date, there does not appear to be a published pure rotational laboratory spectrum of this species.
    
    \subsection{\ce{HNCHCN} (E-cyanomethanimine)}
	\label{HNCHCN}
	
	\ce{HNCHCN} was detected in GBT 100-m telescope observations of Sgr B2 as reported by \citet{Zaleski:2013bc}.  Nine transitions between 9.6--47.8~GHz were identified and assigned based on laboratory spectroscopy presented in the same manuscript.
	
	\subsection{\ce{SiH3CN} (silyl cyanide)}
	\label{SiH3CN}
	
	The tentative detection of \ce{SiH3CN} was reported by \citet{Agundez:2014gm} in IRAM 30-m observations of IRC+10216 through the identification of three weak emission features that they assign to the $J = 9\rightarrow8$, $10\rightarrow9$, and $11\rightarrow10$ transitions of \ce{SiH3CN} at 89.5, 99.5, and 109.4~GHz, respectively, based on the laboratory work of \citet{Priem:1998en}.  The detection was confirmed by \citet{Cernicharo:2017gc} through the detection of additional, higher-frequency transitions in this source.
	
    
    \begin{center}
        \large
        \textsc{\textbf{Seven-Atom Molecules}}
        \label{7atoms}
        \normalsize
    \end{center}        
	
	\subsection{\ce{CH3CHO} (acetaldehyde)}
	\label{CH3CHO}
	
	The detection of the $J_{K_a,K_c} = 1_{1,1}\rightarrow1_{1,0}$ transition of \ce{CH3CHO} at 1065~MHz was reported by \citet{Gottlieb:1973vc} in NRAO 140-ft observations of Sgr B2 and Sgr A.  The assignment was made based on the laboratory work of \citet{Kilb:1957bv} and \citet{Souter:1970bc}.  \citet{Fourikis:1974et} reported a confirming observation of the $2_{1,1}\rightarrow2_{1,2}$ transition in Parkes 64-m observations of Sgr B2 at 3195~MHz.
	
	\subsection{\ce{CH3CCH} (methyl acetylene)}
	\label{CH3CCH}
	
	\citet{Buhl:1973tp} reported the detection of the $J_K = 5_0\rightarrow4_0$ transition of \ce{CH3CCH} at 85.4~GHz in NRAO 36-foot telescope observations of Sgr B2.  Although no reference to the source of the frequency is given, it was presumably obtained from the laboratory work of \citet{Trambarulo:1950hi}.  Subsequent confirming transitions were observed at higher frequencies by \citet{Hollis:1981kl} and \citet{Kuiper:1984rp}.
    
    \subsection{\ce{CH3NH2} (methylamine)}
	\label{CH3NH2}
	
	The detection of \ce{CH3NH2} was simultaneously reported by \citet{Fourikis:1974db} and \citet{Kaifu:1974nq}.  \citet{Kaifu:1974nq} used the Mitaka 6-m and NRAO 36-ft telescopes to observe \ce{CH3NH2} toward Sgr B2 and Orion.  The $a$, $J_{K_a,K_c} = 5_{1,5}\rightarrow5_{0,5}$ and $s$, $4_{1,4}\rightarrow4_{0,4}$ transitions at 73 and 86~GHz, respectively, were assigned based on the laboratory work of \citet{Takagi:1973kq}.  Using the same laboratory work, \citet{Fourikis:1974db} reported the detection of the $a$, $2_{0,2}\rightarrow1_{1,0}$ transition at 8.8~GHz using the Parkes 64-m telescope, also toward Sgr B2 and Orion.
	
	\subsection{\ce{CH2CHCN} (vinyl cyanide)}
	\label{CH2CHCN}
	
	\ce{CH2CHCN}, also known as acrylonitrile, was detected by \citet{Gardner:1975ef} in Parkes 64-m observations of Sgr B2.  The $J_{K_a,K_c} = 2_{1,1}\rightarrow2_{1,2}$ transition at 1372~MHz was identified and assigned based on the laboratory work of \citet{Gerry:1973ju}.  Several additional confirming transitions were later observed in NRAO 140-ft telescope observations of TMC-1 by \citet{Matthews:1983df}.
	
	\subsection{\ce{HC5N} (cyanodiacetylene)}
	\label{HC5N}
	
	\citet{Avery:1976eq} reported the detection of \ce{HC5N} in Algonquin Radio Observatory 46-m telescope observations of Sgr B2.  They observed and assigned the $J = 4\rightarrow3$ transition at 10651~MHz based on the laboratory work of \citet{Alexander:1976ed}.  Later that year, \citet{Broten:1976do} observed the $1\rightarrow0$ and $8\rightarrow7$ transitions with the same facility, also in Sgr B2.
	
	\subsection{\ce{C6H} (hexatriynyl radical)}
	\label{C6H}
	
	The first detection of the \ce{C6H} radical was made by \citet{Suzuki:1986bc} in Nobeyama 45-m observations of TMC-1.  The authors observed three sets of doublet transitions at 23.6, 40.2, and 43.0~GHz which they assigned to the hyperfine-split, $\Lambda$-doubled $J = 17/2\rightarrow15/2$, $29/2\rightarrow27/2$, and $31/2\rightarrow29/2$ transitions, respectively, of \ce{C6H} based on a comparison to the quantum chemical work of \citet{Murakami:1987ys}.  The detection was confirmed the following year with the measurement of the laboratory rotational spectrum by \citet{Pearson:1988qm}.
	
	\subsection{\ce{c-C2H4O} (ethylene oxide)}
	\label{c-C2H4O}
	
	\citet{Dickens:1997fb} reported the detection of \ce{c-C2H4O} in Nobeyama 45-m, Haystack 140-ft, and SEST 15-m telescope observations of Sgr B2.  Nine transitions between 39.6--254.2~GHz were identified and assigned based on the laboratory rotational spectroscopy performed by \citet{Hirose:1974wq}.
	
	\subsection{\ce{CH2CHOH} (vinyl alcohol)}
	\label{CH2CHOH}
	
	\citet{Turner:2001jk} reported the detection of both \emph{syn}- and \emph{anti}-vinyl alcohol in NRAO 12-m observation of Sgr B2.  The \emph{syn} conformer of \ce{CH2CHOH} is the more stable, but only two lines, the $J_{K_a,K_c} = 2_{1,2}\rightarrow1_{0,1}$ and $3_{1,3}\rightarrow2_{0,2}$ transitions at 86.6 and 103.7~GHz, respectively, were identified and assigned based on the laboratory work of \citet{Kaushik:1977je}.  In contrast, five transitions of the \emph{anti} conformer were identified between 71.8--154.5~GHz based on the laboratory work of \citet{Rodler:1985dd}.
	
	\subsection{\ce{C6H^-} (hexatriynyl anion)}
	\label{C6H-}
	
	The first molecular anion to be detected in the ISM, \citet{McCarthy:2006tk} described both the observation and laboratory spectroscopic identification of \ce{C6H^-}.  More than a decade prior, \citet{Kawaguchi:1995tx} had noted the presence of a series of unidentified, harmonically spaced emission features in their Nobeyama 45-m survey of IRC+10216 dubbed B1377.  Based on their laboratory work, \citet{McCarthy:2006tk} successfully assigned the carrier of these transitions to \ce{C6H^-}.  They also identified the $J = 4\rightarrow3$ and $5\rightarrow4$ transitions at 11.0 and 13.8~GHz, respectively, in GBT 100-m telescope observations of TMC-1.
	
	\subsection{\ce{CH3NCO} (methyl isocyanate)}
	\label{CH3NCO}
	
	The detection of \ce{CH3NCO} was reported by \citet{Halfen:2015ez} in ARO 12-m and SMT observations of Sgr B2.  They observed 17 uncontaminated emission lines of \ce{CH3NCO} between 68--116~GHz and assigned them based on laboratory work described in the same manuscript.  The following year, \citet{Cernicharo:2016fm} reported the detection of the molecule in IRAM 30-m and ALMA data toward Orion, as well as extending the laboratory spectroscopy.
	
	\subsection{\ce{HC5O} (butadiynylformyl radical)}
	\label{HC5O}
	
	\citet{McGuire:2017ud} reported the detection of \ce{HC5O} in GBT 100-m observations of TMC-1.  Four hyperfine-resolved components of the $J = 17/2\rightarrow15/2$ transition at 21.9~GHz were detected and assigned based on the laboratory work of \citet{Mohamed:2005ku}.
	
    
    \begin{center}
        \large
        \textsc{\textbf{Eight-Atom Molecules}}
        \label{8atoms}
        \normalsize
    \end{center}      
    
    \subsection{\ce{HCOOCH3} (methyl formate)}
	\label{HCOOCH3}
	
	\citet{Brown:1975jh} reported the detection of \emph{cis}-\ce{HCOOCH3} in Parkes 64-m telescope observations of Sgr B2.  Based on laboratory work presented in the same paper, they identified and assigned the $A$-state $J_{K_a,K_c} = 1_{1,0}\rightarrow1_{1,1}$ transition of \emph{cis}-\ce{HCOOCH3} at 1610.25~MHz.  Several months later, \citet{Churchwell:1975ps} confirmed the detection of the $A$-state line and additionally observed the $E$-state line of the same transition at 1610.9~MHz using the Effelsberg 100-m telescope.  Nearly three decades later, \citet{Neill:2012fr} reported the tentative identification of the higher-energy \emph{trans} conformer of \ce{HCOOCH3} in GBT 100-m telescope observations of Sgr B2, based on laboratory spectroscopy presented in the same work.  A total of seven transitions between 9.1--27.4~GHz were observed.
	
	\subsection{\ce{CH3C3N} (methylcyanoacetylene)}
	\label{CH3C3N}
	
	The detection of \ce{CH3C3N} was reported by \citet{Broten:1984xp} in NRAO 140-ft observations of TMC-1.  A total of seven components of the $J = 5\rightarrow4$ through $8\rightarrow7$ transitions of \ce{CH3C3N} between 20.7--33.1~GHz were resolved and assigned based on the laboratory work of \citet{Moises:1982wx}. 
	
	\subsection{\ce{C7H} (heptatriynylidyne radical)}
	\label{C7H}
	
	\citet{Guelin:1997uy} reported the detection of the \ce{C7H} radical in IRAM 30-m telescope observations of IRC+10216.  Based on the laboratory spectroscopy of \citet{Travers:1996gx}, they identified and assigned five rotational transitions between 83--86.6~GHz.
	
	\subsection{\ce{CH3COOH} (acetic acid)}
	\label{CH3COOH}
	
	The detection of \ce{CH3COOH} was reported by \citet{Mehringer:1997vk} in BIMA and OVRO observations of Sgr B2.  The $J_{K_a,K_c} = 8_{*,8}\rightarrow7_{*,7}$ $A$ and $E$-state lines of \ce{CH3COOH} around 90.2~GHz, and the $9_{*,9}\rightarrow8_{*,8}$ $E$-state line at 100.9~GHz were identified and assigned based on the work of \citet{Tabor:1957ck} and \citet{Wlodarczak:1988qz}. 
	
	\subsection{\ce{H2C6} (hexapentaenylidene)}
	\label{H2C6}
	
	\citet{Langer:1997hk} reported the detection of \ce{H2C6} in Goldstone 70-m telescope observations of TMC-1.  The $J_{K_a,K_c} = 7_{1,7}\rightarrow6_{1,6}$ and $8_{1,8}\rightarrow7_{1,7}$ transitions of \ce{H2C6} at 18.8 and 21.5~GHz, respectively, were identified and assigned based on the laboratory work of \citet{McCarthy:1997ee}.
	
	\subsection{\ce{CH2OHCHO} (glycolaldehyde)}
	\label{glycolaldehyde}
	
	The detection of \ce{CH2OCHO} was reported by \citet{Hollis:2000gn} in NRAO 12-m observations of Sgr B2.  A total of five transitions between 71.5--103.7~GHz were identified and assigned based on the laboratory work of \citet{Marstokk:1973hl}.  Despite numerous claims in the astronomical literature, glycolaldehyde is not a true sugar, and is instead a diose, a 2-carbon monosaccharide, making it the simplest sugar-related molecule.
	
	\subsection{\ce{HC6H} (triacetylene)}
	\label{HC6H}
	
	\citet{Cernicharo:2001mw} reported the detection of \ce{HC6H} in ISO observations of CRL 618.  The $\nu_{11}$ fundamental bending mode of \ce{HC6H} at 16.1~$\mu$m was observed and assigned based ont he laboratory work of \citet{Haas:1994bt}.
	
	\subsection{\ce{CH2CHCHO} (propenal)}
	\label{CH2CHCHO}
	
	A tentative detection of \ce{CH2CHCHO} was reported by \citet{Dickens:2001uc} in NRAO 12-m and SEST 15-m observations of G327.3-0.6 and Sgr B2.  They tentatively assigned six emission lines between 88.5--107.5~GHz to \ce{CH2CHCHO}, based on the laboratory work of \citet{Winnewisser:1975kq}.  The detection was confirmed by \citet{Hollis:2004oc} in GBT 100-m telescope observations of Sgr B2.  They observed and assigned the $J_{K_a,K_c} = 2_{1,1}\rightarrow1_{1,0}$ and $3_{1,3}\rightarrow2_{1,2}$ transitions at 18.2 and 26.1~GHz, respectively, based on the laboratory work of \citet{Blom:1984is}.  \ce{CH2CHCHO} is also referred to as acrolein.
	
	\subsection{\ce{CH2CCHCN} (cyanoallene)}
	\label{CH2CCHCN}
	
	\citet{Lovas:2006pq} reported the detection of \ce{CH2CCHCN} in GBT 100-m observations of TMC-1.  Four transitions between 20.2--25.8~GHz with partially resolved hyperfine structure were identified and assigned based on the laboratory work of \citet{Bouchy:1973db} and \citet{Schwahn:1986ei}.
	
	\subsection{\ce{NH2CH2CN} (aminoacetonitrile)}
	\label{NH2CH2CN}
	
	The detection of \ce{NH2CH2CN} was reported by \citet{Belloche:2008jy} in IRAM 30-m, PdBI, and ATCA observations of Sgr B2. A total of 51 emission features between 80--116~GHz were identified and assigned to transitions of \ce{NH2CH2CN} based on the laboratory work of \citet{Bogey:1990ej}.  As part of their work, \citet{Belloche:2008jy} refined the spectroscopic fit for the molecule based on the work of \citet{Bogey:1990ej} and references therein.
	
	\subsection{\ce{CH3CHNH} (ethanimine)}
	\label{CH3CHNH}
	
	\citet{Loomis:2013fs} reported the detection of \ce{CH3CHNH} in GBT 100-m observations of Sgr B2.  More than two dozen transitions of the molecule between 13.0--47.2~GHz were identified and assigned based on laboratory spectroscopy work presented in the same manuscript.
	
	\subsection{\ce{CH3SiH3} (methyl silane)}
	\label{CH3SiH3}
	
	The detection of \ce{CH3SiH3} was reported by \citet{Cernicharo:2017gc} in IRAM 30-m observations of IRC+10216.  Ten transitions between 80--350~GHz were identified and assigned based on the laboratory work of \citet{Meerts:1982fn} and \citet{Wong:1983iy}.
	
    
    \begin{center}
        \large
        \textsc{\textbf{Nine-Atom Molecules}}
        \label{9atoms}
        \normalsize
    \end{center}      
    
    \subsection{\ce{CH3OCH3} (dimethyl ether)}
	\label{CH3OCH3}
	
	\citet{Snyder:1974op} reported the detection of \ce{CH3OCH3} in NRAO 36-ft, and Maryland Point Observatory NRL 85-ft observations of Orion.  The $J_{K_a,K_c} = 6_{0,6}\rightarrow5_{1,5}$, $2_{1,1}\rightarrow2_{0,2}$, and $2_{2,0}\rightarrow2_{1,1}$ transitions at 90.2, 31.1, and 88.2~GHz, respectively, were observed and assigned based on the laboratory work of \citet{Kasai:1959ed} and \citet{Blukis:1963hm}.
	
	\subsection{\ce{CH3CH2OH} (ethanol)}
	\label{CH3CH2OH}
	
	The detection of \emph{trans}-\ce{CH3CH2OH} was reported by \citet{Zuckerman:1975jx} in NRAO 36-foot telescope observations of Sgr B2.  The $J_{K_a,K_c} = 6_{0,6}\rightarrow5_{1,5}$, $4_{1,4}\rightarrow3_{0,3}$, and $5_{1,5}\rightarrow4_{0,4}$ transitions at 85.3, 90.1, and 104.8~GHz, respectively, were identified and assigned based on the laboratory work of \citet{Takano:1968kz}.  The \emph{gauche} substates of \ce{CH3CH2OH} were later identified as the carriers of 14 previously unidentified emission features in the Nobeyama 45-m observations of Orion presented in \citet{Ohishi:1988co} based on the laboratory work of \citet{Pearson:1997ki}.
	
	\subsection{\ce{CH3CH2CN} (ethyl cyanide)}
	\label{CH3CH2CN}
	
	\citet{Johnson:1977ky} reported the detection of \ce{CH3CH2CN} in NRAO 36-foot observations of Orion, and weakly in Sgr B2.  Two dozen emission features between 88--116~GHz were identified and assigned to \ce{CH3CH2CN} in Orion.  The foundational laboratory spectroscopy was reported by \citet{Heise:1974et}, \citet{Mader:1974ko}, and \citet{Laurie:1959eq}.  Based on this work, additional higher-frequency lines were measured and assigned in \citet{Johnson:1977ky}.
	
	\subsection{\ce{HC7N} (cyanotriacetylene)}
	\label{HC7N}
	
	Reports of the detection of \ce{HC7N} were made nearly simultaneously by \citet{Little:1978hn} (4 January) and \citet{Kroto:1978fr} (1 February), although the former acknowledges an earlier report by \citet{Kroto:1977kq} (13 June).  A few months later (5 October), \citet{Winnewisser:1978bd} also reported the detection of \ce{HC7N}. 
	
	The observations of \citet{Kroto:1977kq} and \citet{Kroto:1978fr} were conducted toward TMC-1 using both the Algonquin 46-m and Haystack 36.6-m telescopes.  \citet{Kroto:1977kq} reported the observation of the $J = 9\rightarrow8$ transition at 10.2~GHz, while \citet{Kroto:1978fr} additionally observed the $21\rightarrow20$ transition at 23.7~GHz.  The assignments were based on laboratory work carried out by a subset of the authors of \citet{Kroto:1978fr}, which they would later report in \citet{Kirby:1980ld}.
	
	\citet{Little:1978hn} used the SRC Appleton Laboratory 25-m telescope to observe the $22\rightarrow21$ transition at 24.8~GHz in both TMC-1 and TMC-2.  The rotational constants required for the detection were provided by H. W. Kroto.
	
	The work of \citet{Winnewisser:1978bd} was presumably carried out independently, as there were no references to the previous three manuscripts in this paper. They reported the detection of the $21\rightarrow20$ transition in IRC+10216.  Although the facility used for the observations is not named in the main portion of the manuscript, the authors acknowledge the operators of Effelsberg, and cite a beam size of 40$^{\prime\prime}$ near 24~GHz, which is reasonably close to that of a 100-m telescope.  There is also no reference given to the source of the rotational transition frequency, nor is a frequency actually given.
	
	\subsection{\ce{CH3C4H} (methyldiacetylene)}
	\label{CH3C4H}
	
	The detection of \ce{CH3C4H} was nearly simultaneously reported by \citet{Walmsley:1984od} (22 March) and \citet{MacLeod:1984er} (15 July).  Later that year, \citet{Loren:1984bc} (1 November) also reported an independent detection, having received word of the earlier work during the publication process.
	
	\citet{Walmsley:1984od} reported the detection in observations of TMC-1 using the Effelsberg 100-m telescope. They observed and assigned the $K = 0$ and $K = 1$ components of the $J = 6\rightarrow5$ and $5\rightarrow4$ transitions at 24.4 and 20.3~GHz, respectively.  Their assignments were based on the laboratory work of \citet{Heath:1955fj}, although they suggest the astronomical observations might be used to refine those measurements moving forward.
	
	Both \citet{MacLeod:1984er} and \citet{Loren:1984bc} observed the same transitions in TMC-1, the former using the Haystack 36.6-m and NRAO 140-foot telescopes, while the latter used solely the NRAO 140-foot.
	
	\subsection{\ce{C8H} (octatriynyl radical)}
	\label{C8H}
	
	\citet{Cernicharo:1996kd} reported the detection of the \ce{C8H} radical in IRAM 30-m observations of IRC+10216, as well as in the archival Nobeyama 45-m observations of \citet{Kawaguchi:1995tx}.  They identified and assigned ten transitions between 31.1--83.9~GHz based on comparison to calculated rotational constants from \citet{Pauzat:1991bx}.  The detection and assignment was confirmed in an accompanying letter by the laboratory work of \citet{McCarthy:1996wx}.
	
	\subsection{\ce{CH3CONH2} (acetamide)}
	\label{acetamide}
	
	The detection of \ce{CH3CONH2} was reported toward Sgr B2 by \citet{Hollis:2006uc} using the GBT 100-m telescope.  A total of eight transitions between 9.3--47.4~GHz were identified and assigned based on the laboratory work of \citet{Suenram:2001dx} and \citet{Ilyushin:2004jj}.
	
	\subsection{\ce{C8H^-} (octatriynyl anion)}
	\label{C8H-}
	
	\citet{Brunken:2007yd} and \citet{Remijan:2007vp} simultaneously reported the detection of \ce{C8H^-}.  Both detections were made with the GBT 100-m telescope; the former described the detection in TMC-1 while the latter was toward IRC+10216.  \citet{Remijan:2007vp} detected the $J = 22\rightarrow21$ and $35\rightarrow34$ through $38\rightarrow37$ transitions between 25.7--44.3~GHz based on the laboratory work of \citet{Gupta:2007vc}.  \citet{Brunken:2007yd} observed the $11\rightarrow10$ through $13\rightarrow12$, and the $16\rightarrow15$ transitions between 12.8--18.7~GHz, based on the same laboratory work.
	
	\subsection{\ce{CH2CHCH3} (propylene)}
	\label{CH2CHCH3}
	
	The detection of \ce{CH2CHCH3}, also referred to as propene, was reported by \citet{Marcelino:2007pu} in IRAM 30-m observations of TMC-1.  A total of thirteen rotational transitions between 84.2--103.7~GHz were identified and assigned based on the laboratory work of \citet{Wlodarczak:1994fb} and \citet{Pearson:1994kt}.
	
	\subsection{\ce{CH3CH2SH} (ethyl mercaptan)}
	\label{CH3CH2SH}
	
	\citet{Kolesnikova:2014fb} reported the detection of \ce{CH3CH2SH} in IRAM 30-m observations of Orion.  A total of 77 unblended lines between 80--280~GHz were identified and assigned based on laboratory spectroscopy presented in the same manuscript.  Subsequent attempts to identify the molecule in Sgr B2(N) by \citet{Muller:2016kd} were unsuccessful.
	
	\subsection{\ce{HC7O} (hexadiynylformyl radical)}
	\label{HC7O}
	
	\citet{McGuire:2017ud} reported the tentative detection of the \ce{HC7O} radical in GBT 100-m observations of TMC-1.  Based on the laboratory work of \citet{Mohamed:2005ku}, a weak detection of the $J = 35/2\rightarrow33/2$ transition with partially resolved hyperfine structure was observed near 19.2~GHz.  The detection was confirmed in \citet{Cordiner:2017dq} by the observation of several additional weak features, without hyperfine structure, corresponding to the $33/2\rightarrow31/2$ transitions near 18.1~GHz.  A composite average of these lines, plus the data from \citet{McGuire:2017ud}, provided a 9.5$\sigma$ detection.
	
    
    \begin{center}
        \large
        \textsc{\textbf{Ten-Atom Molecules}}
        \label{10atoms}
        \normalsize
    \end{center}    
    
    \subsection{\ce{(CH3)2CO} (acetone)}
	\label{acetone}
	
	\citet{Combes:1987re} reported the detection of \ce{(CH3)2CO} in IRAM 30-m, NRAO 140-ft, and NRAO 12-m observations of Sgr B2.  A total of 11 transitions between 18.7--112.4~GHz were identified and assigned based on the laboratory work of \citet{Vacherand:1986kh}.
	
	\subsection{\ce{HO(CH2)2OH} (ethylene glycol)}
	\label{HOCH2CH2OH}
	
	The detection of the $g^{\prime}Ga$ conformer of \ce{HO(CH2)2OH} was reported by \citet{Hollis:2002ke} in NRAO 12-m observations of Sgr B2.  Four transitions between 75.1--93.0~GHz were identified and assigned based on the laboratory work of \citet{Christen:1995eu}.  The higher-energy $aGg^{\prime}$ conformer was later detected by \citet{Rivilla:2017ce} in GBT 100-m, IRAM 30-m, and SMA observations of G31.41+0.31.  
	
	\subsection{\ce{CH3CH2CHO} (propanal)}
	\label{CH2CH2CHO}
	
	\citet{Hollis:2004oc} reported the detection of \ce{CH3CH2CHO} in GBT 100-m observations of Sgr B2.  A total of five transitions between 19.2--22.2~GHz were identified and assigned based on the laboratory work of \citet{Butcher:1964jbb}.  \ce{CH3CH2CHO} is also referred to as propionaldehyde.
	
	\subsection{\ce{CH3C5N} (methylcyanodiacetylene)}
	\label{CH3C5N}
	
	The detection of \ce{CH3C5N} was reported by \citet{Snyder:2006vi} in GBT 100-m telescope observations of TMC-1.  Ten transitions between 18.7--24.9~GHz were identified and assigned to the $K=0$ and $K=1$ components of the $J_K = 12_*\rightarrow11_*$ through $16_*\rightarrow15_*$ transitions of \ce{CH3C5N} based on the laboratory work of \citet{Alexander:1978bd} and \citet{Chen:1998hg}.
	
	\subsection{\ce{CH3CHCH2O} (propylene oxide)}
	\label{CH3CHCH2O}
	
	\citet{McGuire:2016ba} reported the detection of \ce{CH3CHCH2O} in GBT 100-m and Parkes 64-m observations of Sgr B2.  The $J_{K_a,K_c} = 1_{1,0}\rightarrow1_{0,1}$, $2_{1,1}\rightarrow2_{0,2}$, and $3_{1,2}\rightarrow3_{0,3}$ transitions at 12.1, 12.8, and 14.0~GHz, respectively, were identified and assigned based on laboratory spectroscopy presented in the same manuscript.  \ce{CH3CHCH2O} was the first chiral molecule detected in the interstellar medium.
	
	\subsection{\ce{CH3OCH2OH} (methoxymethanol)}
	\label{CH3OCH2OH}
	
	The detection of \ce{CH3OCH2OH} was reported by \citet{McGuire:2017gy} in ALMA observations of NGC 6334I.  More than two dozen largely unblended transitions between 239--349~GHz were identified and assigned based on the laboratory work of \citet{Motiyenko:2017dw}, which was in press at the time.
	
    
    \begin{center}
        \large
        \textsc{\textbf{Eleven-Atom Molecules}}
        \label{11atoms}
        \normalsize
    \end{center}
    
    \subsection{\ce{HC9N} (cyanotetraacetylene)}
	\label{HC9N}
	
	\citet{Broten:1978iu} reported the detection of \ce{HC9N} in TMC-1 using the Algonquin Radio Observatory 46-m and NRAO 140-foot telescopes.  The $J = 18\rightarrow17$ and $25\rightarrow24$ transitions at 10.5 and 14.5~GHz, respectively, were identified on the basis of theoretical calculations described in the same manuscript.  The detection was later confirmed by the laboratory spectroscopy performed by \citet{Iida:1991rt}.
	
	\subsection{\ce{CH3C6H} (methyltriacetylene)}
	\label{CH3C6H}
	
	The detection of \ce{CH3C6H} was reported by \citet{Remijan:2006yd} in GBT 100-m observations of TMC-1.  Ten spectral lines corresponding to the $K=0$ and $K=1$ components of the $J_K = 12_*\rightarrow11_*$ through $16_*\rightarrow15_*$ transitions between 18.7--24.9~GHz were identified and assigned based on the laboratory work of \citet{Alexander:1978bd}.
	
	\subsection{\ce{CH3CH2OCHO} (ethyl formate)}
	\label{CH3CH2OCHO}
	
	\citet{Belloche:2009ki} reported the detection of \emph{trans}-\ce{CH3CH2OCHO} in IRAM 30-m observations of Sgr B2.  A total of 41 unblended transitions of \emph{trans}-\ce{CH3CH2OCHO} between 80--268~GHz were identified and assigned based on the laboratory data of \citet{Medvedev:2009fs}.  The subsequent detection of the \emph{gauche}-conformer was reported by \citet{Tercero:2013fz} in IRAM 30-m observations of Orion.  They identified and assigned 38 unblended transitions between 80--281~GHz based on the laboratory work of \citet{Demaison:1984if}.
	
	\subsection{\ce{CH3COOCH3} (methyl acetate)}
	\label{CH3COOCH3}
	
	The detection of \ce{CH3COOCH3} was reported by \citet{Tercero:2013fz} in IRAM 30-m observations of Orion.  A total of 215 unblended transitions of \ce{CH3COOCH3} between 80--281~GHz were identified and assigned based on the laboratory work of \citet{Tudorie:2011br}.
	
    
    \begin{center}
        \large
        \textsc{\textbf{Twelve-Atom Molecules}}
        \label{12atoms}
        \normalsize
    \end{center}	
	
	\subsection{\ce{C6H6} (benzene)}
	\label{C6H6}
	
	\citet{Cernicharo:2001mw} reported the detection of \ce{C6H6} in ISO observations of CRL 618.  The $\nu_4$ bending mode of \ce{C6H6} near 14.8~$\mu$m was identified and assigned based on the laboratory work of \citet{Lindenmayer:1988kx}.
	
	\subsection{$n$-\ce{C3H7CN} ($n$-propyl cyanide)}
	\label{n-C3H7CN}
	
	The detection of $n$-\ce{C3H7CN}, also referred to as $n$-butyronitrile, was reported by \citet{Belloche:2009ki} in IRAM 30-m observations of Sgr B2.  A total of 50 unblended transitions of the \emph{anti}-conformer of $n$-\ce{C3H7CN} were identified and assigned based on an extensive re-analysis of the literature spectra presented in the same manuscript.  
	
	\subsection{$i$-\ce{C3H7CN} (\emph{iso}-propyl cyanide)}
	\label{i-C3H7CN}	
	
	\citet{Belloche:2014jd} reported the detection of $i$-\ce{C3H7CN} in ALMA observations of Sgr B2 between 84--111~GHz, based on the laboratory work of \citet{Muller:2011bz}. This was the first branched carbon-chain molecule detected in the ISM.
	
    
    \begin{center}
        \large
        \textsc{\textbf{Thirteen-Atom Molecules}}
        \label{13atoms}
        \normalsize
    \end{center}	
	
	\subsection{\ce{C6H5CN} (benzonitrile)}
	\label{C6H5CN}	
	
	\citet{McGuire:2018it} reported the detection of \ce{C6H5CN} in GBT 100-m observations of TMC-1.  A total of nine transitions, some with resolved $^{14}$N hyperfine structure, between 18.4--23.2~GHz were identified and assigned on the basis of laboratory work described in the same paper, as well as that of \citet{Wohlfart:2008hg}.
	
    
    \begin{center}
        \large
        \textsc{\textbf{Fullerene Molecules}}
        \label{fullerenes}
        \normalsize
    \end{center}	
    
    \subsection{\ce{C60} (buckminsterfullerene)}
	\label{C60}
	
	\citet{Cami:2010fi} reported the detection of \ce{C60} in \emph{Spitzer} observations of Tc~1.  Strong emission features from \ce{C60} at 7.0, 85, 17.4, and 18.9~$\mu$m were identified and assigned based on the work of \citet{Martin:1993cj} and \citet{Fabian:1996hd}.
	
    \subsection{\ce{C60^+} (buckminsterfullerene cation)}
	\label{C60+}
	
	\citet{Berne:2013ow} reported the detection of \ce{C60^+} in \emph{Spitzer} observations toward NGC 7023.  Two emission features at 7.1 and 6.4~$\mu$m were assigned to \ce{C60^+} based on the laboratory matrix-isolation work of \citet{Kern:2013jk}.  A number of additional features were assigned based on theoretical calculations performed and described in \citet{Berne:2013ow}.  Subsequent laboratory work by \citet{Campbell:2015hp} confirmed the presence of \ce{C60^+} in the ISM by measuring the gas-phase spectrum of \ce{C60^+}.  Based on those results, they determined that \ce{C60^+} was the carrier of the 9632 and 9577~\AA~diffuse interstellar bands, the first definitive molecular assignment of a carrier of one of these bands.  Follow-up observational and laboratory studies by \citet{Walker:2015ey} and \citet{Campbell:2016hl} solidified the identification.
	
    \subsection{\ce{C70} (rugbyballene)}
	\label{C70}	
	
	\citet{Cami:2010fi} reported the detection of \ce{C70} in \emph{Spitzer} observations of Tc~1. They identified and assigned four emission features between 12--22~$\mu$m to \ce{C70} based on the work of \citet{Nemes:1994jm}, \citet{vonCzarnowski:1995iv}, and \citet{Stratmann:1998dr}.
	
\section{Tentative Detections}
\label{tentative}

This section contains those molecules for which a detection has been classified by the authors as tentative.  Those molecules once viewed as tentative that have since been confirmed are listed in \S\ref{known}.

    \subsection{\ce{C2H5N} (aziridine)}
	\label{C2H5N}
	
	\citet{Dickens:2001ku} reported the tentative detection of the $J_{K_a,K_c} = 5_{5,0}\rightarrow4_{4,1}$ transition of \ce{C2H5N}, also known as ethylenimine, at 226~GHz toward Sgr B2, G34.3+.2, G10.47+0.03, and G327.3-0.6 in observations made with the SEST 15-m, NRAO 12-m, and NASA Deep Space Network 70-m telescopes.  The laboratory data were reported by \citet{Thorwirth:2000ii}.  
	
	\subsection{\ce{SiH} (silicon hydride)}
	\label{SiH}
	
	\citet{Schilke:2001fr} reported the tentative detection of two features in a CSO line survey of Orion that were coincident with transitions of SiH.  Two groups of hyperfine transitions in the $^2\Pi _{1/2}$ state at 625 and 628~GHz fell within range of the observations.  The lower set is significantly blended with other emitting species, whereas the higher-frequency set appears unblended, and may show partially resolved hyperfine structure.  The laboratory frequencies were from \citet{Brown:1985cr}. 
	
	\subsection{\ce{FeO} (iron monoxide)}
	\label{FeO}
	
	\citet{Walmsley:2002ud} identified an absorption feature in IRAM 30-m observations of Sgr B2 at 153~GHz that they tentatively assign to the $J = 5\rightarrow4$, $\Omega = 4$ transition of FeO, based on the laboratory work of \citet{Allen:1996te}.  Later, \citet{Furuya:2003jh} used the Nobeyama Millimeter Array to confirm the detection of this line, but no additional transitions were observed.  Most recently, \citet{Decin:2018ju} reported a tentative detection of the $11\rightarrow10$, $\Omega = 4$ transition in ALMA observations of R Dor at 337~GHz.  
	
	\subsection{\ce{OCN-} (cyanate anion)}
	\label{OCNm}
	
	There is a long history of attempts to identify the carrier of the ``XCN" feature in astrophyiscal ice observations near 2167~cm$^{-1}$.  On the basis of extensive laboratory work (see, e.g., \citealt{Schutte:1997hd}), \citet{vanBroekhuizen:2005ff} claim an identification of \ce{OCN-} in ices along numerous sightlines to low-mass YSO's using the Very Large Telescope (VLT).  Only a single absorption feature could be ascribed to \ce{OCN-}.  As this is a single feature detection, and no gas-phase detection has been claimed, \ce{OCN-} has been listed as a tentative interstellar species in this census.
	
	\subsection{\ce{C6H5OH} (phenol)}
	\label{C6H5OH}
	
	\citet{Kolesnikova:2013hw} reported the laboratory rotational spectroscopy of \ce{C6H5OH}, and identified a number of coincidences between transitions of \ce{C6H5OH} and unassigned lines in IRAM 30-m observations of Orion between 80--280~GHz \citep{Tercero:2012cv}. 
	
	\subsection{\ce{NO+} (nitrosylium cation)}
	\label{NOp}
	
	\citet{Cernicharo:2014eq} reported the tentative detection of a single line ($J = 2\rightarrow1$) of \ce{NO+} at 238~GHz toward B1-b using the IRAM 30-m telescope.  The assignment was made based on laboratory rotational spectroscopy described in the same paper.
    
    \subsection{\ce{(NH2)2CO} (urea)}
    \label{urea}
    
    \citet{Remijan:2014jb} reported on the evidence for a tentative detection of \ce{(NH2)2CO} in observations of Sgr B2.  The data, covering 100--250~GHz, were obtained from the BIMA, CARMA, NRAO 12-m, SEST 15-m, and IRAM 30-m telescopes.  The laboratory data were from \citet{Brown:1975wo}, \citet{Kasten:1986fm}, and  \citet{Kretschmer:1996hf}. A single set of physical parameters reproduced the features across all observational sets within a factor of two, and interferometric maps of individual transitions showed a consistent spatial structure.  The lack of a sufficiently large set of unblended transitions precluded a definitive detection.  
    
    \subsection{\ce{MgCCH} (magnesium monoacetylide)}
    \label{MgCCH}
    
    \citet{Agundez:2014gm} reported the tentative detection of \ce{MgCCH} in IRAM 30-m observations of IRC+10216.  They identified and assigned two sets of features corresponding to $l$-doubled transitions of the $^2\Sigma ^+$ ground electronic state of \ce{MgCCH} at 89 and 99~GHz based on the laboratory work of \citet{Brewster:1999kr}. 
    
    \subsection{\ce{NCCP} (cyanophosphaethyne)}
    \label{NCCP}
    
    \citet{Agundez:2014gm} reported the tentative detection of \ce{NCCP} in IRAM 30-m observations of IRC+10216.  They identified and tentatively assigned three signals in their spectra to the $J = 15\rightarrow14$, $16\rightarrow15$, and $18\rightarrow17$ transitions of \ce{NCCP} at 81, 87, and 97~GHz, respectively, based on the laboratory work of \citet{Bizzocchi:2001ij}.
    
    \subsection{\ce{C2H5OCH3} (t-ethyl methyl ether)}
    \label{C2H5OCH3}
    
    \citet{Charnley:2001dc} indicated a tentative detection of \ce{C2H5OCH3} in NRAO 12-m and BIMA observations of Orion-KL and Sgr B2(N).  A detection of \ce{C2H5OCH3} was then later reported by \citet{Fuchs:2005fw} in IRAM 30-m and SEST 15-m observations of W51e2.  This detection was disputed by \citet{Carroll:2015dp}, who re-observed both W51e2 and Sgr B2(N) using the ARO 12-m and GBT 100-m telescopes, and failed to detect \ce{C2H5OCH3}, setting an upper limit column density substantially lower than the reported value of \citet{Fuchs:2005fw}.  Evidence supporting the tentative detection in Orion-KL was later reported by \citet{Tercero:2015dl} using IRAM 30-m and ALMA observations of the region, although the authors claim the detection should still be viewed as tentative.  The laboratory spectroscopy was performed by \citet{Hayashi:1975up} and \citet{Fuchs:2003yd}.

    \subsection{\ce{CH3NHCHO} (n-methylformamide)}
    \label{CH3NHCHO}
    
    A tentative detection of \ce{CH3NHCHO} was reported toward Sgr B2(N2) by \citet{Belloche:2017bb} using ALMA observations in the range 84 -- 114~GHz.  The laboratory rotational spectroscopy are provided in the same reference.  
    
\section{Disputed Detections}
\label{disputed}

This section contains those molecules for which a detection has been claimed and subsequently disputed in the literature.  Those species which were disputed, but later confirmed are listed in \S\ref{known}.  Species for which a tentative detection has been claimed and also disputed are not listed.  

    \subsection{\ce{HC11N} (cyanopentaacetylene)}
    \label{HC11N}
    
    \citet{Bell:1997vj} reported a detection of \ce{HC11N} in NRAO 140-foot observations of TMC-1.  They assigned two emission features at 12848.728~MHz and 13186.853~MHz to the $J = 38\rightarrow37$ and $39\rightarrow38$ transitions of \ce{HC11N} based on the laboratory work of \citet{Travers:1996gs}.  The detection was later disputed by \citet{Loomis:2016jsa} who re-observed TMC-1 using the 100-m GBT.  They targeted six transitions of \ce{HC11N}, including those originally reported by \citet{Bell:1997vj}, and detected no signal at any of these frequencies.  They suggest correlator artifacts may have been responsible for the signals seen in \citet{Bell:1997vj}.  Subsequently, \citet{Cordiner:2017dq} confirmed the findings of \citet{Loomis:2016jsa}, setting a yet lower upper limit with more sensitive data.
    
    \subsection{\ce{NH2CH2COOH} (glycine)}
    \label{glycine}
    
    \citet{Kuan:2003yt} reported a detection of 27 lines of \ce{NH2CH2COOH} in observations toward Orion, Sgr B2, and W51 using the NRAO 12-m telescope.  Each source contained 13--16 of the 27 features; three features were seen in common between the sources.  The frequencies were obtained from the rotational constants measured and reported in \citet{Suenram:1980cs} and \citet{Lovas:1995it}.  The detection was later disputed by \citet{Snyder:2005tr}, who claimed to have identified a number of issues with the analysis.  Notably, \citet{Snyder:2005tr} argue that the frequency predictions used by \citet{Kuan:2003yt} were extrapolated too far above the measured laboratory transitions.  \citet{Snyder:2005tr} used laboratory frequencies from \citet{Ilyushin:2005iu} that covered the observed frequency range, and examined archival NRAO 12-m and SEST 15-m observations of Orion, Sgr B2, and W51.  They report a number of ``missing'' transitions of \ce{NH2CH2COOH} that they predicted should have been strongly detected, if the detection of \citet{Kuan:2003yt} held.  Subsequent searches by \citet{Jones:2007fa}, \citet{Cunningham:2007kl}, and \citet{Belloche:2008jy} using both interferometers and single-dish facilities failed to detect \ce{NH2CH2COOH}, and set upper limits to the column density lower than that claimed in the detection of \citet{Kuan:2003yt} (see \citealt{Belloche:2008jy} \S4.4 for a detailed discussion).
    
\section{Species Detected in External Galaxies}
\label{exgals}

A remarkable fraction (65/204; $\sim$32\%)  of the known interstellar and circumstellar molecules have also been detected in observations of external galaxies.  A list of these species is given in Table~\ref{exgal_mols}.  For most species, the provided reference is to the earliest claim of a detection in the literature.  In some cases, additional references are provided for context.  Molecules for which a tentative detection in external galaxies have been claimed are indicated.  In the case of fullerene molecules, there has been some debate in the literature over the claimed identification of these molecules  \citep{Duley:2012dv}, and thus these are not included in this table at this time.  

A few notable absences stand out.  The detection of \ce{H2Cl+} \citep{Muller:2014je} hints that the detections of \ce{HCl} and \ce{HCl+} may be achievable, assuming the extragalactic abundances follow those seen in our galaxy where these species are within factors of 1--3 of each other \citep{Monje:2013fk}.  Similarly, the presence of \ce{SH+} would indicate that \ce{SH} is a likely candidate, given its marginally higher abundance in galactic sightlines (see, e.g., \citealt{Neufeld:2012gz}).  The very recent detections of the much larger species \ce{HCOOCH3} and \ce{CH3OCH3} \citep{Qiu:2018gd,Sewiio:2018jx} suggest a reservoir of complex molecules may be also observable.

The complete list of external galaxies in which these detections were made are included in the Supplementary Information Python database.  By far the largest contributors to these detections are NGC 253 (24 molecules), the line of sight to PKS 1830-211 (15 molecules), and M82 (14 molecules).  

\begin{table*}
\centering
\caption{List of molecules detected in external galaxies with references to the first detections.  Tentative detections are indicated, and some extra references are occasionally provided for context.}
\begin{tabular*}{\textwidth}{l @{\extracolsep{\fill}} l @{\extracolsep{\fill}} l @{\extracolsep{\fill}} l @{\extracolsep{\fill}} l @{\extracolsep{\fill}} l @{\extracolsep{\fill}} l @{\extracolsep{\fill}} l @{\extracolsep{\fill}} l @{\extracolsep{\fill}} l @{\extracolsep{\fill}}}
\hline\hline
\multicolumn{2}{c}{2 Atoms}&\multicolumn{2}{c}{3 Atoms}&\multicolumn{2}{c}{4 Atoms}&\multicolumn{2}{c}{5 Atoms}\\
Species	&	Ref.	&	Species	&	Ref.	&	Species	&	Ref.	&	Species	&	Ref.	\\
\hline
\ce{CH}	&	46	&	\ce{H2O}	&	3	&	\ce{NH3}	&	16	&	\ce{HC3N}	&	20, 13	\\
\ce{CN}	&	13	&	\ce{HCO+}	&	41	&	\ce{H2CO}	&	6	&	\ce{CH2NH}	&	25	\\
\ce{CH+}	&	15	&	\ce{HCN}	&	37	&	\ce{HNCO}	&	32	&	\ce{NH2CN}	&	18	\\
\ce{OH}	&	45	&	\ce{OCS}	&	22	&	\ce{H2CS}	&	18	&	\ce{H2CCO}	&	25	\\
\ce{CO}	&	36	&	\ce{HNC}	&	13	&	\ce{C2H2}	&	19	&	\ce{C4H}	&	25	\\
\ce{H2}	&	42	&	\ce{H2S}	&	10	&	\ce{HOCO+}	&	1, 18	&	\ce{c-C3H2}	&	39	\\
\ce{SiO}	&	23	&	\ce{N2H+}	&	23	&	\ce{l-C3H}	&	25	&	\ce{CH2CN}	&	25	\\
\ce{CS}	&	11	&	\ce{C2H}	&	13	&	\ce{H3O+}	&	47	&	\ce{H2CCC}	&	25	\\
\ce{SO}	&	14, 33	&	\ce{SO2}	&	17	&	\ce{c-C3H}$^{\P}$	&	18	\\
\ce{NS}	&	17	&	\ce{HCO}	&	38, 5	\\
\ce{NO}	&	17	&	\ce{HCS+}	&	26	\\
\ce{NH}	&	8	&	\ce{HOC+}	&	43	\\
\ce{SO+}	&	25	&	\ce{C2S}	&	18	\\
\ce{CO+}	&	4	&	\ce{NH2}	&	27	\\
\ce{HF}	&	48, 35, 24	&	\ce{H3+}	&	7	\\
\ce{CF+}	&	30	&	\ce{H2O+}	&	44	\\
\ce{OH+}	&	48, 35, 9	&	\ce{H2Cl+}	&	28	\\
\ce{SH+}	&	31	\\
\ce{ArH+}	&	29	\\
\hline\hline
\end{tabular*}
\begin{tabular*}{\textwidth}{l @{\extracolsep{\fill}} l @{\extracolsep{\fill}} l @{\extracolsep{\fill}} l @{\extracolsep{\fill}} l @{\extracolsep{\fill}} l @{\extracolsep{\fill}} l @{\extracolsep{\fill}} l @{\extracolsep{\fill}} l @{\extracolsep{\fill}} l @{\extracolsep{\fill}} l @{\extracolsep{\fill}} l @{\extracolsep{\fill}}}
\multicolumn{2}{c}{6 Atoms}&\multicolumn{2}{c}{7 Atoms}&\multicolumn{2}{c}{8 Atoms}&\multicolumn{2}{c}{9 Atoms}&\multicolumn{2}{c}{12 Atoms}\\
Species	&	Ref.	&	Species	&	Ref.	&	Species	&	Ref.	&	Species	&	Ref.	&	Species	&	Ref.	\\
\hline
\ce{CH3OH}	&	12	&	\ce{CH3CHO}	&	25	&	\ce{HCOOCH3}	&	40	&	\ce{CH3OCH3}	&	34, 40	&	\ce{C6H6}	&	2	\\
\ce{CH3CN}	&	21	&	\ce{CH3CCH}	&	21	&	\ce{HC6H}	&	2	\\
\ce{NH2CHO}	&	26	&	\ce{CH3NH2}	&	25	\\
\ce{HC4H}	&	2	&	\ce{HC5N}$^{\P}$	&	1	\\
\hline
\end{tabular*}
\justify
$^{\P}$Tentative detection\\\textbf{References:} [1] \citet{Aladro:2015bv} [2] \citet{BernardSalas:2006ck} [3] \citet{Churchwell:1977nh} [4] \citet{Fuente:2006ba} [5] \citet{GarciaBurillo:2002iy} [6] \citet{Gardner:1974jh} [7] \citet{Geballe:2006be} [8] \citet{GonzalezAlfonso:2004ed} [9] \citet{GonzalezAlfonso:2013hi} [10] \citet{Heikkila:1999my} [11] \citet{Henkel:1985ik} [12] \citet{Henkel:1987bq} [13] \citet{Henkel:1988wm} [14] \citet{Johansson:1991hn} [15] \citet{Magain:1987uh} [16] \citet{Martin:1979dh} [17] \citet{Martin:2003bh} [18] \citet{Martin:2006eh} [19] \citet{Matsuura:2002do} [20] \citet{Mauersberger:1990bc} [21] \citet{Mauersberger:1991yt} [22] \citet{Mauersberger:1995yn} [23] \citet{Mausersberger:1991nh} [24] \citet{Monje:2011gg} [25] \citet{Muller:2011hn} [26] \citet{Muller:2013ie} [27] \citet{Muller:2014cj} [28] \citet{Muller:2014je} [29] \citet{Muller:2015ik} [30] \citet{Muller:2016ds} [31] \citet{Muller:2017ki} [32] \citet{NguyenQRieu:1991dg} [33] \citet{Petuchowski:1992rm} [34] \citet{Qiu:2018gd} [35] \citet{Rangwala:2011dd} [36] \citet{Rickard:1975fm} [37] \citet{Rickard:1977hn} [38] \citet{Sage:1995hf} [39] \citet{Seaquist:1986wo} [40] \citet{Sewiio:2018jx} [41] \citet{Stark:1979ka} [42] \citet{Thompson:1978kc} [43] \citet{Usero:2004bx} [44] \citet{Weiss:2010ez} [45] \citet{Weliachew:1971nd} [46] \citet{Whiteoak:1980en} [47] \citet{vanderTak:2007ku} [48] \citet{vanderWerf:2010ix} \\
\label{exgal_mols}
\end{table*}
    
\section{Species Detected in Interstellar Ices}
\label{ices}

Observations of ices require a background illuminating source for absorption, limiting the number of sight lines that are available for study.  Further complications arise when comparing with laboratory spectra, as the peak positions, linewidths, and intensities of molecular ices features are known to be sensitively dependent on temperature, crystal structure of the ice (or lack thereof), and mixing or layering with other species (see, e.g., \citealt{Ehrenfreund:1997kj}, \citealt{Schutte:1999yc}, and \citealt{Cooke:2016hw}).  As a result, only a handful of species (\ce{H2O}, \ce{CO}, \ce{CO2}, \ce{CH4}, \ce{CH3OH}, and \ce{NH3}) have been definitively detected in interstellar ices.  

The first molecular ice detection was reported by \citet{Gillett:1973iq} who observed an absorption feature at 3.1~$\mu$m toward Orion-KL which they attribute to \ce{H2O} based on comparison to the laboratory work of \citet{Irvine:1968fb}.  \citet{Soifer:1979gu} then reported the detection of \ce{CO} at 4.61~$\mu$m in absorption toward W33A, based on the laboratory work of \citet{Mantz:1975ir}.  

The laboratory work of \citet{dHendecourt:1986wx} was then used to detect a further four species.  As mentioned in \S\ref{CO2}, \ce{CO2} was detected by \citet{deHendecourt:1989iu}, based on their own laboratory work, in absorption at 15.2~$\mu$m toward several IRAS sources.  Also discussed previously (\S\ref{CH4}), \ce{CH4} was simultaneously detected in the gas- and solid-phases by \citet{Lacy:1991fb} toward NGC 7538 IRS 9.  A feature was attributed to \ce{CH4} at 7.7$\mu$m.  That same year, \citet{Grim:1991jh} identified and assigned an absorption feature at 3.53~$\mu$m in UKIRT observations toward W33A to \ce{CH3OH}.  Finally, the detection of \ce{NH3} was reported by \citet{Lacy:1998do} who assigned an absorption feature at 1110~cm$^{-1}$ toward NGC 7538 IRS 9.

Several more species have been tentatively identified due to the coincidence of a spectral feature with laboratory data for a molecule under certain temperature or mixture conditions.  \citet{Palumbo:1995nc} and \citet{Palumbo:1997nd} identified an absorption feature at 4.90~$\mu$m toward a number of sources that corresponded to OCS when mixed with \ce{CH3OH}, based on their own laboratory work.  As discussed in \S\ref{OCNm},  \citet{vanBroekhuizen:2005ff} claim an identification of \ce{OCN-} based on a lengthy history of laboratory work (e.g., \citealt{Schutte:1997hd}).

Finally, a number of additional molecular carriers have been suggested or tentatively assigned, but have not yet been definitively confirmed.  These possible interstellar molecules are discussed in some detail in a review article by \citet{Boogert:2015fx}.

\section{Species Detected in Protoplanetary Disks}
\label{ppds}

Compared to the molecular clouds from which they form, the detected molecular inventory of protoplanetary disks is relatively sparse.  A list of species detected in disks, including isotopologues, is given in Table~\ref{ppd_mols}, along with early and/or representative detection references.  The detections of \ce{H2D+} and \ce{HDO} that were reported by \citet{Ceccarelli:2004gf} and \citet{Ceccarelli:2005gq}, but were later disputed by \citet{Guilloteau:2006bw}, are not included. 

The (relatively) low number of gas-phase species so far detected in disks can likely be attributed to a combination of factors.

\begin{enumerate}
    \item Often narrow ranges of physical environments conducive to a large gas-phase abundance.  For example, many complex molecules are likely locked into ices in the cold, shielded mid-plane of the disk and are liberated into the gas phase for detection in only certain, narrow regions of the disk \citep{Walsh:2014jq}.  
    \item Conversely, gas-phase molecules that reach the upper layers of the disk are subject to the harsh, PDR-like radiation environment of the star and the resulting photodestructive processes \citep{Henning:2010uc}.
    \item The angular extent of these sources on the sky is small (of order arcseconds; \citealt{Brogan:2015cg}), largely degrading the utility of the most sensitive single dish facilities due to beam dilution.  As a result of the overall low abundance, the required surface brightness sensitivity likewise limits the effectiveness of many small and mid-scale interferometers.  As a result, the most complex and low-abundance species are being detected exclusively with ALMA, and are already pushing the limits of the instrument \citep{Loomis:2018bt}.
\end{enumerate}

Detection and analysis of the most complex molecules seen to date, \ce{CH3OH} and \ce{CH3CN} \citep{Walsh:2016bq,Oberg:2015dh}, have benefited from velocity stacking and newly-developed matched-filtering techniques to extract useful signal-to-noise ratios \citep{Loomis:2018bt}.  While these techniques are likely to produce a number of new detections of lower-abundance molecules in the coming years, astrochemical models combined with sensitivity analyses suggest that the total number of molecules detectable with complexity greater than \ce{CH3OH} and \ce{CH3CN} may be small \citep{Walsh:2017nw}.

\begin{table*}
\centering
\caption{List of molecules, including rare isotopic species, detected in protoplanetary disks, with references to representative detections.  The earliest reported detection of a species in the literature is provided on a best-effort basis.  Tentative and disputed detections are not included (see text).}
\begin{tabular*}{\textwidth}{l @{\extracolsep{\fill}} l @{\extracolsep{\fill}} l @{\extracolsep{\fill}} l @{\extracolsep{\fill}} l @{\extracolsep{\fill}} l @{\extracolsep{\fill}} l @{\extracolsep{\fill}} l @{\extracolsep{\fill}} l @{\extracolsep{\fill}} l @{\extracolsep{\fill}}}
\hline\hline
\multicolumn{2}{c}{2 Atoms}&\multicolumn{2}{c}{3 Atoms}&\multicolumn{2}{c}{4 Atoms}&\multicolumn{2}{c}{5 Atoms}&\multicolumn{2}{c}{6 Atoms}\\
Species	&	Ref.	&	Species	&	Ref.	&	Species	&	Ref.	&	Species	&	Ref.	&	Species	&	Ref.	\\
\hline
\ce{CN}	&	10, 21	&	\ce{H2O}	&	5, 30, 19	&	\ce{NH3}	&	29	&	\ce{HCOOH}	&	12	&	\ce{CH3CN}	&	24	\\
\ce{C^{15}N}	&	18	&	\ce{HCO+}	&	10, 21	&	\ce{H2CO}	&	10	&	$c$-\ce{C3H2}	&	28	&	\ce{CH3OH}	&	35	\\
\ce{CH+}	&	34	&	\ce{DCO+}	&	8	&	\ce{C2H2}	&	22	&	\ce{CH4}	&	14	\\
\ce{OH}	&	23, 30	&	\ce{H^{13}CO+}	&	36, 8	&	\ce{HC3N}	&	7	\\
\ce{CO}	&	2	&	\ce{HCN}	&	10, 21	\\
\ce{^{13}CO}	&	31	&	\ce{DCN}	&	27	\\
\ce{C^{18}O}	&	9	&	\ce{H^{13}CN}	&	17	\\
\ce{C^{17}O}	&	32, 16	&	\ce{HC^{15}N}	&	17	\\
\ce{H2}	&	33	&	\ce{HNC}	&	10	\\
\ce{HD}	&	3	&	\ce{N2H+}	&	26, 11	\\
\ce{CS}	&	25, 4, 15	&	\ce{N2D+}	&	20	\\
\ce{C^{34}S}	&	1	&	\ce{C2H}	&	10	\\
\ce{SO}	&	13	&	\ce{CO2}	&	6	\\
\hline
\end{tabular*}
\justify
\textbf{References:} [1] \citet{ArturdelaVillarmois:2018ix} [2] \citet{Beckwith:1986ej} [3] \citet{Bergin:2013em} [4] \citet{Blake:1992ef} [5] \citet{Carr:2004dp} [6] \citet{Carr:2008if} [7] \citet{Chapillon:2012er} [8] \citet{vanDishoeck:2003ip} [9] \citet{Dutrey:1994iu} [10] \citet{Dutrey:1997uc} [11] \citet{Dutrey:2007ps} [12] \citet{Favre:2018kf} [13] \citet{Fuente:2010kj} [14] \citet{Gibb:2013ek} [15] \citet{Guilloteau:2012eo} [16] \citet{Guilloteau:2013ca} [17] \citet{Guzman:2015dd} [18] \citet{HilyBlant:2017da} [19] \citet{Hogerheijde:2011ih} [20] \citet{Huang:2015ii} [21] \citet{Kastner:1997tj} [22] \citet{Lahuis:2006yd} [23] \citet{Mandell:2008ky} [24] \citet{Oberg:2015dh} [25] \citet{Ohashi:1991kn} [26] \citet{Qi:2003dn} [27] \citet{Qi:2008kq} [28] \citet{Qi:2013ht} [29] \citet{Salinas:2016fo} [30] \citet{Salyk:2008im} [31] \citet{Sargent:1987cu} [32] \citet{Smith:2009ep} [33] \citet{Thi:1999dt} [34] \citet{Thi:2011ii} [35] \citet{Walsh:2016bq} [36] \citet{vanZadelhoff:2001ko} \\
\label{ppd_mols}
\end{table*}

\section{Species Detected in Exoplanetary Atmospheres}
\label{exoplanets}

Although exoplanet atmospheres have now been observed for some time \citep{Charbonneau:2002yw}, only a small handful of molecules have been detected in these environments. That number, and the ability to robustly measure molecules in exoplanetary atmospheres, is expected to increase somewhat with the launch of the \emph{James Webb Space Telescope} \citep{Schlawin:2018cq}.  Table~\ref{exoplanet_mols} lists those molecules for which a consensus seems to have been reached in the literature as being detected.  There is extensive debate in the literature as to the robustness of many claimed detections (see \citealt{Madhusudhan:2016dl} for a thorough overview).  Thus, references are provided both to some early detections, for historical context, and to some more recent literature.  The reference list is intended to be representative, not exhaustive.

\begin{table}
\centering
\caption{List of molecules detected in exoplanetary atmospheres, with references to representative detections.  Tentative and disputed detections are not included.}
\begin{tabular}{p{1in} l}
\hline\hline
Species	&	References	\\
\hline
\ce{CO}	&	9, 1, 7, 2	\\
\ce{TiO}	&	4, 11, 10	\\
\ce{H2O}	&	14, 3, 5, 6, 8	\\
\ce{CO2}	&	12, 9, 7	\\
\ce{CH4}	&	13, 1, 12, 2	\\
\hline
\end{tabular}
\justify
\textbf{References:} [1] \citet{Barman:2011op} [2] \citet{Barman:2015dy} [3] \citet{Deming:2013ge} [4] \citet{Haynes:2015cf} [5] \citet{Kreidberg:2014hi} [6] \citet{Kreidberg:2015er} [7] \citet{Lanotte:2014gl} [8] \citet{Lockwood:2014dv} [9] \citet{Madhusudhan:2011kw} [10] \citet{Nugroho:2017jw} [11] \citet{Sedaghati:2017ft} [12] \citet{Stevenson:2010hf} [13] \citet{Swain:2008fm} [14] \citet{Tinetti:2007hy} \\
\label{exoplanet_mols}
\end{table}
\section{Analysis \& Discussion}
\label{statistics}

As of publication, a total of 204 individual molecules have been detected in the ISM.  Here, several graphical representations of the census results presented above are shown and briefly discussed.

\subsection{Cumulative Detection Rates}
\label{cumulative_rates}

\begin{figure}[b!]
\includegraphics[width=\columnwidth]{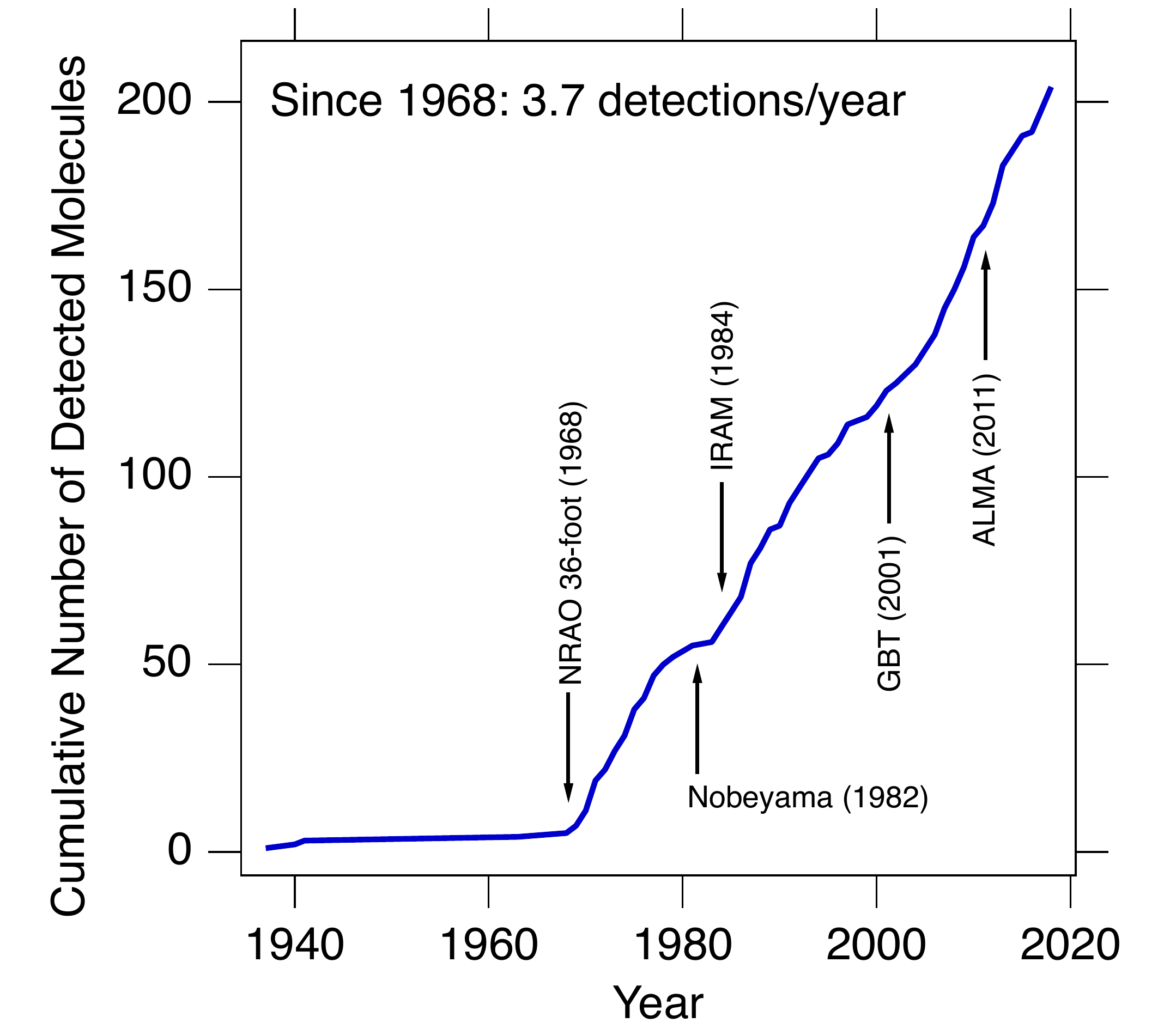}
\caption{Cumulative number of known interstellar molecules over time.  After the birth of molecular radio astronomy in the 1960s, there have been on average 3.7 new detections per year ($R^2$ = 0.991 for a linear fit beginning in 1968). The commissioning dates of several major contributing facilities are noted with arrows.}
\label{cumulative_detects}
\end{figure}

The cumulative number of known interstellar molecules with time is presented in Figure~\ref{cumulative_detects}, as well as the commissioning dates of a number of key observational facilities.  Although the first molecules were detected between 1937--1941 (see \S\ref{CH}, \ref{CN}, and \ref{CH+}), and OH in 1963 (\S\ref{OH}), it wasn't until the late 1960s with the development of high-resolution radio receivers optimized for line observations that detections began in earnest.  Since 1968, the rate of detections has remained remarkably constant at $\sim$3.7 new molecules per year.  

Notably, beginning in 2005 there is some evidence that this detection rate has increased.  A fit of the data solely between 2005~--~2018 yields a rate of $\sim$5.4 molecules per year.  This would appear to be consistent with the advent of the GBT as a molecule-detection telescope, and a significant uptick in the rate of detections from IRAM, as discussed further below.

\begin{figure}
\includegraphics[width=\columnwidth]{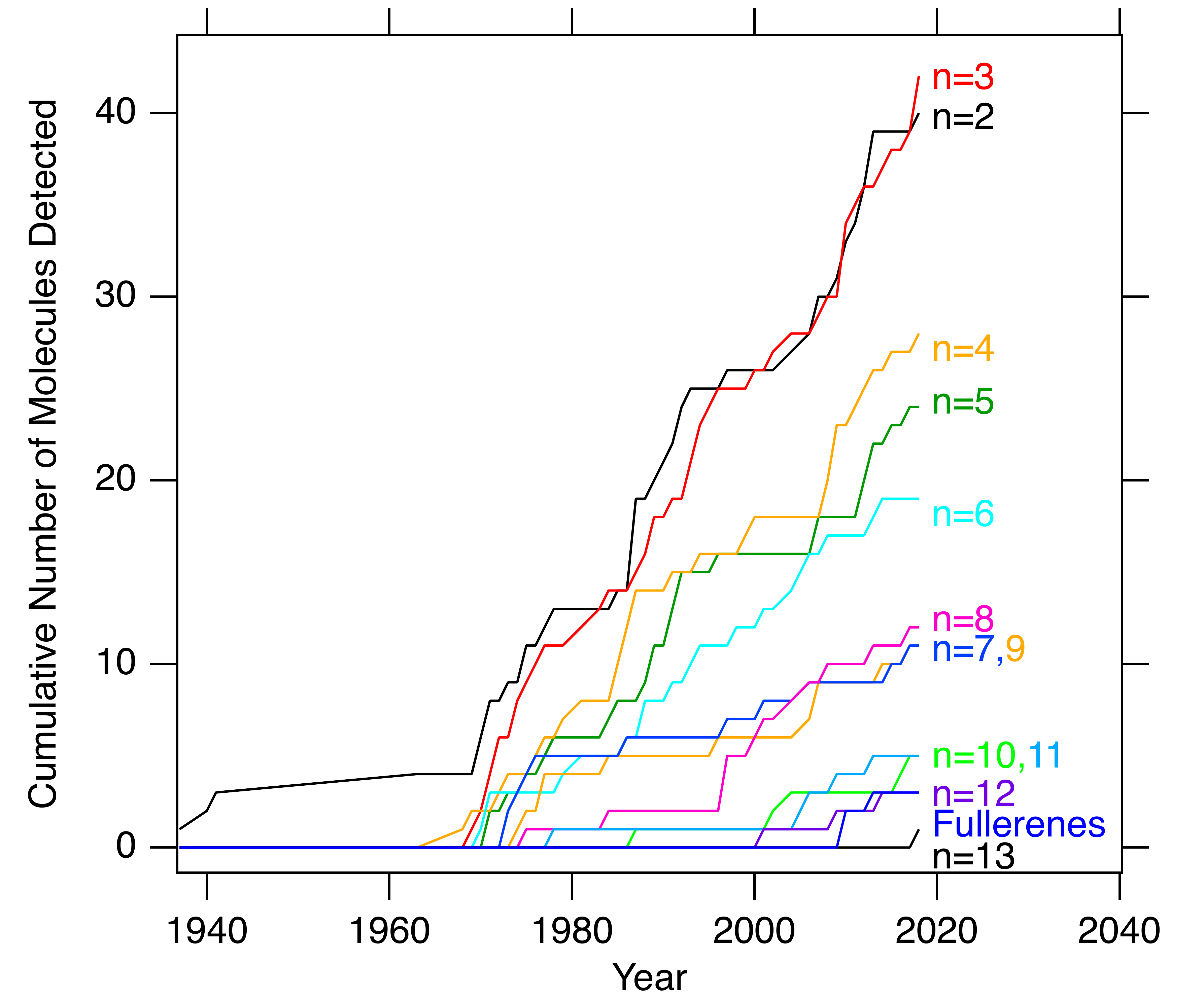}
\caption{Cumulative number of known interstellar molecules with 2--13 atoms, as well as fullerene molecules, as a function of time. The traces are color coded by number of atoms, and labeled on the right. }
\label{cumulative_by_atoms}
\end{figure}

\begin{figure}
\includegraphics[width=\columnwidth]{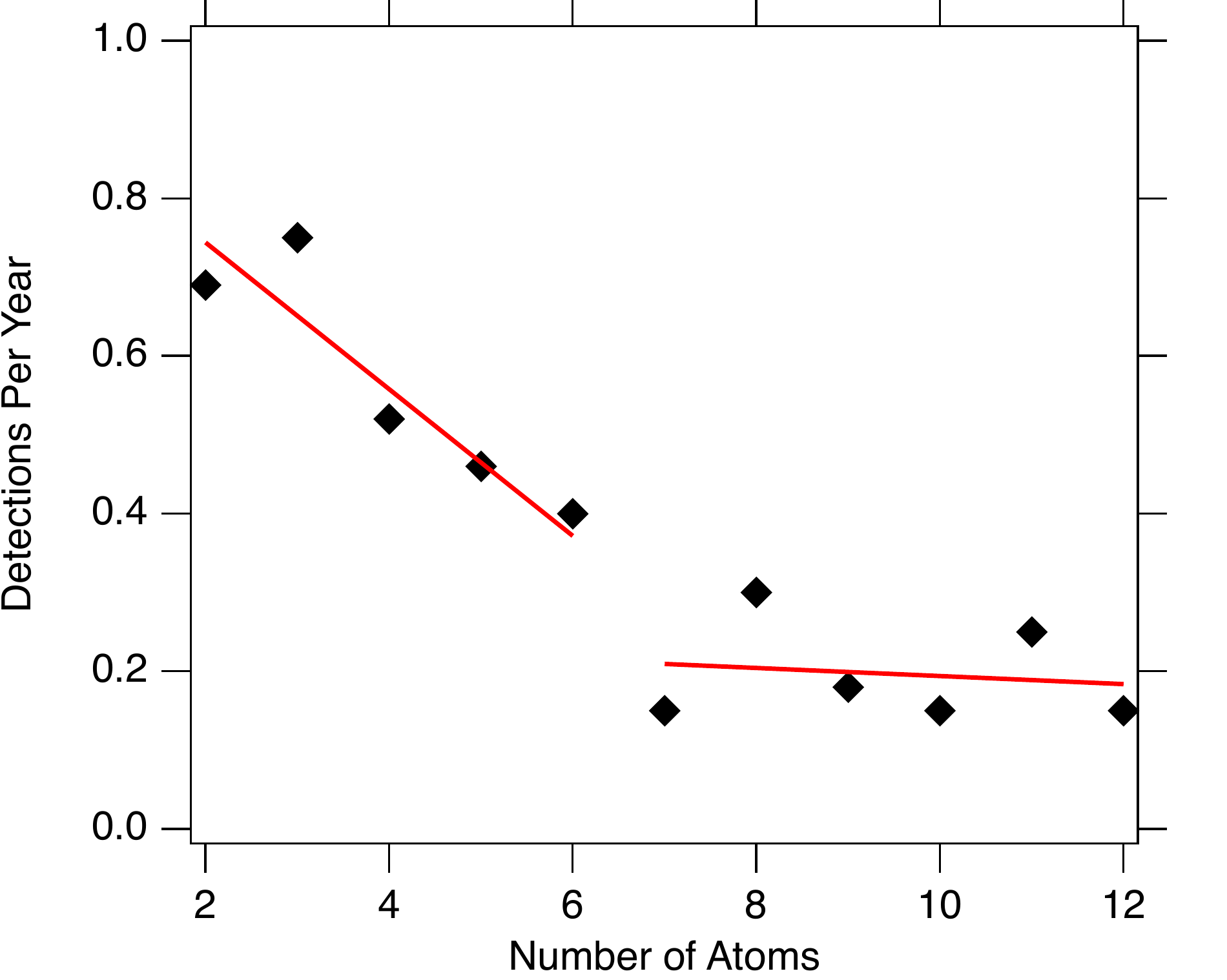}
\caption{Number of detections per year for a molecule with a given number of atoms.  The data can be visually divided into a gradually decreasing trend from two to six atom molecules, while molecules with seven to twelve atoms have a steady average detection rate of $\sim$0.2 per year.  }
\label{rates_by_atoms_figure}
\end{figure}

Figure \ref{cumulative_by_atoms} displays the same data as Figure~\ref{cumulative_detects}, broken down by the cumulative detections of molecules containing \emph{n} atoms, while Table~\ref{rates_by_atoms_table} and Figure~\ref{rates_by_atoms_figure} display the rates of new detections per year sorted by the number of atoms per molecule.  These rates show a steady decrease between 2--6 atom molecules, at which point the rate appears to flatten to an average of about one new detection every five years.  Of note, molecules with 10 or more atoms did not begin to be routinely detected until the early 2000s.

\begin{table}[h!]
\centering
\caption{Rates (\emph{m}) of detection of new molecules per year, sorted by number of atoms per molecule derived from linear fits to the data shown in Figure~\ref{cumulative_by_atoms} as well as the $R^2$ values of the fits, for molecules with 2--12 atoms.  The start year was chosen by the visual onset of a steady detection rate, and is given for each fit. }
\begin{tabular*}{\columnwidth}{c @{\extracolsep{\fill}}  c @{\extracolsep{\fill}}  c @{\extracolsep{\fill}}  c }
\hline\hline
\# Atoms    &   \emph{m} (yr$^{-1}$)    &   $R^2$       &   Onset Year      \\
\hline
2           &   0.69        &   0.97        &   1968            \\
3           &   0.75        &   0.99        &   1968            \\
4           &   0.52        &   0.98        &   1968            \\
5           &   0.46        &   0.96        &   1971            \\
6           &   0.40        &   0.98        &   1970            \\
7           &   0.15        &   0.93        &   1973            \\
8           &   0.30        &   0.91        &   1975            \\
9           &   0.18        &   0.91        &   1974            \\
10          &   0.15        &   0.67        &   2001            \\
11          &   0.25        &   0.83        &   2004            \\
12          &   0.15        &   0.85        &   2001            \\
\hline
\end{tabular*}
\label{rates_by_atoms_table}
\end{table}

\subsection{Detecting Facilities}
\label{scopes}

Table~\ref{detects_by_scope} lists the total number of detections by each facility listed in \S\ref{known}.  When two or more telescopes were used for a detection, each was given full credit.  In total, 48 different facilities have contributed to the detection of at least one new species.  Acronym definitions for these are given in Table~\ref{abbrevs} or in-text.

\begin{figure}
\includegraphics[width=\columnwidth]{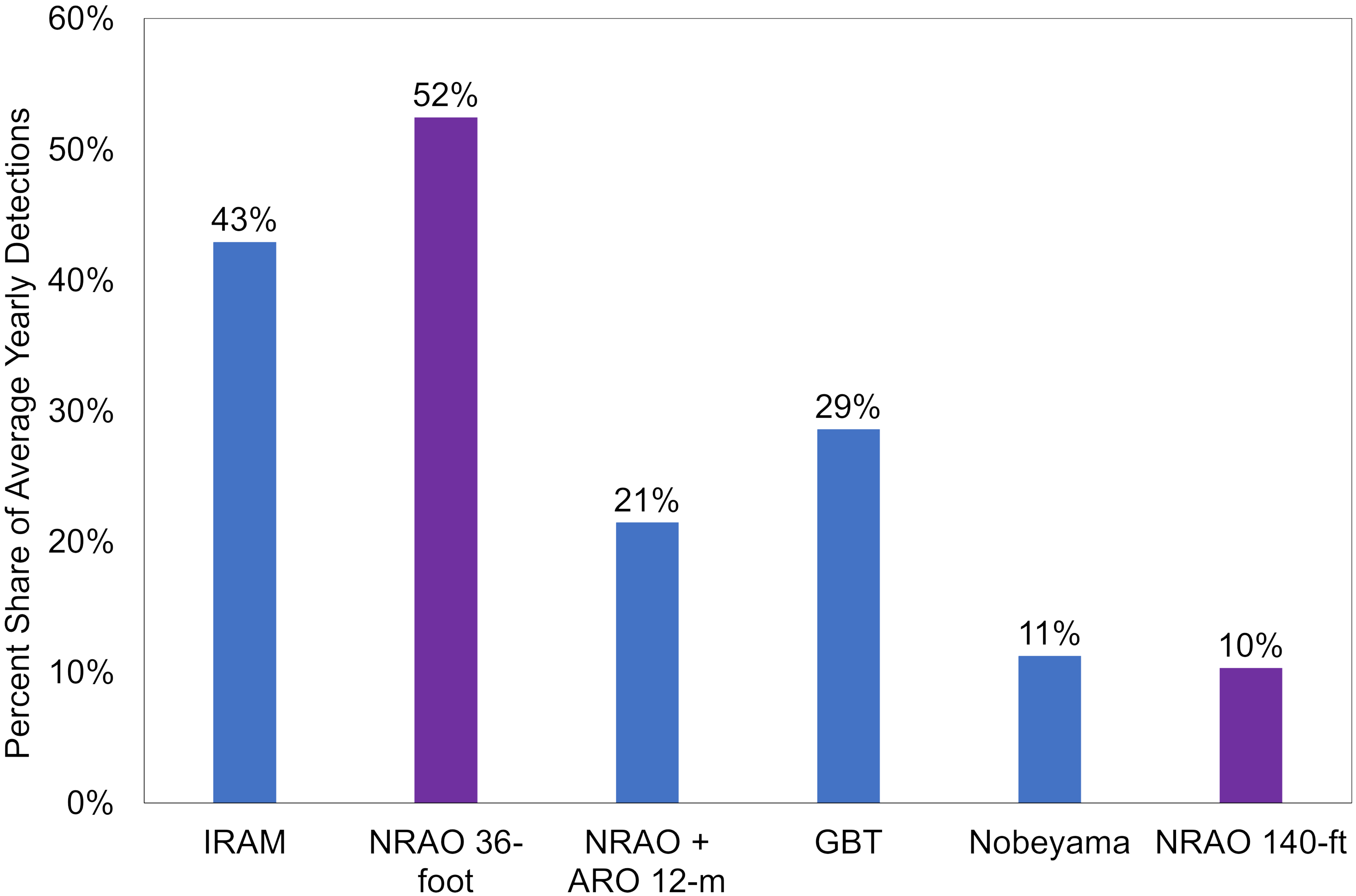}
\caption{Percentage share of the yearly detections (3.7) that a facility contributed (or is contributing) over its operational lifetime.  For example, over its operational career, the NRAO 36-foot antenna was accounting for 52\% of the annual detections.  Facilities no longer in operation are colored purple, current facilities are in blue. }
\label{share_of_detects}
\end{figure}

The Nobeyama 45-m, GBT, ARO/NRAO 12-m, NRAO 36-ft, and IRAM 30-m telescopes have cumulatively contributed to more than half of the new molecular detections.  It is worth examining the impact of a facility on the new detections over its operational lifetime.  For instance, while the IRAM 30-m telescope clearly dominates in the total number of detections, the NRAO 36-ft telescope produced more detections per year of its operational life.  Figure~\ref{share_of_detects} shows this information graphically for the top six contributing facilities; in this case, the NRAO 12-m and ARO 12-m have been combined, despite being operated by different organizations over its lifetime.  During its operation, the NRAO 36-ft contributed to more than half of all new molecular detections.  Of the currently operating facilities, the IRAM 30-m telescope is contributing the largest percentage of new detections (43\%), followed by the GBT 100-m telescope (29\%).

\begin{figure}
\includegraphics[width=\columnwidth]{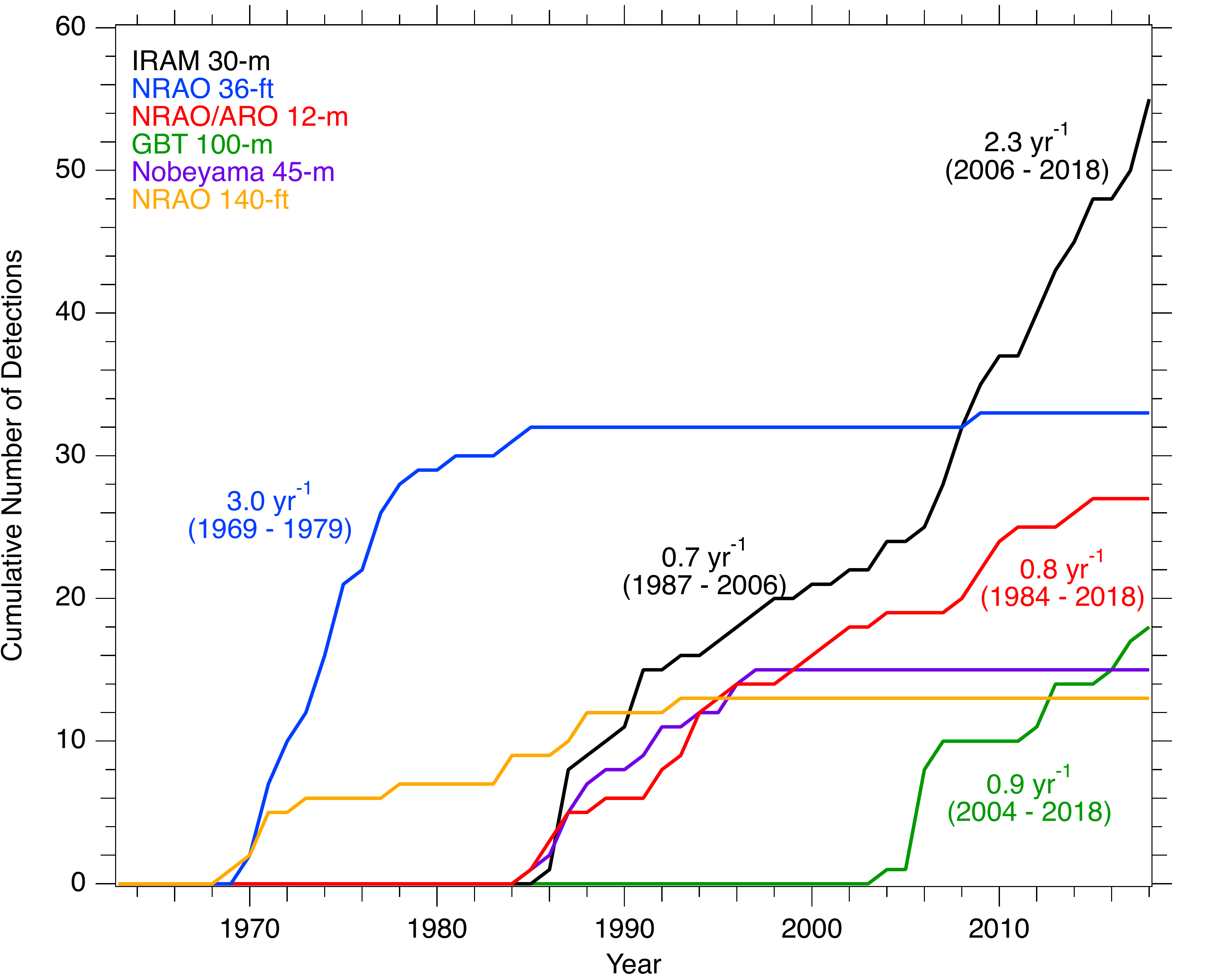}
\caption{Cumulative number of detections per facility by year.  For clarity, only those facilities with more than 10 total detections are shown.  The detection rates for several facilities over selected time periods are highlighted.}
\label{scopes_by_year}
\end{figure}

Figure~\ref{scopes_by_year} presents this data in another way, displaying the cumulative number of detections over time by each facility with more than 10 total detections.  During its prime, the NRAO 36-ft telescope was producing an average of three new detections a year, a rate which has not been matched since, although the IRAM 30-m has shown a significant increase in its detection rate since 2006, more than tripling its yearly detections.  Indeed, if this modern detection rate for the IRAM 30-m is used instead, the percentage share shown in Figure~\ref{share_of_detects} increases to 59\% for that facility.

\begin{table}[h!]
\footnotesize
\centering
\caption{Total number of detections for each facility listed in \S\ref{known}.}
\begin{tabular*}{\columnwidth}{l @{\extracolsep{\fill}}  c @{\extracolsep{\fill}}  l @{\extracolsep{\fill}}  c }
\hline\hline
Facility	&	\# 	&	Facility	&	\# \\
\hline
ALMA	&	3	&	Mayall 4-m	&	1\\
APEX	&	4	&	McMath Solar Telescope	&	1\\
ARO 12-m	&	7	&	Millstone Hill	&	1\\
ATCA	&	1	&	Mitaka 6-m	&	1\\
Aerobee-150 Rocket	&	1	&	Mt Hopkins	&	1\\
Algonquin 46-m	&	3	&	Mt Wilson	&	3\\
BIMA	&	1	&	NRAO 12-m	&	20\\
Bell 7-m	&	8	&	NRAO 140-ft	&	13\\
CSO	&	2	&	NRAO 36-ft	&	33\\
Effelsberg 100-m	&	4	&	NRL 85-ft	&	1\\
FCRAO 14-m	&	4	&	Nobeyama	&	15\\
FUSE	&	1	&	OVRO	&	2\\
GBT	&	18	&	Odin	&	1\\
Goldstone 70-m	&	1	&	Onsala	&	4\\
Hat Creek	&	2	&	Parkes 64-m	&	5\\
Haystack	&	3	&	PdBI	&	2\\
Herschel	&	7	&	SEST 15-m	&	2\\
IRAM	&	55	&	SMA	&	2\\
IRAS	&	1	&	SMT	&	7\\
IRTF	&	2	&	SOFIA	&	1\\
ISO	&	5	&	Spitzer	&	3\\
KPNO 4-m	&	3	&	U Mass 14-m	&	1\\
Kuiper	&	1	&	UKIRT	&	1\\
MWO 4.9-m	&	2	&	Yebes 40-m	&	1\\
\hline
\end{tabular*}
\label{detects_by_scope}
\end{table}

\subsection{Molecular Composition}
\label{elemental}

\begin{figure*}
\includegraphics[width=\textwidth]{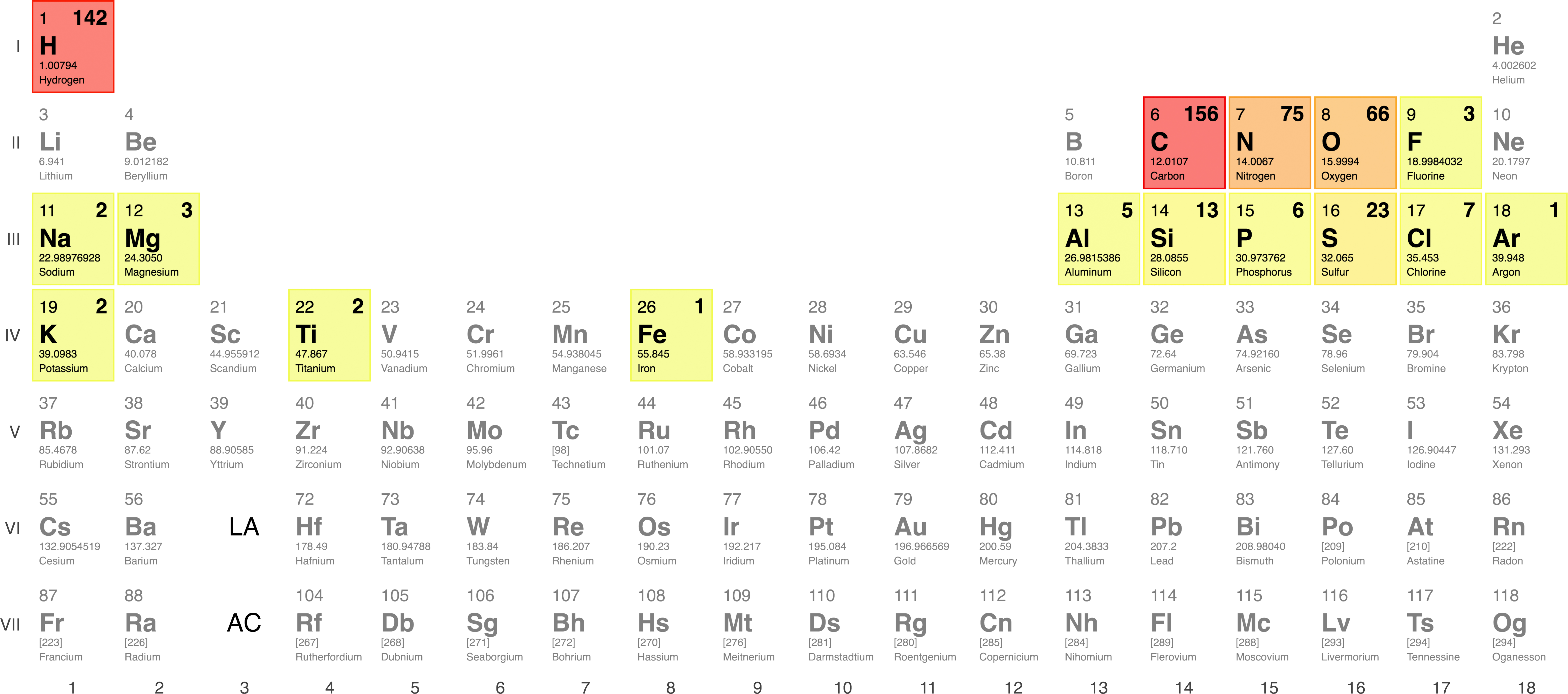}
\caption{Periodic table of the elements, absent the lathanide and actinide series, color coded by number of detected species containing each element.  For those elements with detected ISM molecules, that number is displayed in the upper right of each cell.}
\label{periodic_heatmap}
\end{figure*}

Carbon, hydrogen, nitrogen, and oxygen dominate the elemental composition of interstellar molecules, with sulfur and silicon in a distant second tier.  Indeed, the entirety of the known molecular inventory is constructed out of a mere 16 of the 118 known elements. Figure~\ref{periodic_heatmap} provides a visual representation of this composition. 

\begin{figure}
\includegraphics[width=\columnwidth]{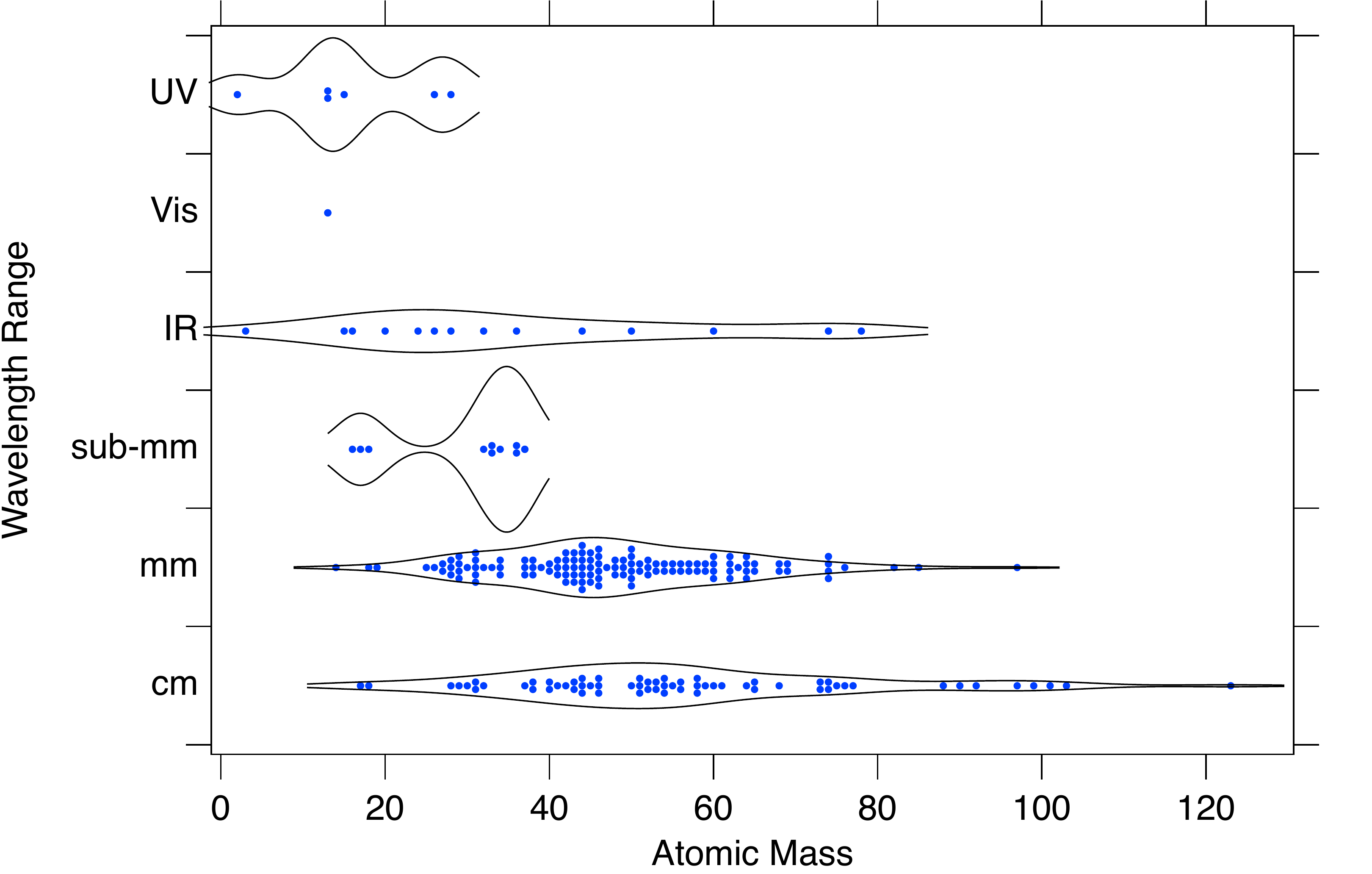}
\caption{Violin plot of the mass of molecules detected in each wavelength range.  In some cases, the initial detection was reported in two or more wavelength ranges (see text), and credit was given to each for this analysis.  The frequency ranges for cm, mm, and sub-mm used were 0--50, 50--300, and 300+~GHz, respectively.}
\label{violin_plot}
\end{figure}

Also of note is the relationship between the mass of the known molecules and the wavelength ranges that have contributed to their detection (Figure~\ref{violin_plot}).  As might be expected, there is little dependence on mass in the number of infrared detections, as the effects of increasing mass on vibrational frequencies are unlikely to shift these modes significantly enough to push them entirely out of the infrared region.  Rotational transitions, however, are heavily dependent on mass. For a given rotational temperature, and to first order, the strongest rotational transitions of a molecule shift to lower frequencies with increasing mass (see \S\ref{detecting:radio} and Appendix~\ref{app:complexity}).  This trend is reflected in the distribution of atomic masses detected by cm, mm, and sub-mm instruments.  There is a marked preference for low-mass species to be seen at high frequencies, and for high-mass species, especially those with mass $>$80~amu, to be detected at the lowest frequencies.  For a more complete discussion of the effects of increased mass and complexity on the wavelength ranges where the strongest transitions occur, see Appendix~\ref{app:complexity}.

\begin{figure}
\includegraphics[width=\columnwidth]{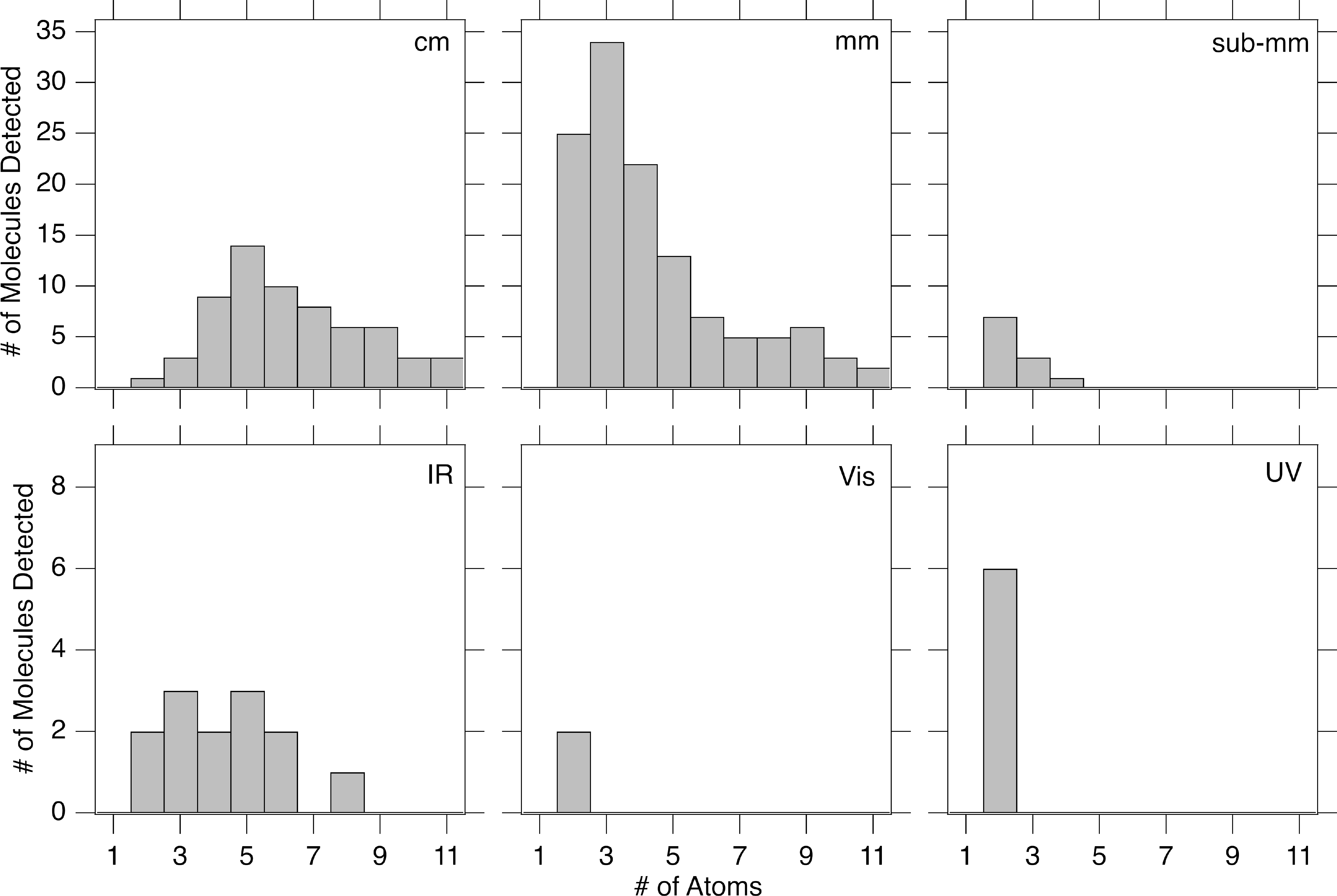}
\caption{Histograms of the number of molecules detected in each wavelength range with a given number of atoms.   In some cases, the initial detection was reported in two or more wavelength ranges (see text), and credit was given to each for this analysis.  The fullerene molecules are not included here.}
\label{atoms_histogram}
\end{figure}

This trend is also borne out in the number of atoms in a molecule detected at each wavelength (Figure~\ref{atoms_histogram}).  The cm range has the peak of its distribution at 5 atoms, but extends heavily to larger numbers.  The mm range shows a lower peak (3) atoms, but also sees a distribution to larger numbers.  As might be expected, the infrared shows a rather flat distribution, as the vibrational transitions of molecules probed here will all generally fall within the IR, regardless of mass or number of atoms.  Interestingly, all molecules detected in the visible and UV are diatomics.

\begin{figure}
\includegraphics[width=\columnwidth]{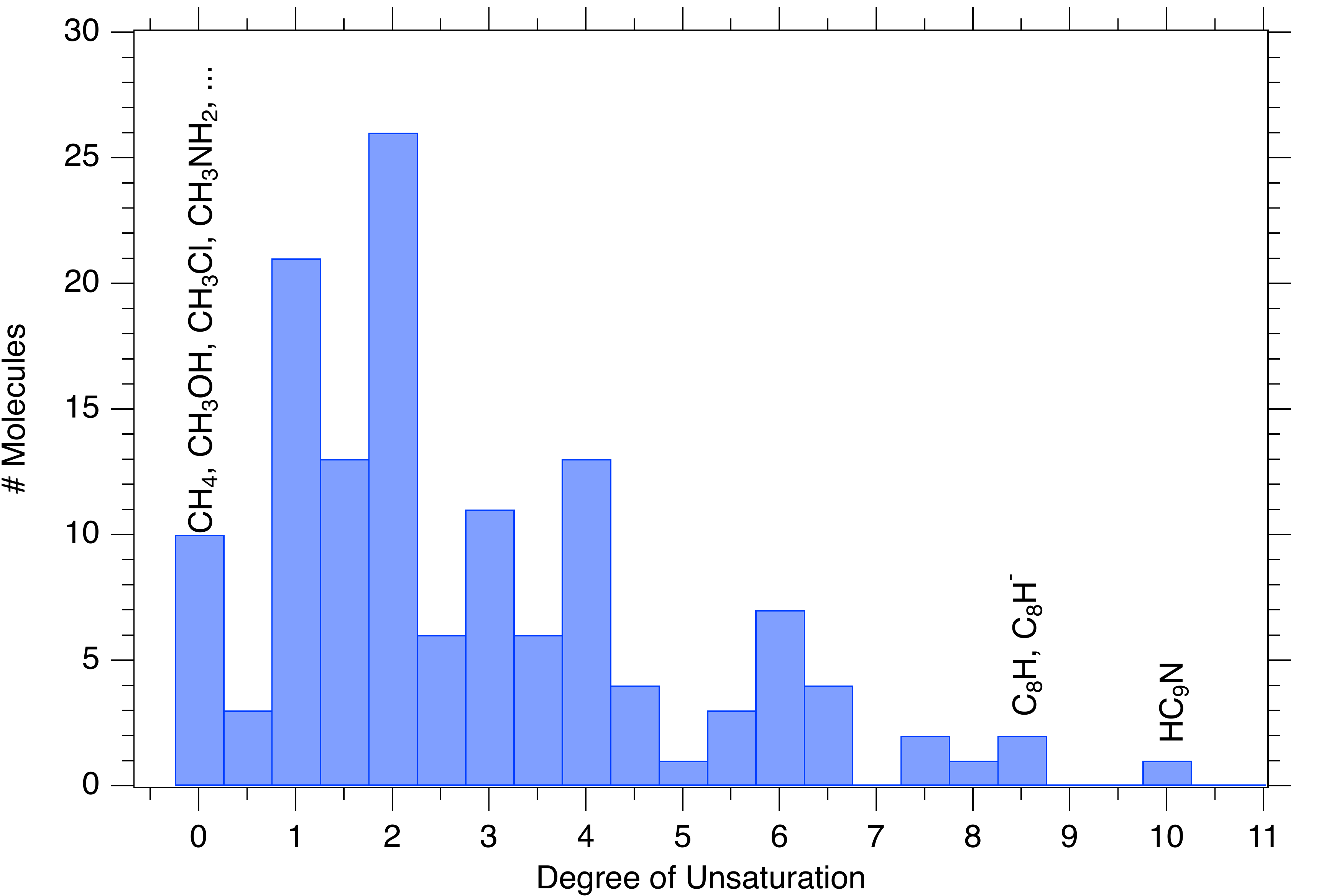}
\caption{Histogram of the degree of unsaturation, or alternatively the number of rings and $\pi$-bonds (see text), for hydrocarbon molecules containing only H, O, N, C, Cl, or F. Only seven interstellar hydrocarbons are fully saturated with no rings or $\pi$ bonds.  Most molecules contain one or two rings or $\pi$ bonds (a double bond is one $\pi$ bond, a triple bond is two $\pi$ bonds).  The fullerene molecules are not shown.}
\label{unsaturation}
\end{figure}

Another interesting metric is the distribution of saturated and unsaturated hydrocarbon molecules detected.  A fully saturated hydrocarbon is one where no $\pi$ bonds (double or triple bonds) and no rings exist, with these electrons fully dedicated to bonding other elements (usually hydrogen).  A simple formula for calculating the Degree of Unsaturation (DU) or equivalently the total number of rings and $\pi$ bonds in a hydrocarbon molecule is given by Equation~\ref{unsat_eq}:
\begin{equation}
    \rm{DU} = 1 + n(\rm{C}) + \frac{n(\rm{N})}{2} - \frac{n(\rm{H})}{2} - \frac{n(\rm{X})}{2}
    \label{unsat_eq}
\end{equation}
where $n$ is the number of each element in the molecule, and X is a halogen (F or Cl). Here, each atom contributes $x - 2$ to the DU, where $x$ is its valence; thus, the divalent elements O and S do not contribute. A visual representation of this distribution for interstellar hydrocarbons is shown as a histogram in Figure~\ref{unsaturation}.  With the exception of the fullerenes (not shown), \ce{HC9N} is the most unsaturated interstellar molecule, while only ten species are fully saturated: \ce{CH4}, \ce{CH3Cl}, \ce{CH3OH}, \ce{CH3SH}, \ce{CH3NH2}, \ce{CH3OCH3}, \ce{CH3CH2OH}, \ce{CH3CH2SH}, \ce{HOCH2CH2OH}, and \ce{CH3OCH2OH}.  The most common DUs are 1.0 and 2.0, representing one or two rings and/or $\pi$ bonds.  In total, 93\% of all interstellar hydrocarbons are unsaturated to some degree.

\begin{figure}
\includegraphics[width=\columnwidth]{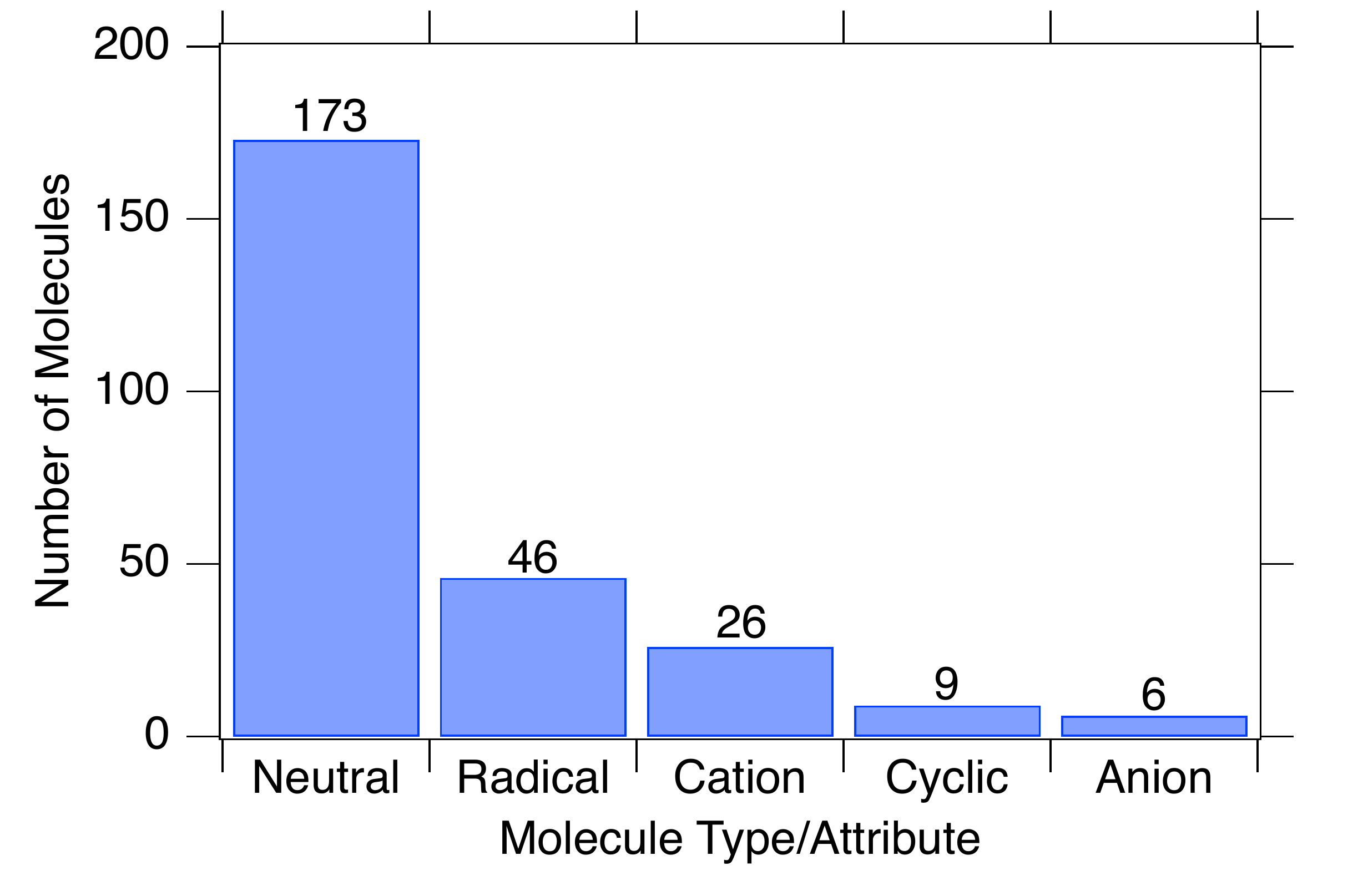}
\caption{Number of known interstellar molecules that are neutral, cationic, anionic, radical species, or cyclic.  Many molecules fall into more than one of these categories (e.g. most radical species have a net neutral charge).}
\label{molecule_types}
\end{figure}

Figure~\ref{molecule_types} provides a visual breakdown of the number of known interstellar molecules that are neutral, cationic, anionic, radical species, or cyclic.  Many molecules fall into more than one of these categories.  An analysis of the rates of detection of these various types of molecules in differing types of interstellar sources is provided in \S\ref{source_types}.

\subsection{Detection Sources, Source Types, and Trends}
\label{source_types}

\begin{figure}
\includegraphics[width=\columnwidth]{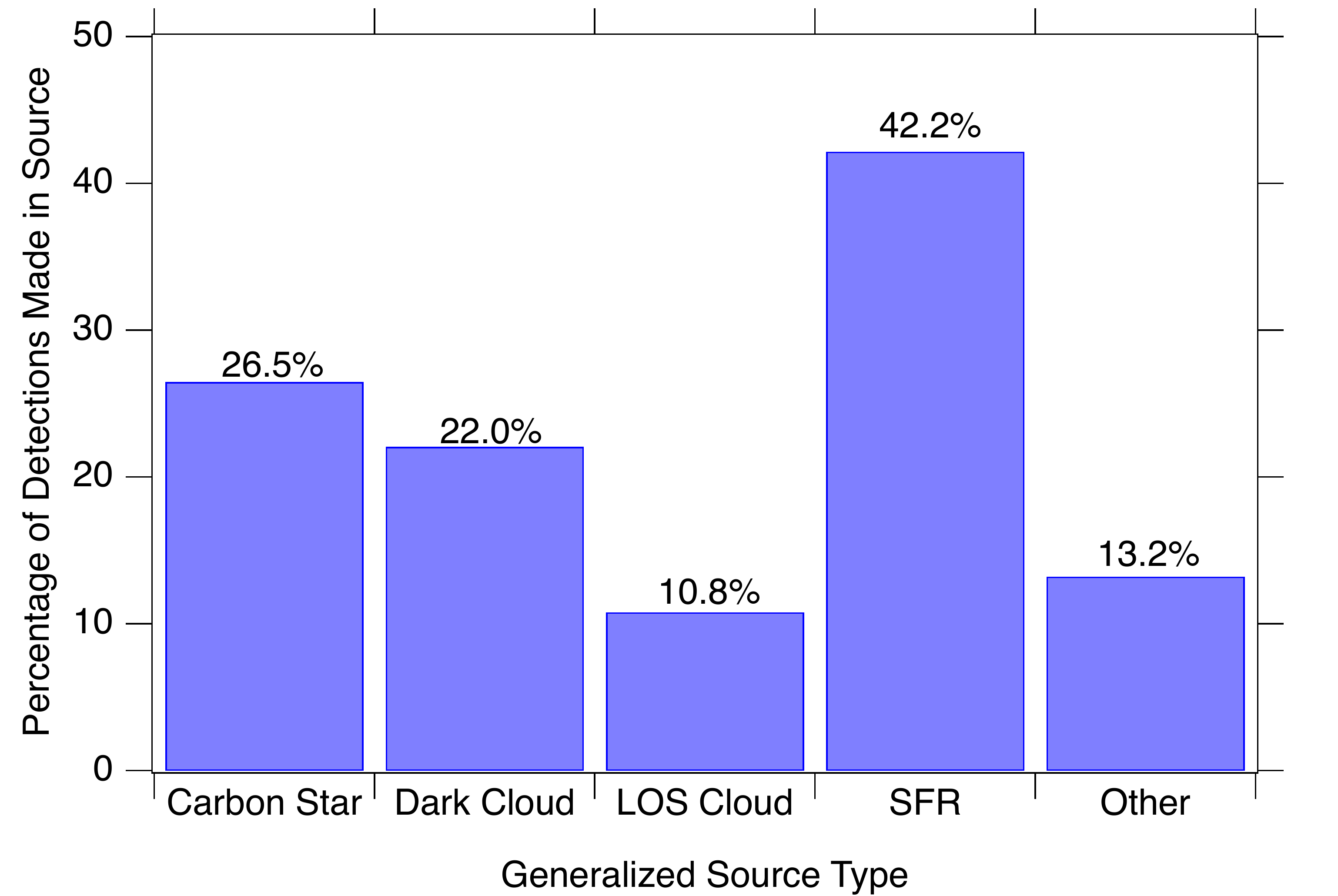}
\caption{Percentage of known molecules that were detected for the first time in carbon stars, dark clouds, LOS clouds, and SFRs (see text).  Some molecules were simultaneously detected in multiple source types (e.g., \ce{C8H-} in TMC-1 and IRC+10216).}
\label{detects_by_source_type}
\end{figure}

More than 90\% of the detections can be readily classified as being made in a source that falls into the generalized type of either a carbon star (e.g., IRC+10216), dark cloud (e.g., TMC-1), a diffuse/translucent/dense cloud along the line of sight to a background source (hereafter `LOS Cloud'), or a star-forming region (SFR; e.g., Sgr B2). Figure~\ref{detects_by_source_type} displays the percentage of interstellar molecules that were detected in each source type.  Note that because some molecules were simultaneously detected in more than one source type, these percentages add to $>$100\%.  The number of detections in each individual source is listed in Table~\ref{detects_by_source}.

\begin{table}[h!]
\centering
\caption{Total number of detections that each source contributed to for the molecules listed in \S\ref{known}.  Detections made in clouds along the line of sight to a background source have been consolidated into `LOS Clouds,' and detections in closely-location regions have been group together as well (e.g. Sgr B2(OH), Sgr B2(N), Sgr B2(S), and Sgr B2(M) are all considered Sgr B2).}
\begin{tabular*}{\columnwidth}{l @{\extracolsep{\fill}}  c @{\extracolsep{\fill}}  l @{\extracolsep{\fill}}  c }
\hline\hline
Source	&	\#	&Source	&	\# \\
\hline
Sgr B2	&	67	&	NGC 7023	&	2\\
IRC+10216	&	51	&	TC 1	&	2\\
TMC-1	&	34	&	W49	&	2\\
Orion	&	24	&	CRL 2688	&	1\\
LOS Clouds	&	22	&	Crab Nebula	&	1\\
L483	&	8	&	DR 21(OH)	&	1\\
W51	&	8	&	Galactic Center	&	1\\
VY Ca Maj	&	5	&	IC 443G	&	1\\
B1-b	&	4	&	IRAS 16293	&	1\\
DR 21	&	4	&	K3-50	&	1\\
NGC 6334	&	4	&	L134	&	1\\
Sgr A	&	4	&	L1527	&	1\\
CRL 618	&	3	&	L1544	&	1\\
NGC 2264	&	3	&	L183	&	1\\
W3(OH)	&	3	&	Lupus-1A	&	1\\
rho Oph A	&	3	&	M17SW	&	1\\
Horsehead PDR	&	2	&	NGC 7027	&	1\\
NGC 2024	&	2	&	NGC 7538	&	1\\
\hline
\end{tabular*}
\label{detects_by_source}
\end{table}

\begin{figure}
\includegraphics[width=\columnwidth]{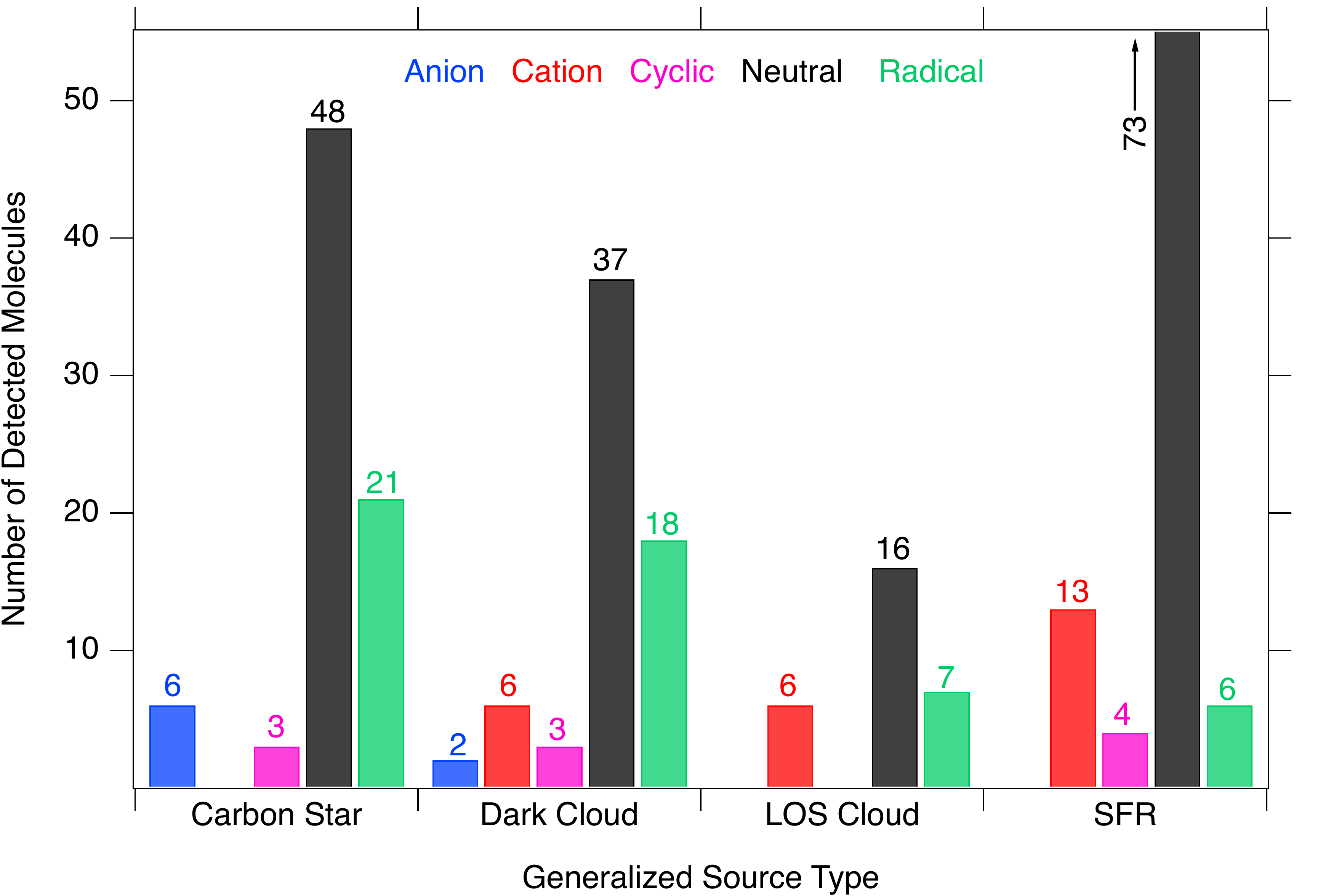}
\caption{Number of known interstellar molecules that are neutral, cationic, anionic, radical species, or cyclic detected in four generalized source types.  A small number of detections that did not fit easily into one of these categories (e.g. \ce{ArH+} in the Crab Nebula supernova remnant) are excluded.  As with Figure~\ref{molecule_types}, species falling into more than one molecule type are counted for each of that type.}
\label{mol_type_by_source_type}
\end{figure}

The distribution of molecules, as categorized by their attributes in Figure~\ref{molecule_types}, across the generalized source types is shown in Figure~\ref{mol_type_by_source_type}.  Immediately obvious from this is that no molecular anions were first detected outside of carbon stars and dark clouds, a fact that remains the case even outside of first detections.  

\begin{figure}
\includegraphics[width=\columnwidth]{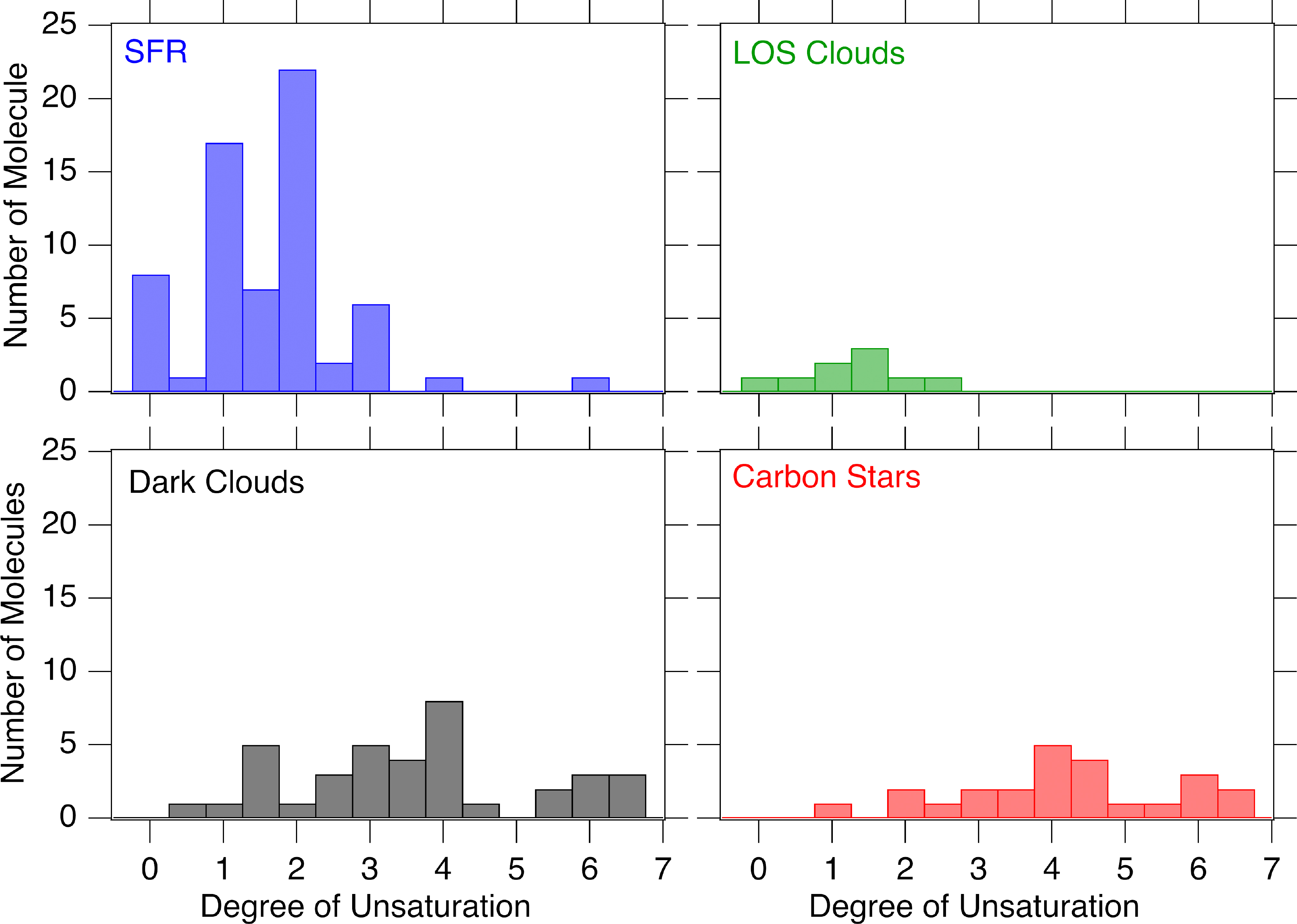}
\caption{Histograms of the degree of unsaturation of molecules across the four generalized source types.}
\label{du_by_source_type}
\end{figure}

\begin{figure}
\includegraphics[width=\columnwidth]{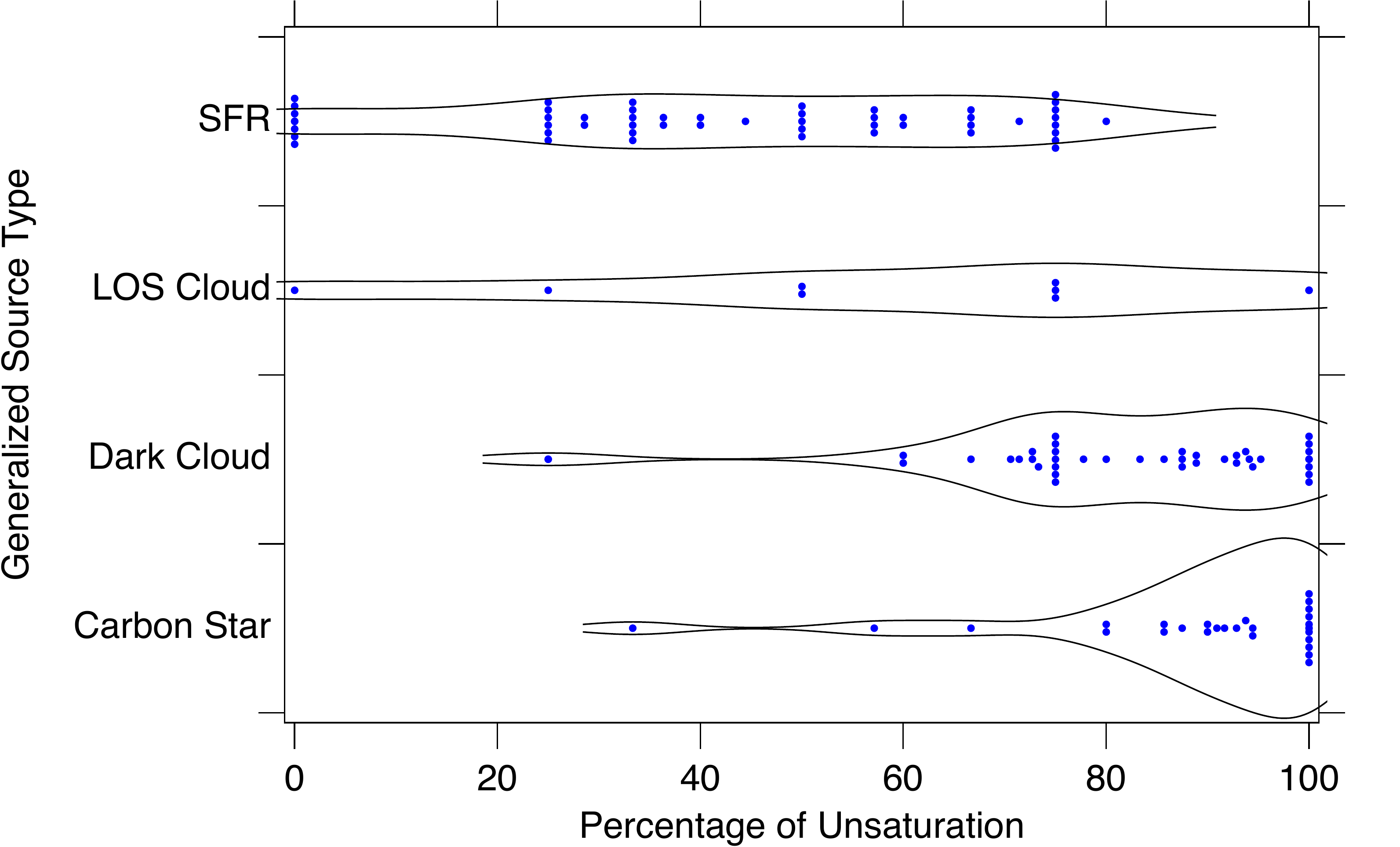}
\caption{Violin plot of the percentage of unsaturation of molecules across the same source types.}
\label{rel_du_by_source_type}
\end{figure}

This may not be surprising when also considering the average degree of unsaturation across these generalized source types (Figure~\ref{du_by_source_type}).  The maximum degree of unsaturation of molecules is molecule-dependent, meaning that the distribution of DU by source type is biased by the length/size of molecules seen there.  This can be mitigated by examining the relative degree of unsaturation - meaning, how close to fully unsaturated the molecules in a source type are.  This is shown in Figure~\ref{rel_du_by_source_type}, where the percentage of unsaturation is the molecule's DU divided by the DU for a fully unsaturated version of that species.

\begin{figure}
\includegraphics[width=\columnwidth]{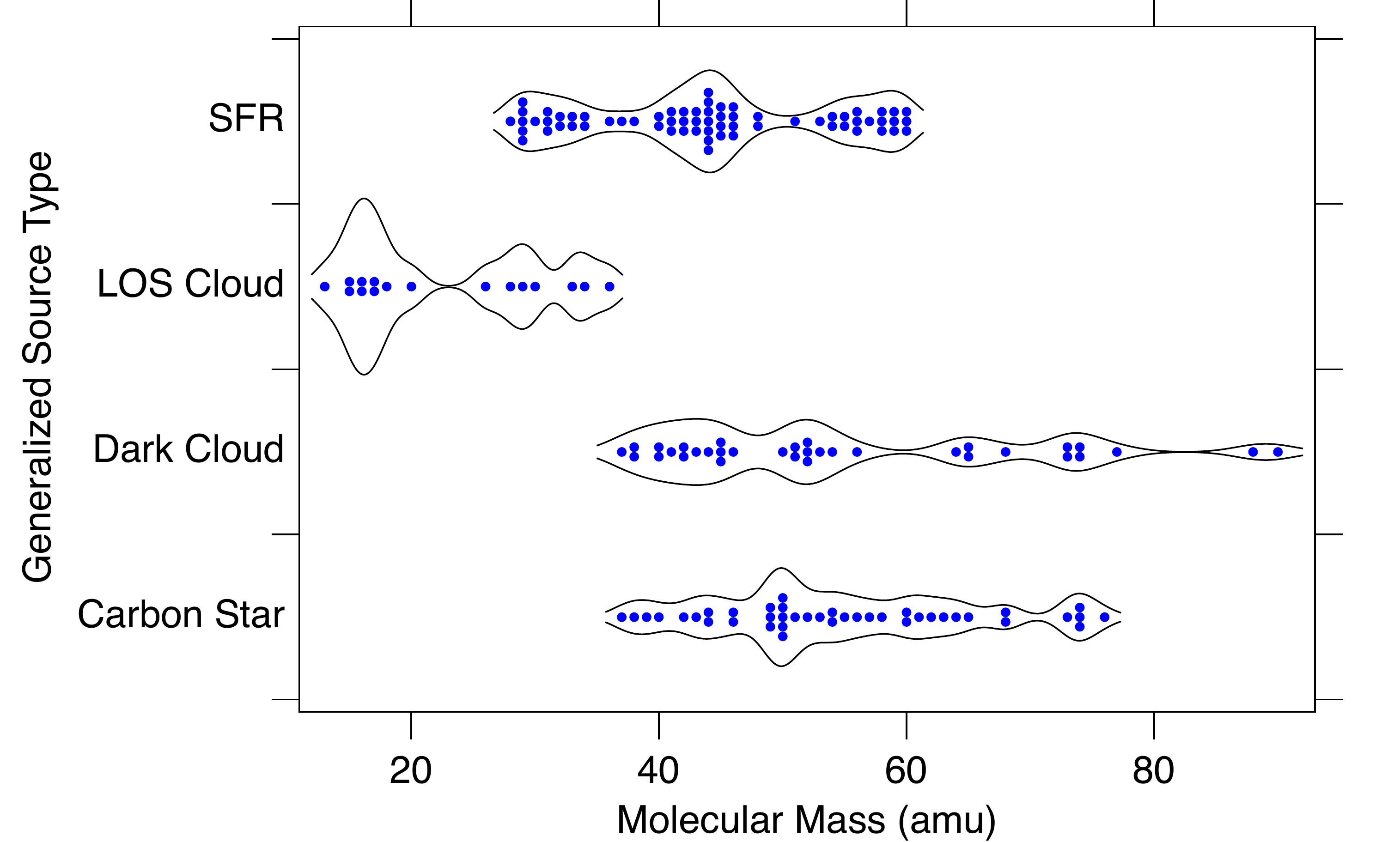}
\caption{Violin plot of the molecular masses of non-fullerene molecules detected in each of the four generalized source types. The highest and lowest 10\% of values (minimum 1) have been dropped from each dataset.}
\label{mass_by_source_type}
\end{figure}

In both cases, and with very few exceptions, molecules detected in carbon stars and dark clouds are on average highly unsaturated compared to other sources, and the six known molecular anions, \ce{CN-}, \ce{C3N-}, \ce{C4H-}, \ce{C5N-}, \ce{C6H-}, and \ce{C8H-} are no exception.  As shown in Figure~\ref{mass_by_source_type}, the molecules detected in these sources also tend to be the most massive.

A possible explanation for this trend is the influence of grain-surface/ice-mantle chemistry in SFRs.  One of the most efficient pathways for increasing the saturation of species in the ISM is through direct hydrogenation on grain surfaces.  A substantial amount of laboratory (e.g., \citealt{Linnartz:2015ec} and \citealt{Fedoseev:2015ef}), quantum chemical (e.g., \citealt{Woon:2002wu}), and chemical modeling (e.g., \citealt{Garrod:2008tk}) work suggests that these hydrogenation processes are efficient in the ISM, often even at low temperatures.  In carbon stars, the bulk of the ice is thought to have been long since evaporated, with any hydrogenation occurring via catalysis on the (mostly) bare grain surfaces \citep{Willacy:2003if}.  In dark clouds, the temperatures are so low that although chemistry is likely occurring in ice mantles, there is no readily apparent mechanism for liberating these molecules into the gas phase for detection.  Some recent work in this area has suggested that cosmic-ray impacts could non-thermally eject this material in these sources \citep{Ribeiro:2015fe}, but this is certainly not as efficient a process as shock-induced \citep{RequenaTorres:2006ki} or thermal desorption that play significant roles in SFRs. 

Looking back at Figure~\ref{mol_type_by_source_type}, a few other trends are apparent.  First, the fraction of species detected in SFRs that are radicals is substantially lower (5\%) than seen in the other environments (11\% to as high as 30\% [dark clouds]).  Radicals tend to be highly chemically reactive species, and perhaps are being depleted as more reaction partners are being liberated from the surface of ice mantles in these regions.  Second, it bears noting that the only source type in which all five types of species discussed here were first detected are dark clouds, perhaps suggesting that these regions are more chemically complex than they may appear at first glance, especially given the prominence of SFRs like Sgr B2 and Orion in the detection of new molecules.

\begin{figure}[t!]
\includegraphics[width=\columnwidth]{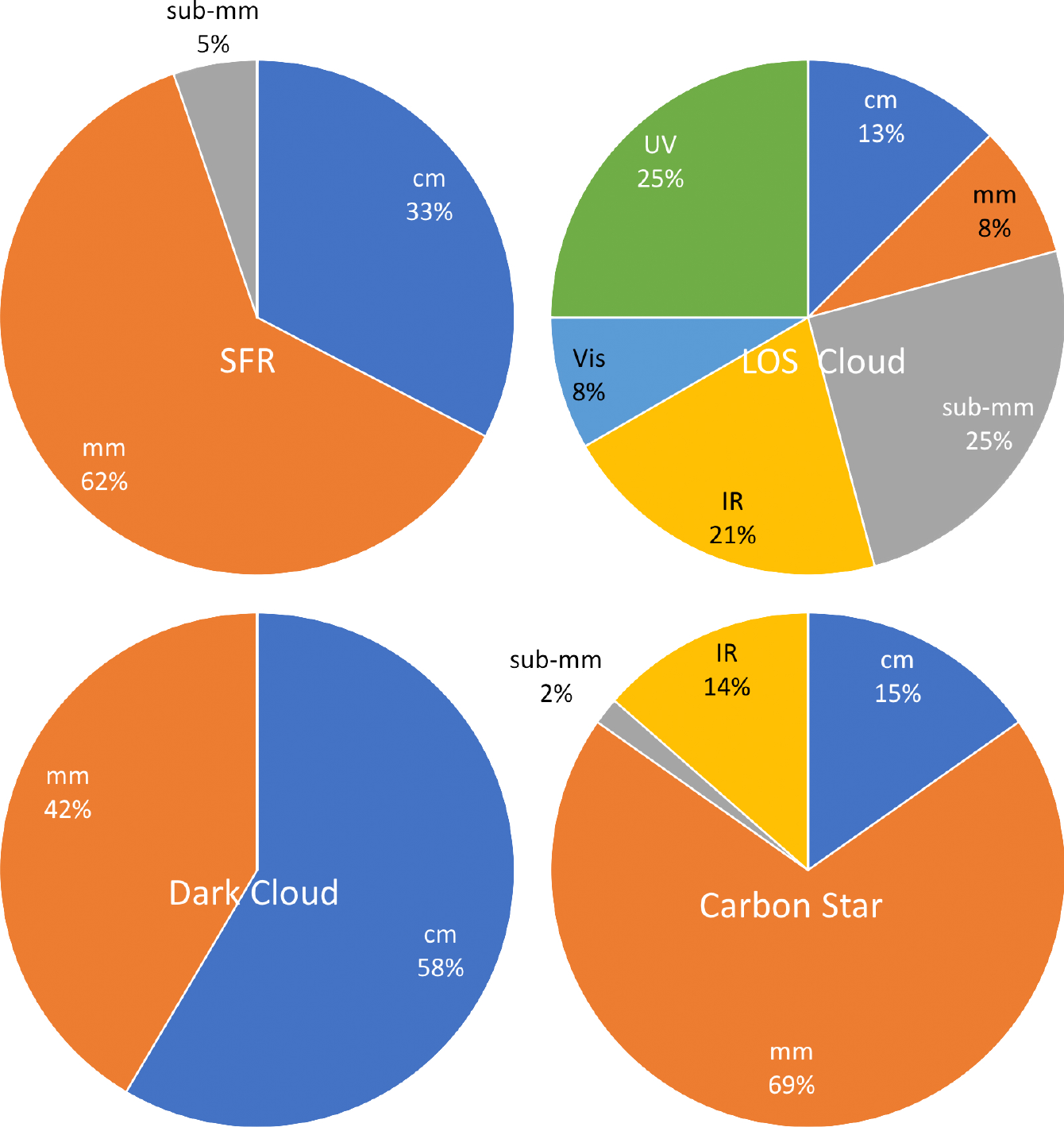}
\caption{Percentage of first detections in each generalized source type that were made at cm, mm, sub-mm, infrared, visible, and ultraviolet wavelengths.}
\label{waves_by_source_type}
\end{figure}

It is also worth examining the wavelength ranges that contributed to first detections in each of these generalized source types (Figure~\ref{waves_by_source_type}).  As discussed in \S\ref{detecting:infrared} and \S\ref{detecting:uvvis}, detections in the IR, visible, and UV portions of the spectrum are nearly always performed in absorption, necessitating both a background source and an optically thin absorbing medium.  LOS clouds satisfy both of these requirements, and thus these regions show the greatest diversity in wavelengths used for detection, whereas SFR and dark clouds, which are often optically thick, have not yet seen a first detection at wavelengths shorter than the far-infrared.

\section{Conclusions}
\label{conclusions}

In summary, 204 individual molecular species have been detected in the ISM.  A further eleven molecules are considered to be tentatively detected, while two species have had their detections disputed.  These detections have been dominated by radio astronomical observations, with the IRAM 30-m, GBT 100-m, Nobeyama 45-m, and NRAO/ARO 12-m telescopes the most prolific extant facilities.  Beginning in 1968, the rate of new detections per year can be well-fit to a linear trend of 3.7 new molecules per year, although there is evidence for an increase in this rate in the last decade, due to the onset of GBT detections and a tripling of the rate of IRAM detections.  A substantial fraction ($\sim$32\%) of known molecules have now been seen in external galaxies, while the numbers of molecules known in protoplanetary disks (23), interstellar ices (6), and exoplanet atmospheres (5) are much smaller due to observational challenges.

\acknowledgments

B.A.M. sincerely thanks the five anonymous referees for their careful reading and suggestions which have substantially improved the quality of this manuscript.  B.A.M. also thanks J. Mangum for helpful discussions regarding the history of NRAO telescope facilities, L.I. Cleeves for discussions of molecules in protoplanetary disks, and A. Burkhardt for a critical reading of the appendix.  The National Radio Astronomy Observatory is a facility of the National Science Foundation operated under cooperative agreement by Associated Universities, Inc. Support for B.A.M. was provided by NASA through Hubble Fellowship grant \#HST-HF2-51396 awarded by the Space Telescope Science Institute, which is operated by the Association of Universities for Research in Astronomy, Inc., for NASA, under contract NAS5-26555. 

\appendix

\renewcommand\thefigure{\thesection\arabic{figure}}   
\renewcommand\thetable{\thesection\arabic{table}}    

\setcounter{figure}{0}    
\setcounter{table}{0} 

\twocolumngrid

\section{Sources of Molecular Complexity and Their Impact on Detectability}
\label{app:complexity}

The purpose of this appendix is to examine a number of the factors which affect the detectability of molecules in the ISM.  When considering a molecule that has not yet been detected in the ISM (or in protoplanetary disks, etc.), examining at each of the factors presented here should provide an informed first-look at the reasons why detection may be challenging, beyond merely a potential low abundance.   The list of factors outlined here is not exhaustive, but instead focuses on those that commonly affect molecules observed in interstellar sources.  Further, while some quantitative measures and analytical formulas will be presented, this discussion is intended to be primarily qualitative in nature.  Some concepts will be presented as zeroth- or first-order approximations. The terminology may not always be quantum mechanically rigorous.

In the last decade, the common astrochemical parlance defining a complex organic molecule (COM) has been any molecule with six or more atoms, with methanol (\ce{CH3OH}) the prototypical simplest COM \citep{Herbst:2009go}.  This definition, while arbitrary, has proven a useful aid in the discussion of structural complexity.  The detectability of molecules, however, is affected by much more than the number of atoms.  Indeed, as will be shown later, a low-abundance linear molecule may be far easier to observe than a higher-abundance, asymmetric-top containing far fewer atoms.  Similarly, a low-abundance, structurally complex molecule may be more readily detected than a highly abundant, structurally simple molecule with very weak transitions. 

As discussed in the main text, the bulk of new detections are made via observation of rotational transitions in the radio, and so that will be the focus of this discussion. Emission or absorption signals from molecules in the cm, mm, and sub-mm regimes almost always arise from rotational transitions as a molecule moves between two rotational energy levels.  The intensity of any given signal is dependent on numerous factors, many of which will be discussed here, but these can largely be broken down into three primary components.
\begin{enumerate}
\item intrinsic, quantum-mechanical properties of the molecule
\item telescope and radioastronomical source properties
\item the absolute number of molecules undergoing that transition.
\end{enumerate}
The molecular, instrumental, and source properties (1 and 2) will vary from line to line, molecule to molecule, and source to source, but the effects of abundance (3), excluding a number of edge cases, are universal.  There are only so many molecules available to undergo a transition, and produce an observable signal, in a source.

\begin{figure}
\includegraphics[width=\columnwidth]{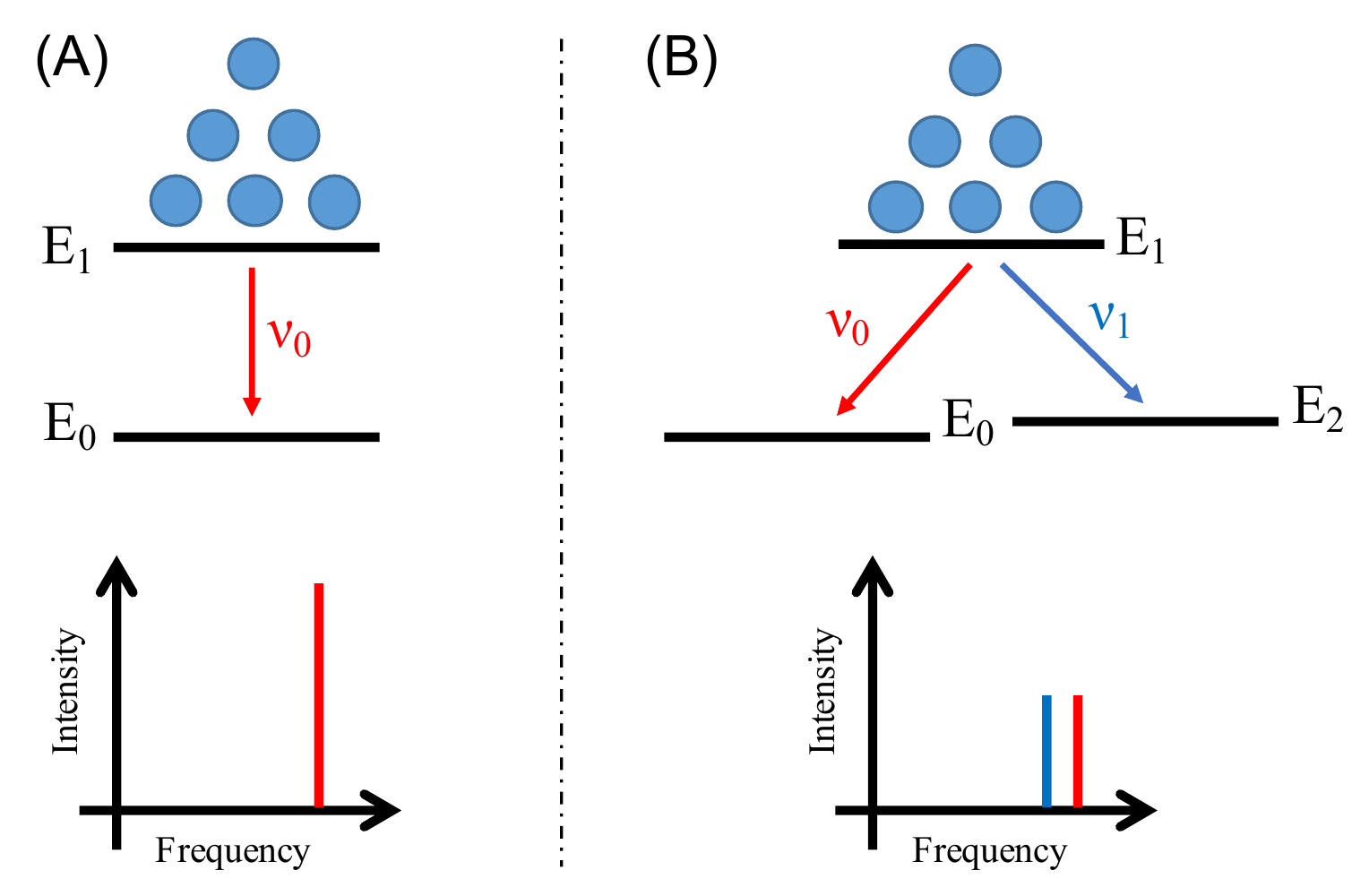}
\caption{Idealized examples of the effects of having additional energy levels to undergo transitions.  In case \textbf{A}, all six molecules can undergo transition~\textcolor{red}{$\nu_0$}.  In case \textbf{B}, about half the molecules will undergo transition \textcolor{red}{$\nu_0$}, but the other half will undergo transition~\textcolor{blue}{$\nu_1$}.  In case \textbf{B}, the intensity of transition~\textcolor{red}{$\nu_0$} is therefore half of what it is in case \textbf{A}.}
\label{v0_v1}
\end{figure}

While this may seem a triviality, the effects are far-reaching.  Figure~\ref{v0_v1} describes two cases for a molecule undergoing a transition and emitting light to be detected by a telescope looking for frequency~\textcolor{red}{$\nu_0$}.  In the first case, where all the population undergoes the transition, the light observed at frequency~\textcolor{red}{$\nu_0$} is twice as great as that of the second case, where the molecule can now undergo a similar transition at a different frequency because of the presence of an additional energy level.  As will become clear in the following sections, the true complexity of a species is largely measured in the number of rotational energy levels the population is distributed over, and undergoing transitions between.  The more levels, the fewer photons that are produced at any given transition frequency, and the more complex the spectrum becomes as additional transition frequencies appear.  While a simple linear molecule like CO may have a few dozen energy levels populated at 300~K, a truly complex molecule may have hundreds of thousands.

Thus, a reasonable measure of the level of complexity is the number of rotational energy levels which can be expected to have a non-trivial fraction of the total number of that molecule in a source.  Because these levels vary in energy, the fraction of the molecules in each state is dependent on the average energy of the population of molecules, which is described by an excitation temperature.  In many cases, this distribution is governed by a Boltzmann distribution, where the fraction ($F_n$) of molecules in any given energy ($E_n$) state $n$ is governed by Equation~\ref{boltzmann1}, and the ratio of population between any two states ($E_n$,~$E_m$) is governed by Equation~\ref{boltzmann2}, at an excitation temperature~$T_{\rm{ex}}$.
\begin{equation}
F_n \propto e^{-\frac{E_n}{kT_{\rm{ex}}}}
\label{boltzmann1}
\end{equation}
\begin{equation}
\frac{F_n}{F_m} = e^{\frac{E_m - E_n}{kT_{\rm{ex}}}}
\label{boltzmann2}
\end{equation}

The rigorous accounting of states with a significant $F_n$ at a given temperature is expressed through the temperature-dependent partition function, which is discussed explicitly in \S\ref{sec:partition}.  While the partition function is often used as a practical proxy for the number of non-trivially populated energy levels, for the purposes of this initial discussion, a simpler approximation can be adopted.  Drawing on Equations~\ref{boltzmann1}~and~\ref{boltzmann2}, the $n^{th}$ energy level of a molecule having energy $E_n = 5kT$ will have $F_n$~=~$e^{-5}$, or $\sim$1\% relative to the ground state ($E_0$~=~$0kT$) population.  Thus, the number of states with energies $<5kT$ is an excellent approximation for the number of states which will have a non-trivial population at any given temperature.  

As discussed above, the number of possible transitions increases with the number of states that have a non-trivial population: more states, more transitions, lower intensity for any given one.  Thus, when examining the individual factors that affect the detectability of a molecule (i.e. the overall intensity of transitions and the number of those transitions), the number of states below $5kT$ can provide a gross approximation of the magnitude of an effect on detectability. 

\subsection{Number of Atoms}
\label{natoms}

The number of atoms in a given molecule is a good first approximation of spectral complexity.   The energy levels of a molecule are determined by the molecular structure and reflected in the moments of inertia.  Consider a simple linear molecule, whose energy levels are given, to zeroth-order, by Equation \ref{linear_energy_app}, where $B$ is the rotational constant (often expressed in units of MHz or cm$^{-1}$), and $J$ is a quantum number representing the total rotational angular momentum \citep{Bernath:2005dw}.
\begin{equation}
E_J = BJ(J+1)
\label{linear_energy_app}
\end{equation}
The rotational constant $B$ is then related to the moment of inertia, $I$, along the molecular axis by Equation \ref{linear_B_app} \citep{Bernath:2005dw}.
\begin{equation}
B = \frac{h^2}{8\pi^2I}
\label{linear_B_app}
\end{equation}

\begin{figure}
\includegraphics[width=\columnwidth]{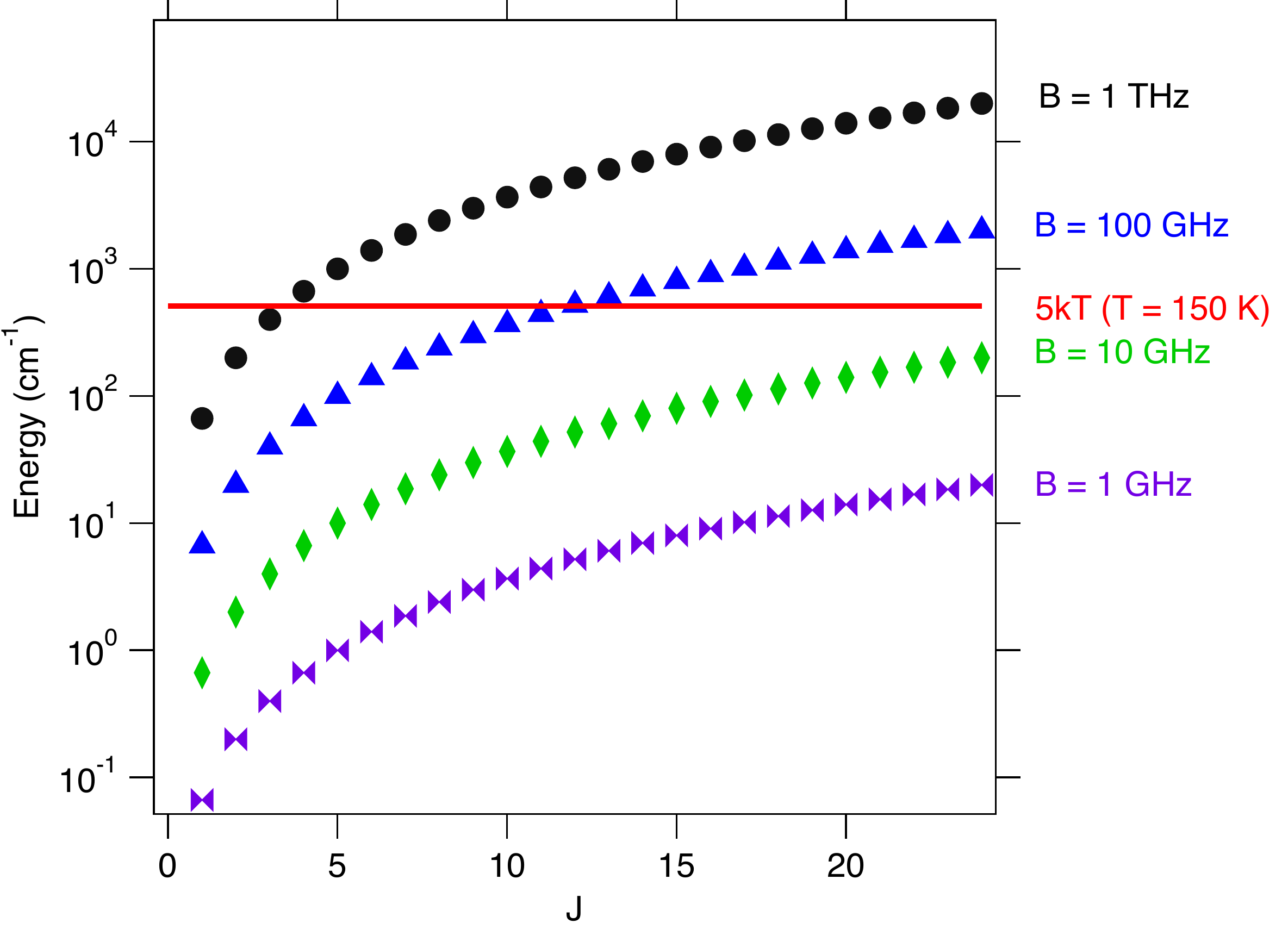}
\caption{Rotational energy levels for molecules with rotational constants $B$ = 1000, 100, 10, and 1 GHz.  The red line is drawn at 521 cm$^{-1}$, the value of $5kT$ at $T~=~150$~K; energy levels below this value are non-trivially populated.}
\label{e_vs_j}
\end{figure}

The moment of inertia is proportional to the amount of torque needed to rotate a molecule, and is determined by the spatial distribution of mass from the axis of rotation, going as the square of the distance.  A longer molecule will therefore, in general, have a larger $I$ (and a correspondingly small $B$) than a shorter molecule with similar constituent atoms.  This means for a given value of $J$, the energy of that level will be lower.  In turn, there are more energy levels accessible $<$5$kT$, and the total population will be spread more thinly among them.  Figure \ref{e_vs_j} shows the number of levels which fall beneath $5kT$ for four hypothetical molecules with rotational constants spanning four orders of magnitude. 

A more specific example is the HC$_\textup{x}$N family of linear molecules, with astronomically-observed constituents as large as HC$_9$N \citep{Broten:1978iu,Loomis:2016jsa}.  Table \ref{tab:hcxn} gives the rotational constants and number of energy levels below $kT$ for these molecules.  It is clear that the number of populated energy levels increases with decreasing $B$, increasing the spectral complexity by a density of lines argument alone.  Additionally, however, consider that for an equivalent population of HCN and \ce{HC7N}, the \ce{HC7N} molecules are distributed over almost an order of magnitude more energy levels.  As a result, the number of molecules undergoing a single rotational transition (and thus producing detectable signal) is overall lower for \ce{HC7N}, resulting in weaker signals.

\begin{table}[htb!]
\centering
\caption{Rotational constants and number of energy levels below $kT$ at 150 K for all HC$_\textup{x}$N species detected in the ISM.  }
\label{tab:hcxn}
\begin{tabular*}{\columnwidth}{l r c l}
\hline
Molecule		&	$B$	&	\# Levels$^{\dagger}$ &	Lab Ref.	\\
                &   (MHz)       &   $<kT$ @ 150 K       &               \\
\hline
HCN			&	44316		&	18					&	1		\\
\ce{HC2N}	&	10986		&	36					&	2		\\
\ce{HC3N}	&	4549			&	58					&	3	\\
\ce{HC4N}	&	2302			&	80					&	4		\\
\ce{HC5N}	&	1331			&	107					&	5	\\
\ce{HC7N}	&	564			&	166					&	6		\\
\ce{HC9N}	&	290			&	231					&	7	\\
\hline
\multicolumn{4}{l}{$^{\dagger}$Hyperfine splitting not considered, for simplicity.}
\end{tabular*}
\justify
[1] \citet{Ahrens:2002iy} [2] \citet{Saito:1984ju} [3] \citet{deZafra:1971cg} [4] \citet{Tang:1999yt} [5] \citet{Alexander:1976ed} [6] \citet{Kirby:1980ld} [7] \citet{McCarthy:2000eo}
\end{table}

\begin{figure*}
\includegraphics[width=\textwidth]{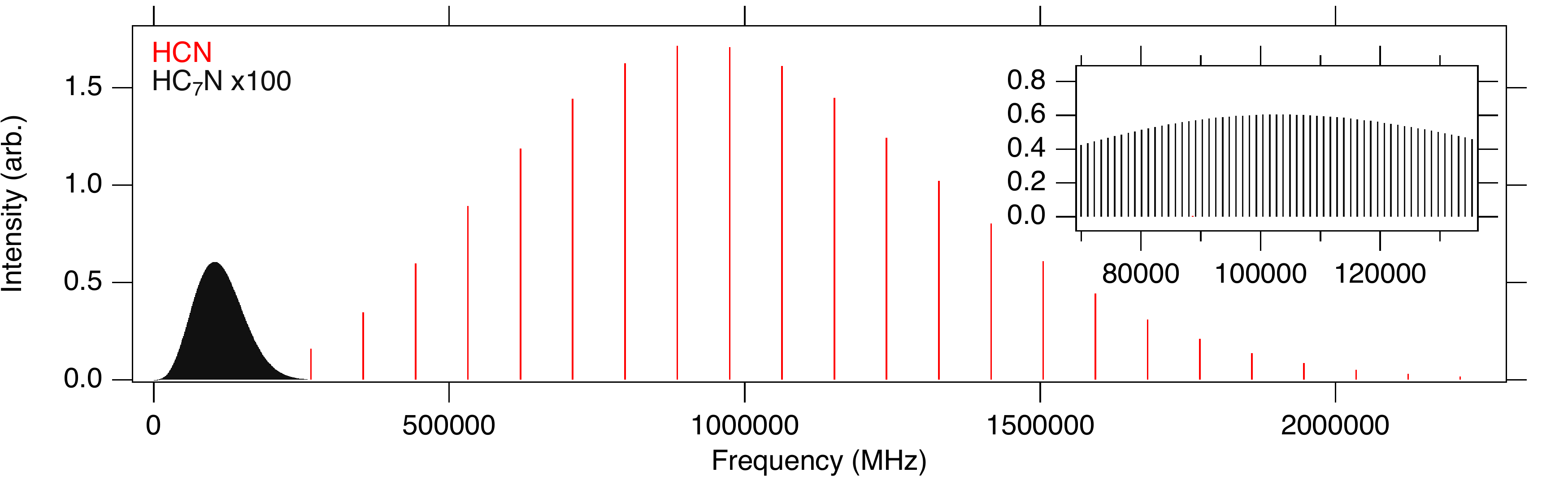}
\caption{Rotational spectra of HCN (red) and \ce{HC7N} (black) at $T = 150$ K for the same number of molecules of each.  Hyperfine has not been considered.  The inset shows a magnified portion of the \ce{HC7N} to show detail as to the density of lines.  In both the main plot and the inset, the intensities of the \ce{HC7N} lines are multiplied by a factor of 100.}
\label{hcn_hc7n}
\end{figure*}

This can be seen most easily by examining the rotational spectra of these molecules at $T = 150$ K.  Figure \ref{hcn_hc7n} shows a comparison of the spectra of HCN and \ce{HC7N} at $T = 150$ K, assuming the same number of molecules for each.  Not only is the \ce{HC7N} spectrum far more spectrally dense, but the intensities are so low that they must be scaled by a factor of 100 to be readily visible on the graph.  The strongest transition of HCN at this temperature is the $J = 10 - 9$, while the strongest transition of \ce{HC7N} is the $J = 91 - 90$.  The strongest HCN transition is nearly 300 times stronger than that of \ce{HC7N}.  A further consequence is that the frequency of a given transition shifts to lower and lower frequency as the number of atoms increases, increasing the moment of inertia, and decreasing the rotational constant.   This is clearly seen in Figure~\ref{hcn_hc7n}, where the strongest HCN transitions are arising in the $\sim$900 GHz region, while for \ce{HC7N}, these are at $\sim$100~GHz.

In general, for families of molecules with similar elemental compositions, adding additional atoms will increase the spectral complexity (line density), decrease the overall intensity of these lines, and shift the lines to lower frequency.

\subsection{Geometry and Symmetry}

The previous discussion focused exclusively on linear molecules as the prime example, but few molecules commonly considered complex by interstellar standards are linear (although certainly this is a debatable position).  When additional atoms are added off of the primary axis, additional complexity can be introduced, dependent on whether additional components to the dipole moment are manifest.  One of the underlying causes of the relative simplicity of linear molecular rotational spectra is the inherent symmetry of a linear molecule.  An allowed rotational transition can only occur when a permanent electric dipole moment is present.  Due to symmetry, linear molecules can only ever possess at most a single component of the electric dipole moment, oriented along the linear axis.  The addition of off-axis atoms introduces, in most cases, a degree of asymmetry.

\begin{figure}
\includegraphics[width=\columnwidth]{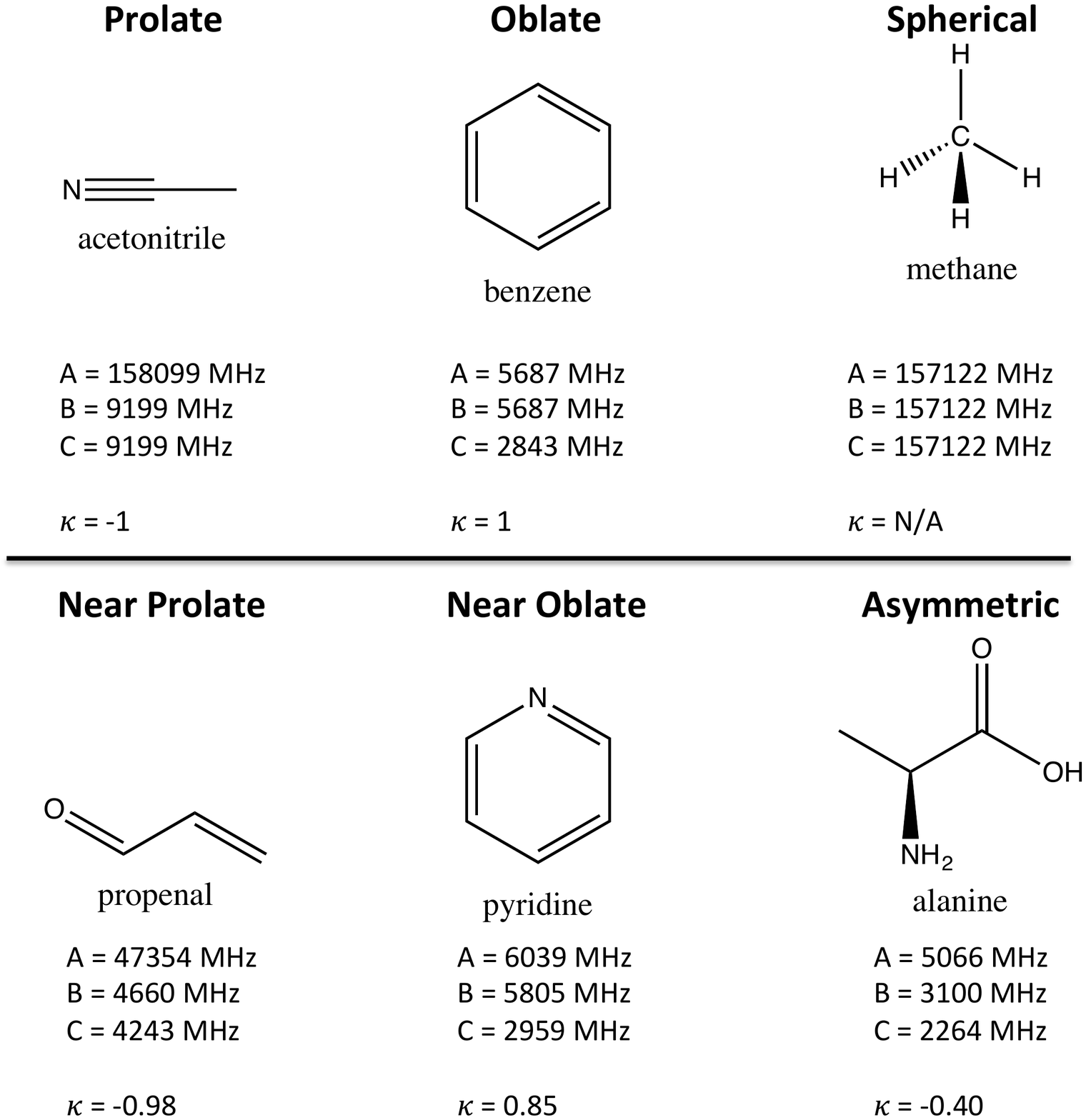}
\caption{Examples of different molecular symmetries demonstrate equivalencies in rotational constants and the corresponding Ray's asymmetry parameters.}
\label{rays_molecules}
\end{figure}

A convenient measure for assessing the degree of asymmetry introduced is $\kappa$, the Ray's asymmetry parameter \citep{Ray:1932yd}, given in Equation \ref{rays_eqn}, where $A$, $B$, and $C$ are the rotational constants of the molecule.
\begin{equation}
\kappa = \frac{2B-A-C}{A-C}
\label{rays_eqn}
\end{equation}
A number of example molecules displaying symmetry, or near symmetry, are shown in Figure \ref{rays_molecules}.  A completely spherically symmetric molecule, such as methane (\ce{CH4}) will have $A = B = C$, and thus $\kappa$ is undefined.  

Among the known interstellar molecules, the most commonly occurring symmetric top species are prolate (commonly thought of as cigar-shaped) with $\kappa = -1$, and $I_a < I_b = I_c$.   Linear molecules are the most common prolate symmetric tops, however any near-linear molecule in which the off-axis mass is symmetrically distributed such as methyl cyanide (acetonitrile, \ce{CH3CN}), resulting in $I_b = I_c$ are also prolate.  As a consequence of this symmetry, no additional permanent dipole components are introduced, and prolate symmetric tops behave largely like linear molecules as complexity increases.

\begin{figure*}
\includegraphics[width=\textwidth]{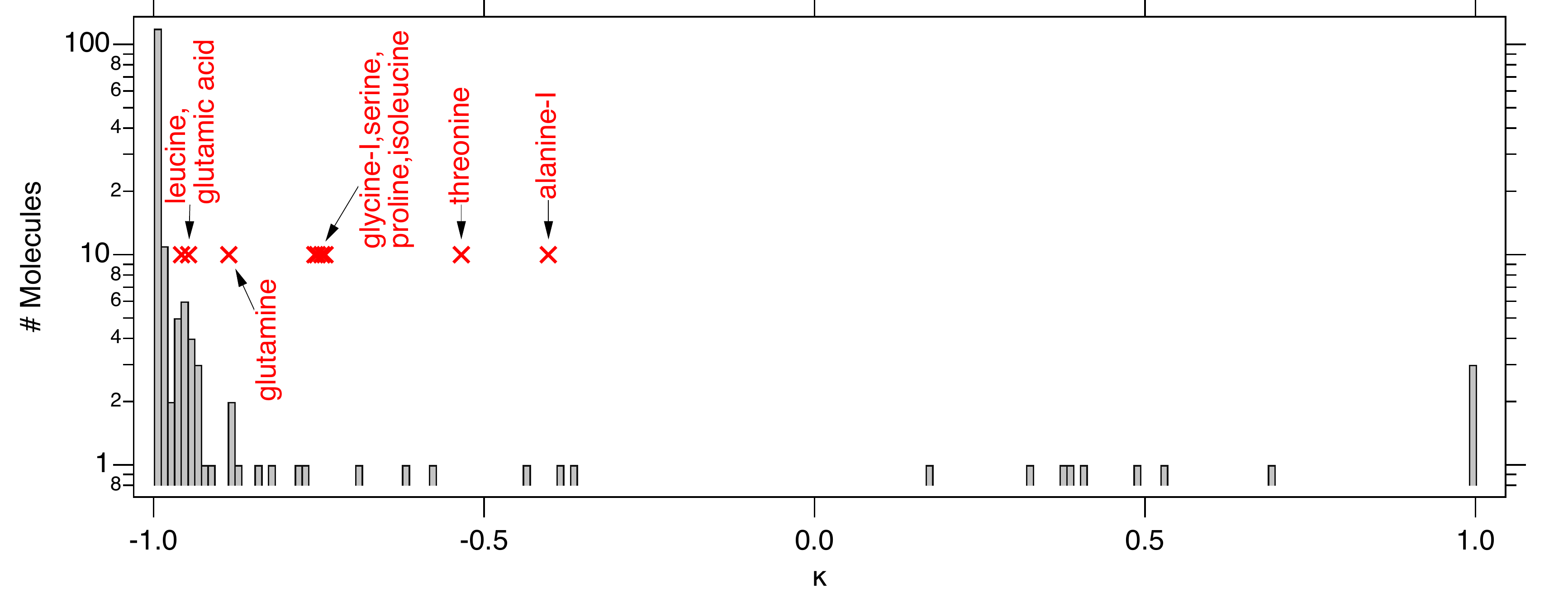}
\caption{Histogram of all detected interstellar molecules by their Ray's Asymmetry Parameter, $\kappa$.  Prolate rotors have $\kappa$ = -1, while oblate rotors have $\kappa$ = 1.  A selection of amino acids are overlaid in red.}
\label{kappas_histo}
\end{figure*}

\begin{figure}
\includegraphics[width=\columnwidth]{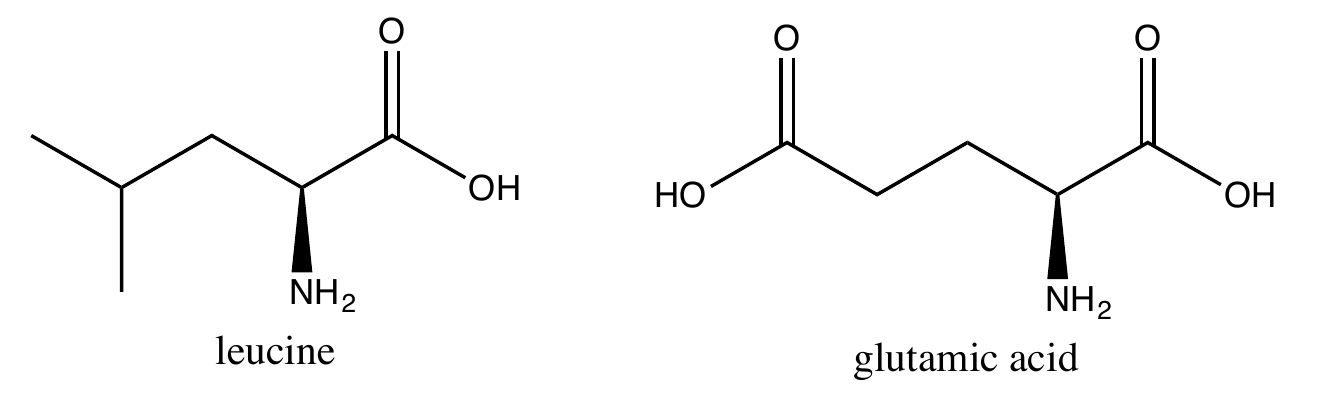}
\caption{Structures of the amino acids leucine and glutamic acid.}
\label{leu_glu}
\end{figure}

Figure \ref{kappas_histo} displays the distribution of known interstellar molecules by their $\kappa$ values.  It is clear that the vast majority ($\sim$90\%) of detected interstellar molecules are prolate or near-prolate. Only 24 molecules have $\kappa > -0.9$.  To first order, therefore, it is reasonable to claim that the more prolate a potential interstellar molecule is, the more likely it is to be detectable.  For the sake of comparison, also plotted on Figure~\ref{kappas_histo} are a number of amino acids, molecules not detected in space, but highly sought and widely considered truly complex.  In the case of glycine and alanine, the simplest amino acids by number of atoms, are both quite asymmetric ($\kappa_{gly} = -0.74$, $\kappa_{ala} = -0.4$), while the most nearly-prolate amino acids, leucine and glutamic acid, contain many more atoms (Figure \ref{leu_glu}).  Thus there is an offsetting balance to be considered in their detectability: the benefits of being (near) prolate vs the drawbacks of increasing number of atoms and structural size.

\subsection{Structural Conformers}

Without altering the overall bonding patterns (and thus changing the molecular make-up), there are often several ways for the atoms within a molecule to be arranged spatially.  A single population of molecules can often be comprised of one or more different arrangements, or \emph{conformers}, of the species.  Because these conformers have distinct moments of inertia, they therefore have distinct rotational spectra.  As a result, the more conformers which a population of molecules can exist in at a given temperature, the fewer molecules are available to undergo any given rotational transition.\\

\begin{figure}
\centering
\includegraphics[width=\columnwidth]{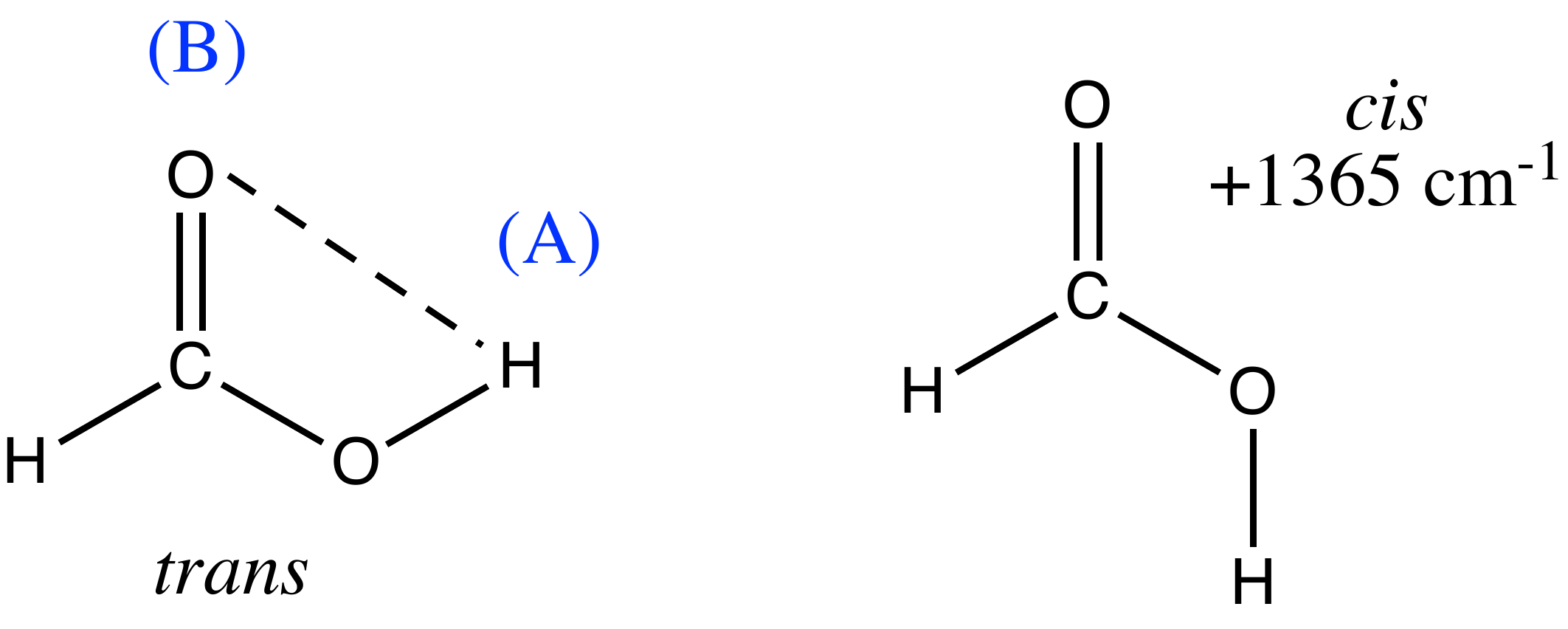}
\caption{The \emph{cis} and \emph{trans} conformers of formic acid.  The hydrogen-bonding interaction of the hydroxyl hydrogen (\textcolor{blue}{A}) with the carbonyl oxygen (\textcolor{blue}{B}) in the \emph{trans} form substantially stabilizes (by 1365 cm$^{-1}$; \citealt{Hocking:1976kv}) the conformer relative to the \emph{cis} form.}
\label{formic_acid}
\end{figure}

The simplest case of conformers is perhaps that between \emph{cis}- and \emph{trans}- arrangements for a small molecule.  Often, one conformer is substantially more stable with respect to the other, and there is little impact on the overall detectability of the species.  A salient example is that of formic acid (HCOOH; Figure \ref{formic_acid}).  Here, the \emph{trans} conformer is greatly stabilized by a hydrogen-bonding interaction.  As a result, while the lower-energy \emph{trans} conformer was first detected in 1971 \citep{Zuckerman:1971de}, and is indeed quite common in regions with any degree of chemical complexity \citep{RequenaTorres:2006ki}, the higher-energy \emph{cis} conformer was only recently discovered and is only seen because of exceptional chemical and physical circumstances \citep{Cuadrado:2016hp}.

\begin{figure}
\includegraphics[width=\columnwidth]{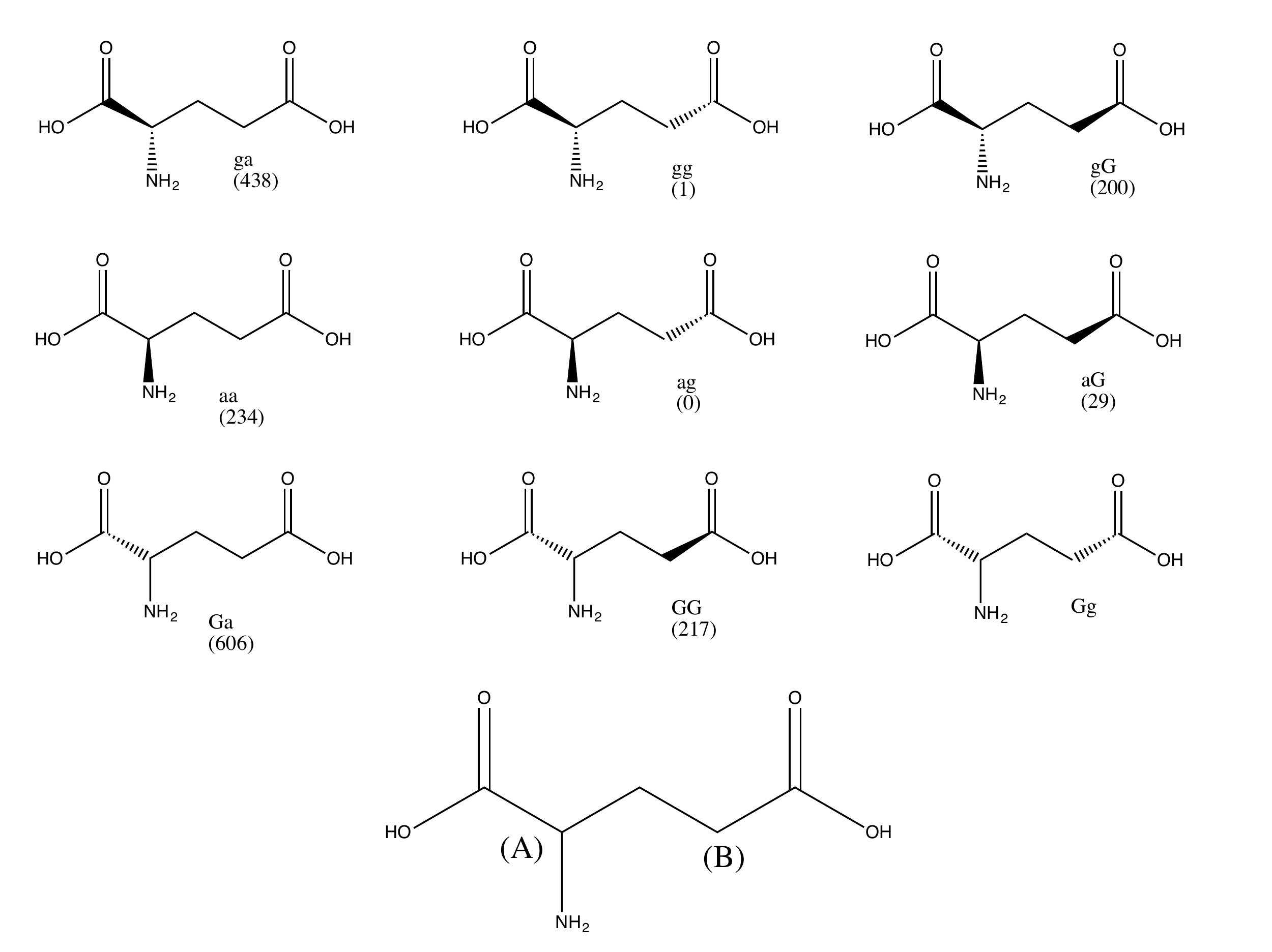}
\caption{Nine lowest-energy conformers of glutamic acid.  Bold angled lines indicate a bond which projects out of the plane of the page, while a dashed line indicates a bond into the plane.  The conformers differ in the relative spatial location of the C(O)OH and NH$_2$ groups around the carbon atoms indicated by the (A) and (B) in the bottom (2D) structure.  Where available, the zero-point corrected energies (in cm$^{-1}$) of the conformers are given, relative to the lowest energy (ag).  Adapted from Figure 1 and Table S1 of \citet{Pena:2012fl}.}
\label{glu_conformers}
\end{figure}

For more complex molecules, such as amino acids, the number of conformers with similar energies is often greatly enhanced.  The nine lowest-energy conformers of glutamic acid, for example, are shown in Figure \ref{glu_conformers} along with their relative energies; all are below 900 cm$^{-1}$.  In this case, it is the spatial arrangement of the C(O)OH and NH$_2$ groups that differ between conformers.  Indeed, dozens of additional higher-energy conformers, in which the relative arrangement changes around other central points, including the hydroxyl-carbonyl hydrogen bonds, also exist.  

From a practical standpoint, the additional conformer of formic acid has no impact on the detectability of the lower-energy \emph{trans} species.  There simply is not enough \emph{cis} to reduce the overall levels of \emph{trans} in a population in a significant way.  On the other hand, an attempt to detect glutamic acid would be severely hindered by the existence of a number of low-energy conformers.  Of these conformers, the ag, gg, and aG are the lowest in energy. When \citet{Pena:2012fl} measured the microwave spectrum of a sample of glutamic acid in the gas phase, they observed all three conformers simultaneously.   Instead of (practically) the entire population existing as the lowest-energy conformer, as in the case of formic acid, here  the overall intensity of any one conformer's spectral signature was diminished, making it harder to detect than if only a single conformer were energetically-favored.

\subsection{Internal Motion}

Considering a single conformer of a molecule, the complexity of the rotational spectrum can be further increased due to the internal structure of that molecule.  In the case of formic acid, the molecule can interconvert between the two conformers by rotation about the C-OH bond.  The barrier to this process is large (4827 cm$^{-1}$; \citealt{Hocking:1976kv}), meaning that it is rare for a full rotation about this bond to occur and this motion is unlikely to have a significant impact on the spectrum.  For functional groups that can undergo internal motion and have only a modest barrier to overcome, however, additional degrees of complexity arise.  

\subsubsection{Rotation}

Perhaps the most common form of internal motion that increases spectral complexity is that of internal rotation, specifically that of methyl (-CH$_3$) functional groups.  A detailed review of the effects of internal rotation on spectra is given by \citet{Lin:1959eb}.  In short, the rotational angular moment of the rotating (or pseudo-vibrating; i.e. torsion) sub-group within the molecule couples with the rotational angular moment of the molecule as a whole, perturbing the energy levels and resulting in new possible transitions.

\begin{figure}
\centering
\includegraphics[width=\columnwidth]{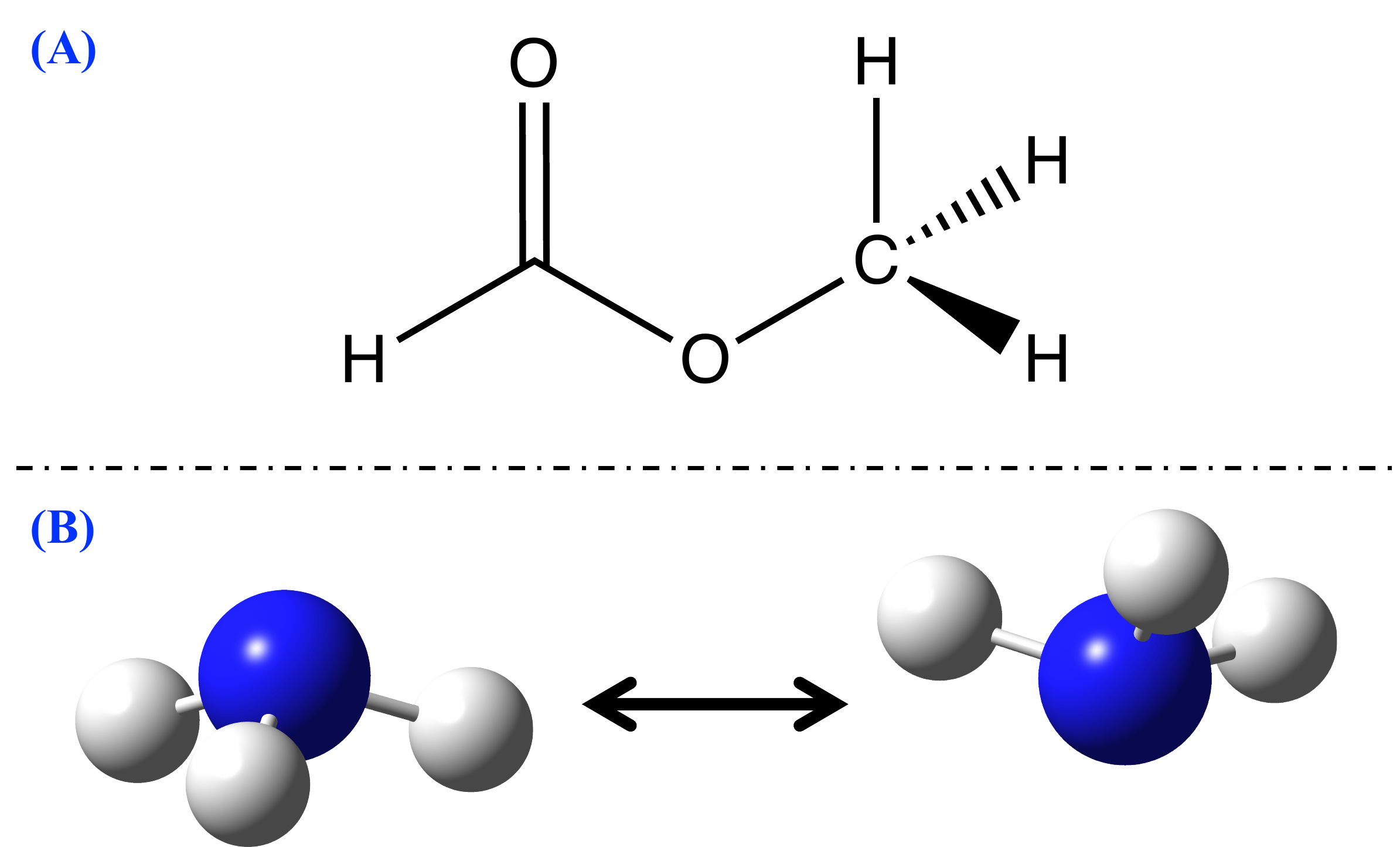}
\caption{Examples of internal motions.  \textbf{(A)} Structure of \emph{cis}-methyl formate.  As the methyl (-CH$_3$) group rotates about the C-O bond, the interaction potential between the hydrogen atoms and the oxygen atoms changes, affecting the coupling of the internal rotational angular momentum to that of the overall angular moment of the system, perturbing energy levels and increasing spectral complexity. \textbf{(B)} Representation of the umbrella-like inversion motion undergone by the \ce{NH3} molecule, which perturbs the standard rotational energy levels and adds complexity to the rotational spectra.}
\label{mf_ammonia}
\end{figure}

As a brief example, the lower-energy \emph{cis} conformer of methyl formate (\ce{CH3OCHO}) is shown in Figure \ref{mf_ammonia}a.  The methyl group is not locked into one orientation, and instead rotates around the C-O bond.  As the hydrogen atoms move, their interaction with the carbonyl (C=O) and ester (C-O-C) oxygen atoms changes, hindering the rotation.  As a result, every rotational energy level is split into three states: a pair of degenerate levels (denoted $E$) and a single non-degenerate state (donated $A$).\footnote{The energy ordering and degree of splitting changes as a function of energy, barrier heights, and other factors beyond the scope of this discussion. See \citet{Lin:1959eb} for a detailed review.}  As a result, the spectrum becomes far more complex, and, because there are now many more states accessible at the same temperature, the overall intensity of any given transition is substantially decreased.  To make matters worse (or better), the degree of this splitting changes with frequency/energy. Some transitions will suffer from this decrease in intensity, but for others, the splitting will be far less than the linewidth and will be unobservable.

\subsubsection{Inversion}

A second common type of motion is inversion.  The underlying principle is the same as for internal rotation: the angular moment of internal functional groups moving within the system couples to the overall angular moment of the system and perturbs the energy levels.  The most common example is seen in ammonia (\ce{NH3}), shown Figure \ref{mf_ammonia}b.  In the case of  \ce{NH3}, the entire molecule inverts in an umbrella motion.  As with the internal rotation, this motion perturbs the standard rigid-rotor energy levels, increasing the number of states accessible, generating more transitions, and reducing individual spectral intensity.  \citet{Townes:1946jt} and \citet{Townes:1975ve} provide excellent summaries of the laboratory and theoretical efforts to characterize this type of internal motion.

\subsection{Nuclear Hyperfine}

\begin{figure}
\includegraphics[width=\columnwidth]{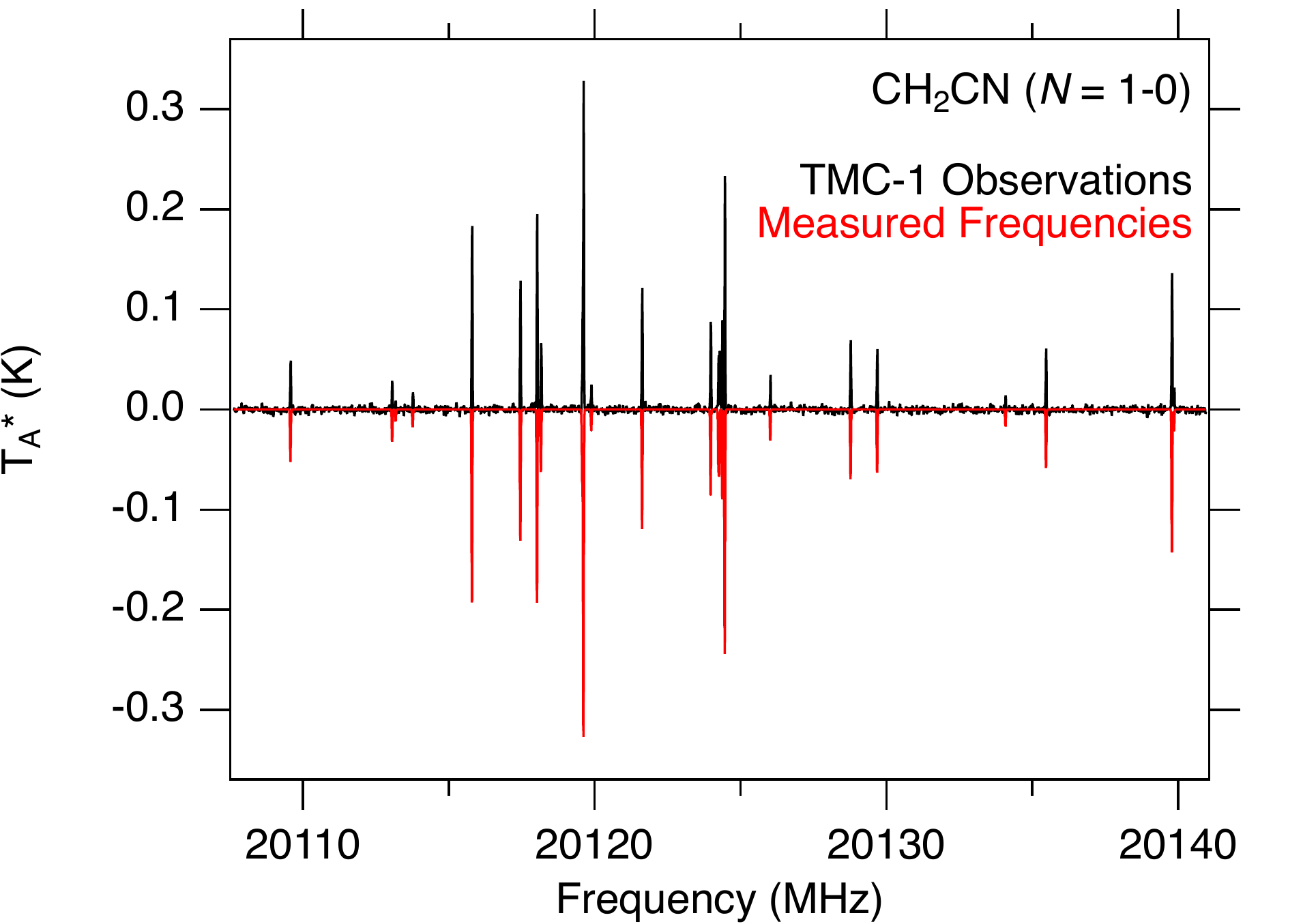}
\caption{Observations of a single rotational transition ($N$~=~1--0) of \ce{CH2CN} in TMC-1 from the \citet{Kaifu:2004tk} survey in \textbf{black}, with the measured laboratory transitions overlaid in \textcolor{red}{red}.  The split lines are the result of the coupling of both $^1$H and $^{14}$N hyperfine splitting of the transition.}
\label{ch2cn}
\end{figure}

A final common type of perturbation is the coupling of nuclear angular momentum (from atoms with non-zero nuclear angular moment) to the overall energy.  This perturbation is often quite small; for example, the splitting from \ce{^{1}H} is rarely resolvable with even high-resolution instruments in the laboratory, much less in interstellar observations.  $^{14}$N hyperfine splitting, on the other hand, is routinely observed in the laboratory, and is often seen in interstellar observations in sources with sufficiently narrow linewidths.  Figure~\ref{ch2cn} shows an example of resolved hyperfine transitions from a single rotational transition \ce{CH2CN} in TMC-1 observations showing both $^1$H and $^{14}$N splitting.  The single rotational transition ($N$~=~1--0) has been split into more than a dozen resolvable lines.  The relative intensities of hyperfine-split transitions are well-known \citep{Townes:1975ve}, and, combined with the structure, can provide unique fingerprints for identification and as probes of temperature and optical depth.

\begin{center}
\textsc{Column Densities and Spectral Intensities}
\end{center}

Given these sources of complexity in the spectrum, it is then useful to consider how the intensity of a spectral line actually changes as a result in a more quantitative way, and the influence of some telescope-specific parameters on the line as well. Given a detection of a molecular line from observations, a column density can be determined, or, alternatively, given an expected column density, a predicted observational intensity can be derived.  A detailed examination of the radiative transfer processes behind these calculations is beyond the scope of this work; the interested reader is referred to the recent work of \citet{Condon:2016tr} and \citet{Mangum:2015wp}.  Here, only the widely-used single-excitation model is described in detail.  A discussion of the effects on detectability when this model breaks down follows.

\subsection{The Single-Excitation Model}
\label{sec:lte}

The most common approach used to analyze observational data is described in detail in \citet{Goldsmith:1999vg}, and is often referred to as the `rotation diagram' method or a `local thermodynamic equilibrium (LTE)' analysis.  Both terms omit the key assumption made in the analysis, which is that the number of molecules in each molecular energy level is assumed to be described exactly by a Boltzmann distribution at a single, uniform excitation temperature, $T_{\rm{ex}}$. 

\citet{Hollis:2004uh} formalized the calculations used to determine column densities and excitation temperatures, and that formalism is adopted here for the purposes of this discussion.  Equation~\ref{hollis_emission} describes a calculation for molecules in emission or absorption, and the parameters used in this equation are given in Table \ref{hollis_params}.  Detailed discussions of several of these parameters are given in \citet{Mangum:2015wp}.

\begin{equation}
N_T = \frac{1}{2}\frac{3k}{8\pi^3}\sqrt{\frac{\pi}{\ln 2}}\frac{Qe^{E_u/T_{\rm{ex}}}\Delta T_A\Delta V}{B\nu S\mu^2\eta_B}\frac{1}{1-\frac{e^{h\nu/kT_{\rm{ex}}} - 1}{e^{h\nu/kT_{\rm{bg}}}-1}}
\label{hollis_emission}
\end{equation}

\begin{table}
\centering
\caption{Definition of parameters in the single-excitation model calculation of column density.}
\label{hollis_params}
\begin{tabular*}{\columnwidth}{l p{1.65in}@{\extracolsep{\fill}}  l}
\hline
Parameter		&	Definition	                                    & Units     \\
\hline
$N_T$		&	Column density				                &   cm$^{-2}$		\\
$T_{\rm{ex}}$		&	Excitation temperature	            &   K			\\
$T_{\rm{bg}}$		&	Background temperature				&   K\\
$k$			&	Boltzmann's constant				        &   J K$^{-1}$\\
$h$			&	Planck's constant				            &   J s	\\
$Q$			&	Partition Function at $T_{\rm{ex}}$		    &   	\\
$E_u$		&	Upper state energy of the transition		&   K   \\
$\Delta T_A$	&	Peak observed intensity of the transition   &   K 	\\
$\Delta V$		&	FWHM of the transition			        &   cm s$^{-1}$	\\
$B$			&	Beam filling factor					        &   \\
$S$			&	Line strength						        &   \\
$\mu^2$		&	Square of transition dipole moment				    &   J cm$^{3}$\\
$\eta_B$		&	Beam efficiency				            &	\\
\hline
\end{tabular*}
\end{table}

A key insight is provided by Equation~\ref{hollis_emission} upon inspection of the $\left(1-\frac{e^{h\nu/kT_{\rm{ex}}} - 1}{e^{h\nu/kT_{\rm{bg}}}-1}\right)$ term.  In the case where $T_{\rm{ex}} < T_{\rm{bg}}$, this term becomes negative.  As a negative value of $N_T$ is unphysical, the value for $\Delta T_A$ must therefore be negative as well, indicating absorption.  

\subsection{Critical Density}
\label{sec:radex}

When the number of molecules in each energy level is not well-described by a Boltzmann distribution, one of the most common causes is that the density of the gas in which the molecule resides has fallen below the density required to thermalize the population.  The distribution of a population of molecules across energy states is a balance between radiative (emission or absorption) processes, and collisional (thermal) processes.  Each transition of a given molecule has a characteristic rate ($A_{ul}$) that governs how quickly it undergoes spontaneous emission of photons, redistributing the population away from Boltzmann equilibrium.  For a population of molecules to be in thermal equilibrium, collisions with other gas molecules must occur frequently enough to out-compete the radiative processes.  In this case, $T_{\rm{ex}}$ becomes equal to the kinetic temperature ($T_{k}$) of the colliding gas, which is usually termed LTE conditions.

The density of gas required to ensure that these collisions dominate the distribution over radiative processes is the critical density ($n_{cr}$), given by Equation~\ref{critical} \citep{Tielens:2005ux} where $A_{ul}$ is the Einstein A coefficient (s$^{-1}$) and $\gamma_{ul}$ is the collisional rate coefficient (cm$^3$ s$^{-1}$) for the transition from upper state $u$ to lower state $l$.
\begin{equation}
n_{cr} = \frac{A_{ul}}{\gamma _{ul}}
\label{critical}
\end{equation}

$A_{ul}$ is directly related to the rate at which a given molecule in state $u$ will spontaneously decay to state $l$ and emit a photon.  Transitions with large $A_{ul}$ undergo emission rapidly.  To maintain a population distribution described by the thermal temperature, a correspondingly higher density -- that results in more frequent collisions -- is therefore required. A generalized approach for multilevel systems with many transitions into and out of a given level can be written as Equation~\ref{critical_gen} \citep{Tielens:2005ux}.
\begin{equation}
n_{cr} = \frac{\Sigma_{l<u}A_{ul}}{\Sigma_{l\neq u}\gamma _{ul}}
\label{critical_gen}    
\end{equation}

The rate at which these collisions occur is related to $\gamma_{ul}$ which is proportional to the cube of the average velocity of the molecules in the gas ($v$; governed by $T_{\rm{k}}$), a cross-sectional area ($\sigma$) related to the size of each collision partner, and their rotational and vibrational excitation.   These cross-sections are almost universally calculated theoretically, but are extraordinarily computationally-expensive, especially with increasing molecular size \citep{Faure:2014iu}.

In general, if the gas density exceeds $n_{cr}$, the relative populations of $u$ and $l$ will be described by a Boltzmann distribution at $T_{\rm{ex}}$ (which is nominally equal to $T_{\rm{k}}$).  Because $n_{cr}$ is distinct for each transition, it is possible for only some of the energy levels of a molecule to be described by one $T_{\rm{ex}}$, while the remaining energy levels are dominated by radiative processes and are not described by $T_{\rm{ex}}$.  

If all values of $A_{ul}$ and $\gamma_{ul}$ are known for a given molecule, then $T_{\rm{k}}$ and a column density can be modeled explicitly without the assumption of a single excitation temperature in what is known as a radiative transfer calculation.  The approximations described in \S\ref{sec:lte} are accurate in the limit that the density is much larger than $n_{cr}$, and a full radiative transfer calculation in such situations will return an equivalent value to that determined by Equation~\ref{hollis_emission} and a single value for $T_{\rm{ex}}$ equal to $T_{\rm{k}}$. A detailed review of radiative transfer calculations is given by \citet{vanderTak:2011td}.

\subsection{Source and Beam Sizes}
\label{sec:dilution}

Often, the region of the sky from which molecular emission (or absorption) is seen is smaller than the telescope beam used for the observations.  The result is a beam-diluted signal that reduces the intensity of the observed emission.  Calculations that do not account for beam dilution would therefore under-predict the column density of a compact source.  The beam filling correction factor, $B$, is given by Equation~\ref{beamdilution}, where $\theta_s$ and $\theta_b$ are the circular Gaussian sizes of the source and the half-power telescope beam, respectively.

\begin{equation}
B = \frac{\theta_s^2}{\theta_s^2 + \theta_b^2}
\label{beamdilution}
\end{equation}

By inspection, $B$ approaches unity as the source size exceeds (fills) the beam size.  Often, for single-dish observations where no reasonable \emph{a priori} assumption can be made regarding the underlying source structure, the emission is assumed to ``fill the beam," and no correction for beam dilution is performed.

A related issue, not explicitly accounted for in these calculations, is the loss of sensitivity in interferometric observations to extended emission as a function of increasing spatial resolution.  The magnitude of this effect is often determined by comparing the total flux observed with a single-dish telescope beam that is assumed or known to contain all of the emission to that recovered by the interferometer.  If the interferometric observations display significantly lower flux, meaning that it has been resolved out, a correction can be made to any column density calculations if the total column (and not just that of the compact sources resolved by the array) is desired.

\subsection{Background Continuum}

The background continuum, $T_{\rm{bg}}$, against which molecules absorb plays a crucial role in the overall intensity of the detected molecular signal.  While the fractional absorption seen for a molecular transition is constant with column density, the absolute observed signal is of course dependent on the magnitude of the background being absorbed against.  Thus, even a large population will present a small signal against a weak background, while, conversely, a very small population can be detected if $T_{\rm{bg}}$ is large.  The background temperature also affects the intensity of emission lines, with the effect of reducing the overall $\Delta T_A$ that is observed.

\subsection{Frequency-dependent Lineshapes}
\label{subsec:velocity}

In interstellar observations, the width of a spectral line is defined, in the non-relativistic limit, in Equation~\ref{velocity}, where $\Delta V$ is the velocity linewidth (km~s$^{-1}$), $\Delta\nu$ is the width of the line in frequency space (MHz), $v_0$ is the reference frequency (typically the central line frequency; MHz), and $c$ is the speed of light (km~s$^{-1}$).  
\begin{equation}
\Delta V = \frac{\Delta\nu}{\nu_0}c
\label{velocity}
\end{equation}
The velocity width of a line is frequency-independent; it is a physical effect of the source.\footnote{The uncertainty principle also dictates a quantum-mechanical width to every line, but in radioastronomical observations this is an immeasurably small contribution in almost all circumstances.}  Because $\Delta V$ is therefore constant, $\Delta\nu$ must increase with increasing $\nu_0$.  Thus, at higher frequencies, the lines are far broader in frequency space.

\begin{figure}
\includegraphics[width=\columnwidth]{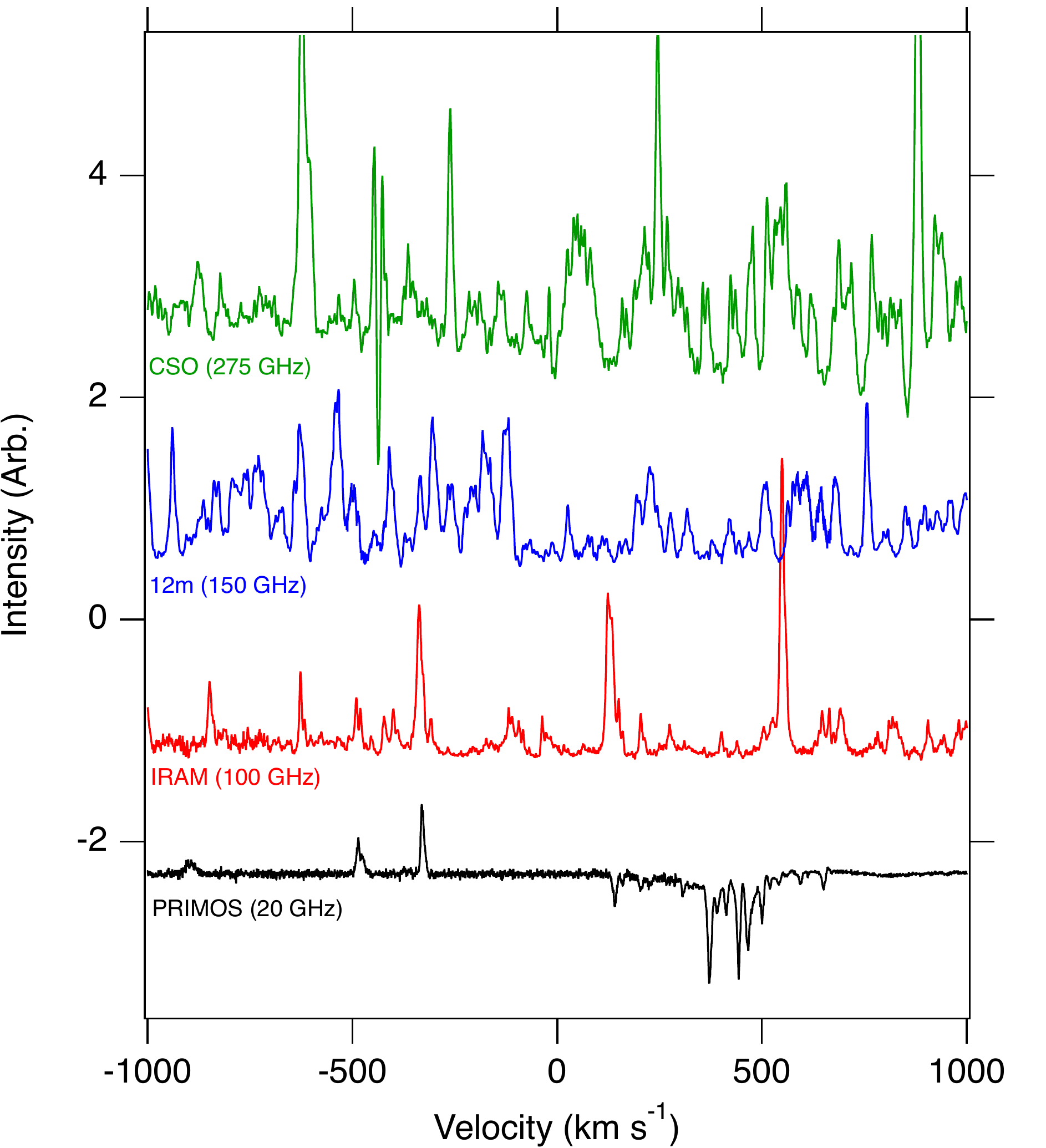}
\caption{Comparison of observational data toward Sgr B2(N) at four different frequencies.  The observations are plotted in velocity space, and the linewidths are all quite similar ($\sim$8--10~km~s$^{-1}$).}
\label{velocity_comp}
\end{figure}

This has a somewhat subtle but profound effect on the ability to detect weak signals in sources with large degrees of molecular complexity.  As a concrete example, we can consider the density of lines in observations of Sgr B2(N) taken at different frequencies from the centimeter through the sub-millimeter.  These are tabulated in Table~\ref{density}, and shown visually in Figure~\ref{velocity_comp}.\footnote{References --  PRIMOS: \citet{Neill:2012fr}; IRAM: \citet{Belloche:2013eba}, NRAO 12 m: \citet{Remijan:2008rt}, CSO: \citet{McGuire:2013bb}. }

\begin{table}
\centering
\caption{Approximate line density in molecular surveys of Sgr B2(N) at different frequencies.}
\begin{tabular}{l | c c c}
\hline\hline
			&				&		\multicolumn{2}{c}{Line Density}					\\
\cline{3-4}
Source		&	Frequency	&		Frequency Space		&		Velocity Space		\\
\hline
PRIMOS		&	20 GHz		&		1 per 10 MHz			&		1 per 150 km/s		\\
IRAM		&	100 GHz		&		1 per 10 MHz			&		1 per 30 km/s		\\
NRAO 12 m	&	150 GHz		&		1 per 10 MHz			&		1 per 20 km/s		\\
CSO			&	275 GHz		&		1 per 10 MHz			&		1 per 10 km/s		\\
\hline
\end{tabular}
\label{density}
\end{table}

Upon visual inspection, it is clear that in velocity space, the spectra are increasingly crowded at higher frequencies, even though the line density in frequency space remains the same.  Indeed, at the highest frequencies, the spectra are nearly line-confusion limited, a condition in which there is spectral line emission in every channel of the data.  This issue sets in when the line density in velocity space becomes equal to the line width.  Once a spectrum is line-confusion limited, there is, to large extent, no additional information to be gained by deeper integration.  All weaker lines from undetected molecules will be buried beneath the lines already visible, thus making it impractical to identify new, rare molecules.

It is important to note that the observations used here are just illustrative.  They were taken with different facilities, having different beam sizes, and to different sensitivities.  Still, the general trend does hold that lower frequency observations, while no less rich in molecular signals, have far more velocity space available to be filled with spectral lines.  Indeed, of the observations shown here, those from PRIMOS at 20 GHz are the deepest and most sensitive ($\sim$2~mK), but still have the lowest line density.  Thus, even once line-confusion is reached in a source at millimeter and sub-millimeter frequencies, substantial discovery space often remains at centimeter wavelengths.

\subsection{Partition Functions}
\label{sec:partition}

The temperature-dependent partition function, $Q(T)$, for a molecule is a representation of the number of energy levels a population is spread out over.  For astronomical purposes, $Q(T)$ is often comprised almost entirely of a rotational component, with small additional contributions from vibrational components (Equation~\ref{total}).
\begin{equation}
Q = Q_{\rm{v}} \times Q_{\rm{r}}
\label{total}
\end{equation}

The rotational partition function, $Q_{\rm{r}}$, is often calculated according to a high-temperature approximation, given by  Equation \ref{Q} where $\sigma$ is a symmetry parameter, $T_{\rm{ex}}$ the excitation temperature (K), and $A$, $B$, and $C$, the rotational constants of the molecule (MHz), also as discussed earlier (c.f. \citealt{Gordy:1984uy}). 
\begin{equation}
Q_{\rm{r}} = \left(\frac{5.34 \times 10^6}{\sigma}\right)\left(\frac{T_{\rm{ex}}^3}{ABC}\right)^{1/2}
\label{Q}
\end{equation}

This approximation offers excellent values down to modestly-low temperatures.  Indeed, for most molecules, deviations from the explicitly-calculated value do not reach 1\% until $\leq$5~K.  In cases where the temperature is low enough for this deviation to be significant, a direct summation of energy levels is required; this process is outlined in \citet{Gordy:1984uy}.   From inspection, it is also clear that larger, more complex molecules, with smaller rotational constants, will have a larger value of $Q_{\rm{r}}$.  This is the same trend discussed earlier in a qualitative fashion.  Finally, it is necessary to note that this approximation does not take into account many of the additional sources of complexity discussed earlier such as internal rotation and nuclear hyperfine splitting, which are not represented in the rotational constants $A$, $B$, and $C$.  This can be addressed either through the addition of degeneracy terms, through explicit state counting, or a combination of both, as described in \citet{Gordy:1984uy}.

The vibrational contribution, $Q_{\rm{v}}$, is given by Equation~\ref{qvib}.
\begin{equation}
Q_{\rm{v}}=\prod\limits_{\substack{i=1}}^{3N-6} \frac{1}{1-e^{-E_i/kT_{\rm{ex}}}}
\label{qvib}
\end{equation}
Here, the energies of each vibrational energy level $E_i$ are considered.  Only when $T_{\rm{ex}}$ is sufficiently high, and there are sufficiently low-lying vibrational states, does $Q_{v}$ make any significant contribution.  This correction factor accounts for the fact that the rotational energy level structure in an excited vibrational state is practically identical to that in the ground state.  Thus, if some non-trivial population of molecules exist in a vibrationally excited state, those levels become accessible as well, spreading out intensity.  Thus, the contribution to the total partition function $Q$ from $Q_{\rm{v}}$ is multiplicative.  Often the vibrational contribution is less than 1\% up to at least 30--40~K, although it can be significant ($\geq$2) at warmer temperatures for some species.  Strictly speaking, an electronic contribution $Q_e$ may also be considered, but is almost never a factor in interstellar detections, at least with radio telescopes.

\begin{center}
\textsc{Generalized Source Types}
\end{center}

\S\ref{statistics} presented four generalized source types -- star-forming regions (SFRs), dark clouds, carbon stars, and line of sight (LOS) clouds -- in which most detections have been made.  The analysis there showed that the types of molecules that are detected for the first time in each of these environments tend to have distinct properties.  Thus, the environment in which a molecular search is conducted can have a significant impact on the detectability from several standpoints.  These include total abundance (related to chemistry and density), excitation temperature (density and kinetic temperature), linewidth and spectral crowding (turbulent vs quiescent regions), and source size (beam dilution).  Below, five of the common types of environments for these searches are described.  While this is by no means an exhaustive list, it covers a large range in physical and chemical conditions, and provides a grounding in the factors which must be considered and can be applied to other situations.

\subsection{Diffuse (LOS) Clouds}

The first detections of interstellar molecules (CH, \ce{CH+}, and CN) were made in the LOS diffuse clouds that pervade the Galaxy \citep{Swings:1937dl,McKellar:1940io,Douglas:1941uc}.  While much of the molecular discovery quickly shifted to SFRs, dark clouds, and carbon stars, new detections in diffuse environments do still occur (see, e.g. the detection of SH by \citealt{Neufeld:2012gz}).

These environments are characterized by low total number densities ($n_{\rm{H}}$~$\sim$10~--~$<$10$^4$~cm$^{-3}$), cool, but not cold kinetic temperatures ($T_{\rm{k}}<100$~K), and enhanced radiation environments over those seen in `standard' dense molecular clouds (for a detailed review, see \citealt{Snow:2006dp}).   The excitation temperature of most molecules observed in these clouds is extremely sub-thermal at $T_{\rm{ex}}$~$\sim$3~K, with the population largely distributed over only the few lowest rotational energy levels. Most sources show moderately to extremely broad spectral absorption features (3--20~km~s$^{-1}$; \citealt{Corby:2017tm}).  The width of the features tends to correspond to the density and compactness of the absorbing gas, with narrow features arising from denser, more compact sources.

The overall angular size of a distinct cloud can vary significantly with distance, but is typically assumed to be larger than the continuum against which it is observed, and is thus at least modestly extended ($>$10$^{\prime\prime}$).  While the general assumption is that the gas is homogeneously distributed across the cloud, there is growing evidence that there may be significant substructure within each complex \citep{Corby:2017tm}.  This can lead to regions of increased density, and decreased temperature and radiation effects.

These physical conditions put constraints on the type and extent of complex chemistry that can occur in these regions.  The harsh radiation environment largely prevents a build up of the icy dust grain layers thought to be needed to make many complex species, although there is some recent evidence showing surprising chemical complexity \citep{Thiel:2017fh}, and the overall low number density results in difficulties achieving detectable abundances of molecules.  Many of the new species detected in these environments are small, simpler precursor molecules which have often reacted away to form more complex molecules in other, more evolved sources.

\subsection{Dark Clouds}

The prototypical dark cloud is TMC-1, a largely homogeneous, extended ($>$60$^{\prime\prime}$), cold ($T_{\rm{kin}}$~$\sim$10~--~20~K), and modestly dense ($n_{\rm{H}}$~$\sim$10$^4$~cm$^{-3}$) source \citep{Bell:1998mw,Hincelin:2011fr,Liszt:2012ft} with extremely narrow linewidths ($\sim$0.3~km~s$^{-1}$; \citealt{McGuire:2017ud}).  The higher density compared to diffuse clouds (with correspondingly high extinction) shields the source, and allows the formation of ices on the surfaces of dust grains.  While saturated complex organic molecules may evolve on these surfaces, the cold kinetic temperatures and quiescent environment tend to force these species to remain on the surface.  As a result, the (detectable) gas-phase inventory tends to be dominated by complex, unsaturated organic molecules formed by gas-phase reactions, typically long-chain carbon molecules like the cyanopolyynes (\ce{HC_{n}N}; \emph{n} = 3, 5, 7, 9).  As discussed in \S\ref{natoms}, these larger molecules will tend to have more transitions at lower frequencies.  At the temperatures in these sources, the Boltzmann peak will tend to fall at low frequencies as well, and has a favorable effect on the otherwise very large partition function (\S\ref{sec:partition}). Thus, these regions tend to see a particularly pronounced number of detections of large, unsaturated molecules at frequencies below $\sim$100~GHz.  Because of the narrow velocity widths, line blending is rarely an issue.

\subsection{Star-Forming Regions}

Upon gravitational collapse, dense molecular clouds begin to form molecular cores, compact, warm ($T_{\rm{kin}}$~$>$100~K), and dense ($n_{\rm{H}}$~$\sim$10$^5$~--~10$^8$~cm$^{-3}$) sources, often associated with a nascent star.  In these regions, as the icy surfaces of the dust grains are heated, or are subjected to shocks, the complex molecular inventories are liberated into the gas phase and become detectable with radio astronomy. As was shown in \S\ref{statistics}, these molecules tend to be more saturated than those seen in dark clouds.  Further, the molecules formed on these surfaces tend to be highly reactive, and can then interact with the previously (largely) isolated gas-phase inventory from the dense cloud stage, depleting those abundances.  The more complex physical environment, including transient events such as shocks and longer-term interactions from protostellar outflows, inject additional energy into the system, driving chemistry not otherwise possible under cooler, more quiescent conditions \citep{Burkhardt:2016bs}.  

The higher densities tend to thermalize the excitation temperatures, pushing the distributions toward LTE.  Single-dish observations of these cores typically fail to resolve a single sub-source, resulting in modestly-broad linewidths (5--15~km~s$^{-1}$; see, e.g. \citealt{WidicusWeaver:2017hf}).  Interferometric observations in these regions, however, can often isolate individuals cores, significantly narrowing the linewidths to a few km~s$^{-1}$ \citep{Belloche:2016fm}.  The large abundances and warm temperatures often result in spectra rapidly approaching (or having reached) line confusion in the mm- and sub-mm regimes with modern facilities, both single dish and interferometric.

\subsection{Evolved (Carbon) Stars}

While complex chemistry in the ISM is dominated by carbon, a number of exotic -- by interstellar standards -- species are also seen in evolved star sources, the definition of which is expanded here to include oxygen-rich stars.  Indeed, every detected molecule with six or more atoms contains a carbon atom; the largest molecule detected without a carbon atom is \ce{SiH4}, which was first detected (perhaps ironically) in the evolved carbon-rich star IRC+10216 \citep{Goldhaber:1984mh}.  The intense physical environments of sources like IRC+10216 and the hypergiant VY Canis Majoris inject heavier atoms like Si, as well as Mg, Fe, Ti, and Al into the gas-phase, where they can be detected as components of molecules such as \ce{SiH4}, MgCN, FeCN, \ce{TiO2}, and AlOH \citep{Ziurys:1995pf,Zack:2011jx,Kaminski:2013gk,Tenenbaum:2010hp}.  While the circumstellar envelope around the stars themselves is compact, molecular emission is often seen in an extended distribution around the star itself \citep{Cernicharo:2013cc}.  The physical conditions change as a function of distance from the star, often in a measurable manner, providing the opportunity to study, for example, dust evolutionary processes within the ISM using a single source \citep{Cernicharo:2011gu}.  Line confusion is rarely an issue in these sources.

\begin{table*}[t!]
    \centering
    \caption{Attributes of \ce{HC5S} and their effects on detectability.}    
    \begin{tabular}{p{2.8in} p{3in}}
        \hline\hline
        \ce{HC5S} ...                      &    Effects         \\
        \hline
        ... is prolate                     &    Most ISM molecules are prolate or near prolate - high symmetry reduces $Q$, generally increasing line intensity \\
        ... is highly unsaturated          &    Highly unsaturated molecules tend to be first detected in carbon stars and dark clouds\\
        ... has two added sources of spectral complexity ($\Lambda$-doubling and resolvable $^1$H splitting)  &      $Q$ will be increased and the line intensities decreased if the splitting is resolved in observations\\
        ... has a rotational constant of $\sim$876~MHz   &       The primary rotational transitions will occur at low frequency (cm-wavelengths)      \\
        ... has no low-lying structural conformers      &       Nearly all the population will be in the ground-state conformer - no decrease in its line intensities     \\
         \hline
    \end{tabular}
    \label{hc5s}
\end{table*}

\begin{center}
\textsc{Bringing it All Together}
\end{center}

With these tools in hand, and the trends and statistics discussed in \S\ref{statistics}, it is possible to make some informed guesses about the likely environments, facilities, and frequency ranges needed to detect a potential interstellar molecule.  The thought process for two such example cases, \ce{HC5S} and $n$-propanol (\ce{CH3CH2CH2OH}), is outlined below.  These discussions presume that the chemistry and elemental abundances in the ISM are favorable enough to produce at least some abundance of a species.

\subsection{Searching for \ce{HC5S}}

\ce{HC5O} (\S\ref{HC5O}) was recently detected in the ISM in observations of TMC-1 at cm-wavelengths with the GBT.  Having often similar bonding characteristics with oxygen and being isovalent, it's logical to assume the S-substituted versions of O-containing compounds may be good interstellar candidates (e.g., \ce{CH3OH}/\ce{CH3SH}, \ce{C3O}/\ce{C3S}, \ce{SiO}/\ce{SiS}).  The rotational spectrum of \ce{HC5S} is known \citep{Gordon:2002kd}.  Table~\ref{hc5s} below lists a few key considerations for the molecule, and the corresponding effects on its likely detectability.

Given the highly unsaturated and C-rich nature of \ce{HC5S}, it is reasonable to expect that a carbon star or dark cloud are the most likely places for a detection (\S\ref{source_types}).  The molecule is to zeroth-order a linear rotor, and so will have a relatively simple spectrum.  The $\Lambda$-doubling splitting is a few MHz and the $^1$H splitting $\sim$0.5~MHz for the lowest of its cm-wave transitions.  The $^1$H splitting quickly collapses, while the $\Lambda$-doubling splitting remains.  These splittings translate to as little as 1~km~s$^{-1}$ ($^1$H) to as much as 250~km~s$^{-1}$ ($\Lambda$-doubling) across the cm-wave region (where the small $B$ constant dictates most transitions will fall).  In a carbon star source such as IRC+10216, linewidths $>$10~km~s$^{-1}$ are routine, and thus only the $\Lambda$-doubling is likely to be resolved; the $^1$H contribution (and reduction in line intensity) could be ignored.  In a dark cloud, where linewidths can be as little as 0.1~km~s$^{-1}$, both splittings are in play.  Molecules detected in carbon stars tend to be modestly warm ($>$50~K), increasing $Q$ and decreasing line intensities by a factor of $\sim$6 over molecules at the temperatures typical of dark clouds (5--15~K).  

\textbf{Carbon Star Search.}  For a detection experiment toward a carbon star, the broader linewidths, and accompanying unresolved hyperfine structure, may slightly offset the increased value of $Q$.  The small $B$ value likely necessitates a cm-wave observation.  Molecules in these sources are seen both in extended emission in the expanding envelope and surrounding gas, and in compact emission nearer the star \citep{Cernicharo:2013cc}, thus a combined search with both an interferometer and a single-dish facility may be required.  

\textbf{Dark Cloud Search.}  For a detection experiment toward a dark cloud, the very narrow linewidths will likely resolve much of the hyperfine structure, reducing the overall intensity.  The decreased value of $Q$ at low temperatures may help offset this, but nevertheless a high sensitivity will likely be required in narrow spectral channels and high resolution, indicating long integration times.  At these even lower temperatures, observations in the cm-wave are more or less mandatory.  Molecules in these sources tend to be quite extended, and thus interferometers will resolve out most signal, requiring a single-dish facility to observe.

\textbf{Verdict?}  A search in either a carbon star or a dark cloud would seem logical, given the unsaturated nature of the molecule.  Insight from a chemical model as to an approximate abundance may help to break the tie.  Any search will need to be done at cm-wavelengths, and given that both source types likely require a single-dish measurement, that would seem a logical place to start.

\subsection{Searching for \ce{CH3CH2CH2OH}}

The detections of methanol (\ce{CH3OH}; \S\ref{CH3OH}) and ethanol (\ce{CH3CH2OH}; \S\ref{CH3CH2OH}), along with their  large abundances, makes the next largest fully-saturated alcohol, propanol (\ce{CH3CH2CH2OH}) a logical target, and it has indeed been recently searched for without success \citep{Muller:2016kd}.  The rotational spectrum of \ce{CH3CH2CH2OH} is known \citep{Kisiel:2010gw}. Table~\ref{propanol} below lists a few key considerations for the molecule, and the corresponding effects on its likely detectability.

\begin{table*}[b!]
    \centering
    \caption{Attributes of \ce{CH3CH2CH2OH} and their effects on detectability.}    
    \begin{tabular}{p{2.8in} p{3in}}
        \hline\hline
        \ce{CH3CH2CH2OH} ...                            &    Effects         \\
        \hline
        ... is nearly prolate ($\kappa$~=~-0.84)        &    Most ISM molecules are prolate or near prolate - high symmetry reduces $Q$, generally increasing line intensity \\
        ... is fully saturated                          &    Fully saturated molecules are most commonly observed in SFRs and LOS clouds\\
        ... is large (12 atoms, 60 amu)                 &    Large molecules are rarely found in LOS clouds, and have large partition functions \\
        ... has an internal methyl rotor                &    $Q$ will be increased and the line intensities decreased if the splitting is resolved in observations\\
        ... has modest rotational constants (5--10~GHz) &    The primary rotational transitions will occur in the mm- and lower sub-mm wavelength ranges      \\
        ... has five structural conformers              &    Some population may be spread out into these other conformers, reducing the overall intensity of the main conformer's lines     \\
        ... has low-lying vibrational states            &    The contribution of these states to $Q$ is likely to be non-trivial \\
         \hline
    \end{tabular}
    \label{propanol}
\end{table*}

The fully saturated nature of the molecule suggests SFRs and LOS clouds are the most likely sources for detection, but the size of the molecule strongly favors SFRs (\S\ref{source_types}).  The linewidths in these sources are modest depending on whether single-dish or array observations are used ($\sim$1--5~km~s$^{-1}$), but these are far broader than the expected internal rotor splitting, so this should not greatly affect the observations.  The temperature in these sources, however, is often quite warm (80--300~K).  This can non-trivially populate both the low-lying vibrational states of the lowest energy conformer, but also populate the other conformers of the molecule as well, all with cumulative decreases in the population of the lowest conformer in its ground vibrational state.  At these temperatures, the strongest transitions will fall around 200~GHz, and will extend into the sub-mm.

\textbf{Verdict?}  Given the expected low line-intensity due to the large partition function and the existence of multiple low-lying vibrational states and conformers, individual lines are expected to be quite weak.  At (sub)mm-wavelengths, single-dish spectra of SFRs are often line-confusion limited, making the identification of weak features challenging.  The compact nature of these sources also causes single-dish observations to suffer from beam dilution effects.  Interferometric observations are likely better suited to these observations, as they generally result in narrower linewidths that permit the observation of weaker lines before line-confusion sets in.  Choosing a target SFR that is on the cooler side would help push the population toward the ground vibrational state and lowest energy conformer.

\section{Python 3 Script}
\label{app:script}

\setcounter{figure}{0}    
\setcounter{table}{0} 

As Supplemental Information, a Python 3 script containing nearly all of the data presented here is available.  The file hosted by the Journal is a snapshot of the script at time of submission.  A live version is accessible at \href{https://github.com/bmcguir2/astromolecule_census}{\url{https://github.com/bmcguir2/astromolecule_census}}.  The preamble of the script contains a substantial block of text outlining the capabilities, restrictions, and cautions for use.  Here, a few simple possible use cases are outlined.

The code is intended for use in an interactive Python 3 environment.  It has been tested in IPython 6.1.0 using Python 3.6.3.  It may work in other Python environments as well, potentially with some minor customization.  It is unlikely to readily work in Python 2.7, and is in fact hard-coded to exit if it detects Python 2.7 in use.

The entirety of the information content of the script is contained in \texttt{Molecule} and \texttt{Source} custom classes.  The \texttt{Molecule} class has 53 non-trivial attributes, while the \texttt{Source} class has 6.  Not all attributes are required, and many are not currently in use, but are in place for future development.  The default value for most is \texttt{None}.

Many utility functions are included with the script.  Depending on when in development they were written, several functions may do the same thing as part of their execution in several different ways.  Some are still in development, may not be fully implemented, or may need additional manipulation of their output before it is useful for its intended purpose.  Users are strongly encouraged to read a function carefully before using, or to write their own.

The \texttt{summary(y)} function returns nearly the entire database's worth of information for a \texttt{Molecule} or \texttt{Source} \texttt{y}.  For example, issuing\\

\begin{figure*}
\includegraphics[width=\textwidth]{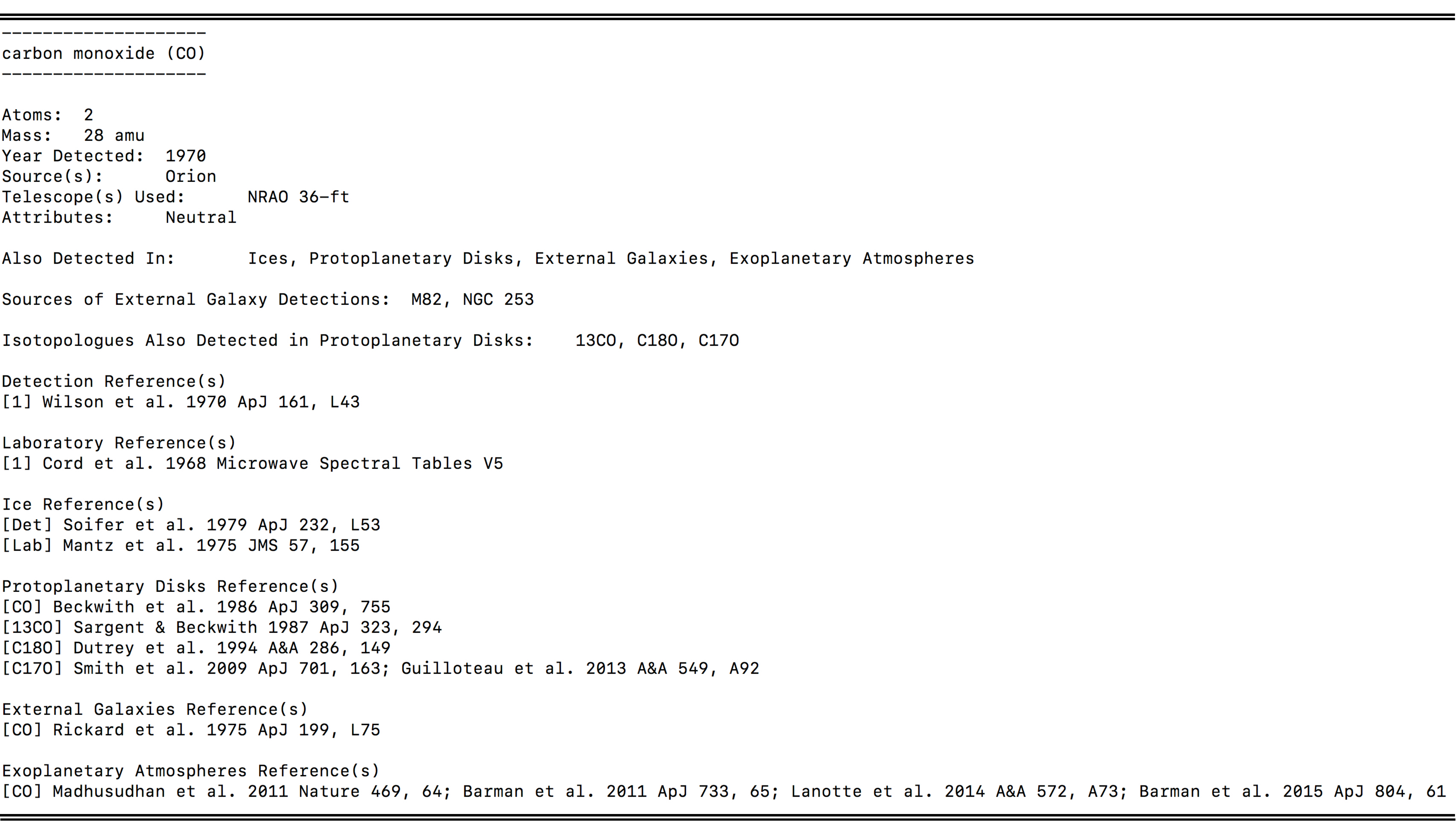}
\caption{Results of issuing the command \texttt{summary(CO)}.}
\label{summary:co}
\end{figure*}

\indent \texttt{>> summary(CO)}\\

\noindent produces the output shown in Figure~\ref{summary:co} for the databases entry for carbon monoxide (CO).  Similarly, for a source like IRC+10216, issuing\\

\begin{figure}
\includegraphics[width=\columnwidth]{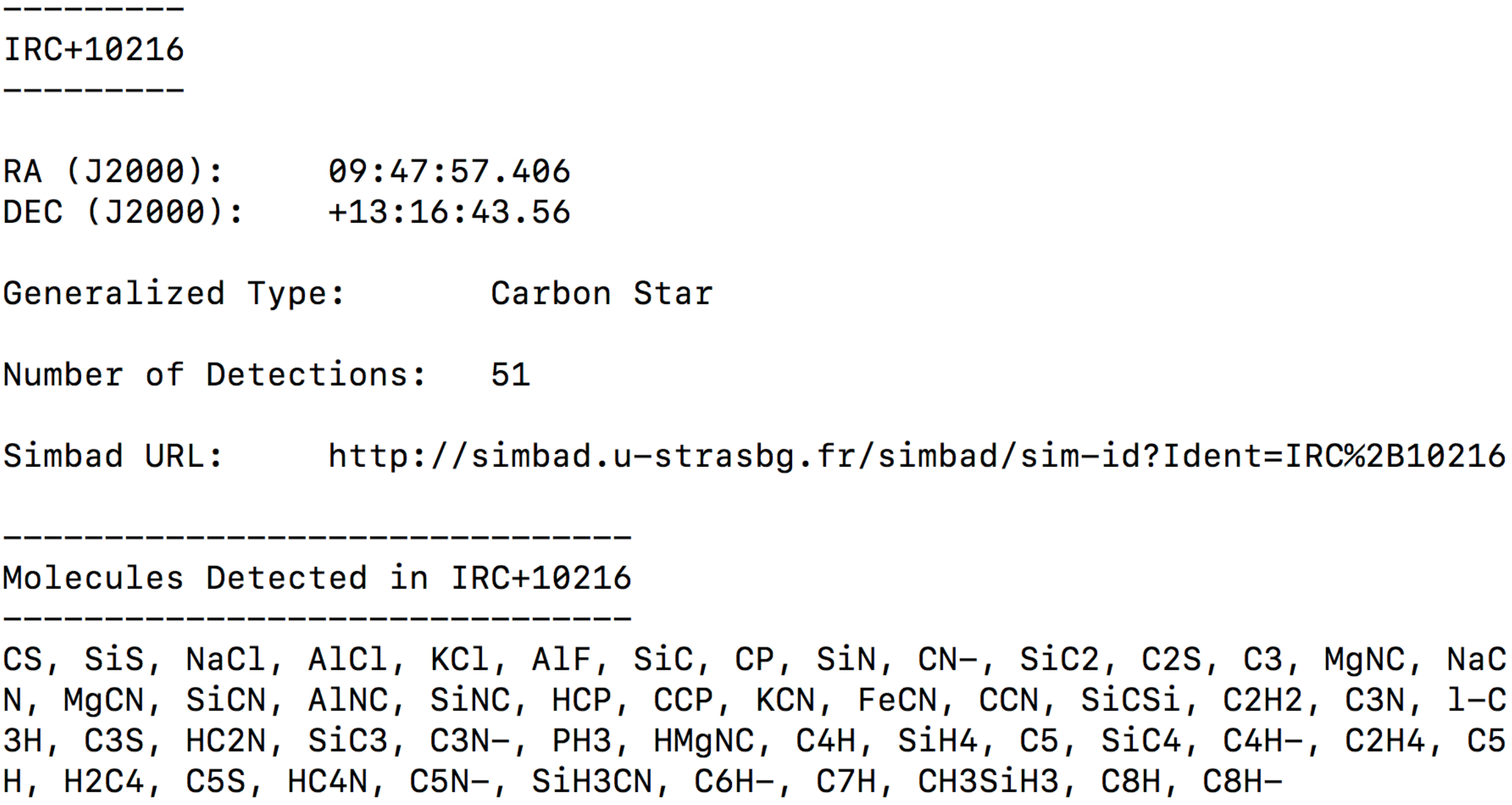}
\caption{Results of issuing the command \texttt{summary(IRC10216)}.}
\label{summary:irc}
\end{figure}

\indent \texttt{>> summary(IRC10216)}\\

\noindent produces the output shown in Figure~\ref{summary:irc}.

The script contains hard-coded python lists that contain all molecules and all sources, with variable names of  \texttt{full\_list} and \texttt{source\_tag\_list}, respectively.  These can be iterated over to find molecules or sources that satisfy a desired set of conditions.  For example, to find all molecules containing sulfur that were first detected in Sgr B2, the loop below could be used.

\begin{verbatim}
for x in full_list:
    if x.S > 0 and `Sgr B2' in x.sources:
        print(x.formula)
\end{verbatim}

\clearpage


\end{document}